\documentclass[twocolumn]{aastex631}

\usepackage{amsmath}
\usepackage{ctable}
\usepackage{graphicx}
\usepackage{natbib}
\usepackage{epsfig}
\usepackage{enumitem}
\usepackage{longtable}

\begin{document}

\title{DUSTY STELLAR SOURCES CLASSIFICATION BY IMPLEMENTING MACHINE LEARNING METHODS BASED ON SPECTROSCOPIC OBSERVATIONS IN THE MAGELLANIC CLOUDS}
\shorttitle{Dusty Stellar Spectral Classiﬁcation Through Machine Learning}
\shortauthors{Ghaziasgar et al.}

\author[0000-0001-5855-846X]{Sepideh Ghaziasgar}
\affiliation{School of Astronomy, Institute for Research in Fundamental Sciences (IPM), P.O. Box 1956836613, Tehran, Iran}

\author[0009-0006-3280-5622]{Mahdi Abdollahi}
\affiliation{School of Astronomy, Institute for Research in Fundamental Sciences (IPM), P.O. Box 1956836613, Tehran, Iran}

\author[0000-0001-8392-6754]{Atefeh Javadi}
\affiliation{School of Astronomy, Institute for Research in Fundamental Sciences (IPM), P.O. Box 1956836613, Tehran, Iran}

\author{Jacco Th. van Loon}
\affiliation{Lennard-Jones Laboratories, Keele University, ST5 5BG, UK}

\author[0000-0003-0356-0655 ]{Iain McDonald}
\affiliation{Jodrell Bank Centre for Astrophysics, Alan Turing Building, University of Manchester, M13 9PL, UK}

\author[0000-0002-0861-7094]{Joana Oliveira}
\affiliation{Lennard-Jones Laboratories, Keele University, ST5 5BG, UK}

\author{Amirhossein Masoudnezhad}
\affiliation{Department of Physics, Sharif University of Technology, P.O. Box 11155-9161, Tehran, Iran}

\author{Habib G. Khosroshahi}
\affiliation{School of Astronomy, Institute for Research in Fundamental Sciences (IPM), P.O. Box 1956836613, Tehran, Iran}
\affiliation{Iranian National Observatory, Institute for Research in Fundamental Sciences (IPM), Tehran, Iran}

\author[0000-0002-5110-1026]{Bernard H. Foing}
\affiliation{Leiden Observatory, Leiden University, P.O.Box 9513, 2300 RA Leiden, Netherlands}

\author{Fatemeh Fazel}
\affiliation{Leiden Observatory, Leiden University, P.O.Box 9513, 2300 RA Leiden, Netherlands}
%
%

\correspondingauthor{Atefeh Javadi}
\email{atefeh@ipm.ir}

\begin{abstract}
Dusty stellar point sources are a significant stage in stellar evolution and contribute to the metal enrichment of galaxies. These objects can be classified using photometric and spectroscopic observations using color-magnitude diagrams (CMD) and infrared excesses in spectral energy distributions (SED). We employed supervised machine learning spectral classification to categorize dusty stellar point sources, including young stellar objects (YSOs) and evolved stars comprising oxygen- and carbon-rich asymptotic giant branch stars (AGBs), red supergiants (RSGs), and post-AGB (PAGB) in the Large and Small Magellanic Clouds, based on spectroscopic labeled data derived from the Surveying the Agents of Galaxy Evolution (SAGE) project, which involved 12 multiwavelength filters and 618 stellar objects. Despite dealing with missing values and uncertainties in the SAGE spectral datasets, we achieved accurate classifications of these sources. To address the challenge of working with small and imbalanced spectral catalogs, we utilized the Synthetic Minority Oversampling Technique (SMOTE), which generates synthetic data points. Subsequently, among all the models applied before and after data augmentation, the Probabilistic Random Forest (PRF) classifier, a tuned Random Forest (RF), demonstrated the highest total accuracy, reaching $\mathbf{89\%}$ based on the recall metric in categorizing dusty stellar sources. In this study, using the SMOTE technique does not improve the accuracy of the best model for the CAGB, PAGB, and RSG classes; it stays at $\mathbf{100\%}$, $\mathbf{100\%}$, and $\mathbf{88\%}$, respectively. However, there are variations in the OAGB and YSO classes. Accordingly, we collected photometrically labeled data with properties similar to the training dataset and classified them using the top four PRF models with an accuracy of more than $\mathbf{87\%}$. We also collected multiwavelength data from several studies to classify them using our consensus model, which integrates four top models to present common labels as the final prediction.
\end{abstract}
\keywords{stars: classiﬁcation --- stars: AGB and post-AGB --- galaxies: spectroscopic catalog --- methods: data analysis --- methods: machine learning}
\section{Introduction}\label{introduction}
Dusty stellar objects are significant in the chemical enrichment of the interstellar medium (ISM) and galaxies, which contain heavy elements required for star and planet formation. These objects are point sources surrounded by gas and dust that are important for a wide range of processes, including the loss of mass from evolved stars \citep{2018A&AR-massloss-agb-hofner} and the formation of these stars. These types of stars are easily detectable due to their high luminosity $(\sim 10^{3.5-5.5}\  \textup{L}_\odot)$ and red colors \citep{karakas-2014PASA...31...30K}, making them stand out in galaxies, especially in the infrared (IR) observations.
The Magellanic Clouds (MCs), comprising the Large and Small Magellanic Clouds (LMC and SMC), are nearby dwarf irregular galaxies, respectively 50 kpc and 60 kpc away \citep{Pietrzy-lmc-2013Natur.495...76P, LMC-SMC-2009A&A...496..399S,Subramanian-2011ASInC...3..144S} with metallicities of 0.5 and 0.2$Z_{\odot}$ \citep{1992ApJ-metalicity-russel} is an excellent location to investigate stellar contributions to dust emission \citep{2015MNRAS.451.3504R}. In addition, since the MCs are close to the Milky Way, their stellar populations and dusty stellar sources can be distinguished with considerable accuracy.
The point sources in these galaxies can be identified photometrically according to their color-magnitude diagrams (CMDs). However, photometric identification alone may not provide sufficient information to distinguish between classes of dusty stars. To enable more accurate stellar classification, infrared spectra obtained by the Infrared Spectrograph (IRS) aboard the {\it Spitzer} Space Telescope \citep{IRS2004ApJS} have been used to classify approximately 1000 point sources across the LMC and SMC \citep{2010PASP..122..683K,2011MNRAS.411.1597W, 2015MNRAS.451.3504R, 2017MNRAS.470.3250J}.
Dusty stellar point sources are classified according to their evolutionary stage and chemical enrichment, comprising young stellar objects (YSO), asymptotic giant branches (AGB; oxygen- and carbon-rich), red supergiants (RSG), and post-asymptotic giant branches (PAGB). While other dusty stellar sources, such as AGNs \citep{Hony2011, Pennock-2022MNRAS} and OB stars \citep{Adams-2013ApJ...771..112A, Sheets-2013ApJ...771..111S}, may potentially have IR excesses, this study's scope is centered around the five dusty stellar classes discussed due to the availability of the IRS spectral labeled data.\\
\indent
The early-evolutionary star is known as a YSO. It can be a valuable probe for studying recent star-formation activity in galaxies \citep{2023ApJ-yso-starformation}, along with their physical conditions and processes. Due to the star formation process, YSOs are surrounded by protostellar envelopes containing gas and dust disks \citep{2016JASS...33..119S}. Therefore, YSOs contribute to our understanding of star formation and allow us to characterize star-forming regions. In addition, the YSOs can have varying luminosities based on their mass and evolution stage, which can be classified based on their spectral class from optical to IR and luminosity \citep{2009ApJ-seale-ys0-class, oliveira-2013MNRAS.428.3001O, 2018Ap&SS-Miettinen-ml-yso}. \\
\begin{figure*}[ht!]
	{\hbox{
		\epsfig{figure=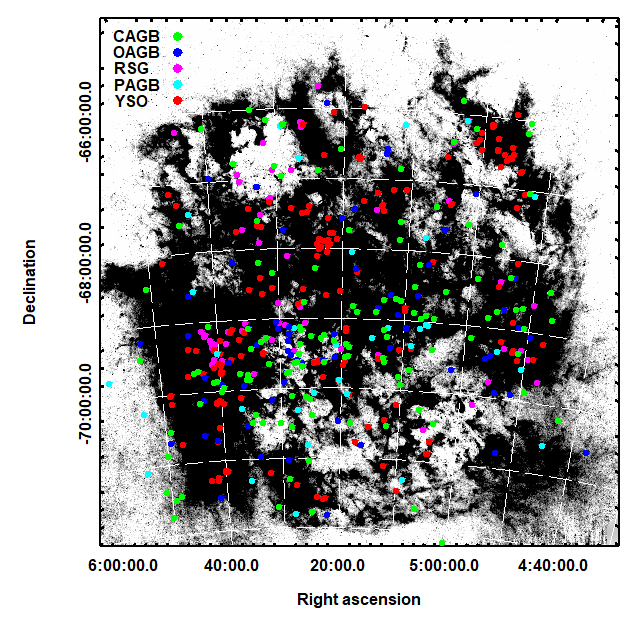,width=90mm,height=90mm}
            \epsfig{figure=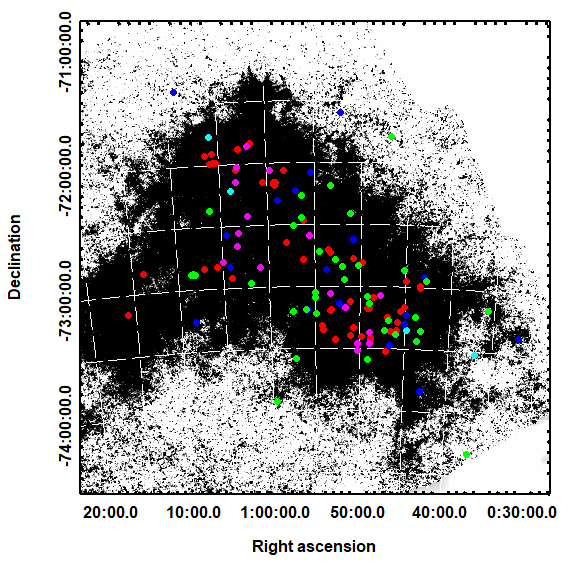,width=90mm,height=90mm}
        \centering
	}}
	\caption{Left panel: Locations of dusty sources in the Large Magellanic Cloud (LMC) based on {\it Spitzer}/Infrared Spectrograph (IRS) data, including sources from both the SAGE-Spec program and archival observations. Right panel: Locations of dusty sources in the Small Magellanic Cloud (SMC) selected from IRS staring-mode archival data within the SAGE-SMC footprint. The background images of the Magellanic Clouds were taken from the {\it Herschel} Space Observatory \citep{2010-Herschel}.}
    
	\label{fig:region}
\end{figure*}
The evolved stars \citep{Boyer2011b} are dust producers, and their role in enriching galaxies is crucial. In the final stage of stellar evolution, low– and intermediate mass (0.8-8 \(\textup{M}_\odot\)) stars evolve into the Asymptotic Giant Branch (AGB) phase, and high mass (M$\ge$ 8 \(\textup{M}_\odot\)) stars enter the Red Supergiants (RSG) phase \citep{2005ARA&A-falk-ahb, 2018A&AR-massloss-agb-hofner}. During the AGB phase, a fraction of the mass of the stars is lost at rates of up to $\sim 10^{-4} \, M_{\odot} \, \text{yr}^{-1}$ \citep{1992ApJ-wood-variable-mass-loss,Srinivasan2009AJ}, leading to the chemical enrichment of the interstellar medium (ISM). Most AGB stars are distinguished as long-period variables (LPVs) with pulsation periods in the range of months to years \citep{1983ApJ-wood-variable, Whitelock-2003MNRAS.342...86W,2009AJ-vijh-variable-sage,2011MNRAS.411..263J, 2012ApJ-lmc-mass-loss}, and they have circumstellar dust envelopes with high mass-loss rates \citep{2011MNRAS.414.3394J,2016ApJ-McDonald, Goldman-2017MNRAS.465..403G,2018A&AR-massloss-agb-hofner,2022IAUS..366..210J}.
According to the chemistry of the photosphere and the outer dust envelope and their carbon-to-oxygen abundance ratios, carbon-rich AGB stars (C-rich or CAGB stars; \( C/O > 1 \)), oxygen-rich AGB stars (O-rich or OAGB stars; \( C/O < 1 \)), and S-type stars \(( C/O \sim 1 \)) can be categorized. Hence, the OAGB stars have more oxygen than carbon, forming silicates and oxides, while CAGB stars, where carbon exceeds oxygen due to the third dredge-up process \citep{2020JApA-kmath-agb-postagb}, produce carbon-based molecules. As a result, these stars differ in spectral characteristics and dust production due to their initial mass and metallicity \citep{2011MNRAS.411..263J, 2011MNRAS.414.3394J, 2013MNRAS.432.2824J,2020ApJ...891...43S, 2021ApJS..256...43S}.\\
\indent
The RSGs evolved from the Main sequence O-types or B-types with initial masses ($\sim$8-30 $\textup{M}_\odot$) in the helium-burning evolutionary phase. During this phase, there is a significant increase in size and a decrease in surface temperature. Therefore, they are at a key stage in stellar evolution, and lastly, their core collapses to produce supernovae, neutron stars, or black holes. \citep{2003AJ-massey-redsuperginat,2010ASPC-Levesque-redsupergiant, yang2020-smc}.\\
\indent
The PAGB stars, in the late stages of their evolution, are luminous, low- or intermediate-mass stars that conclude their AGB branch evolution with a phase of mass loss. Their transition from giant branches to white dwarfs after the oxygen-rich and carbon-rich AGB phases \citep{Kamath2014MNRAS, Kamath2015MNRAS}. Having shed their outer layers, these stars are characterized by unique chemical compositions. These stars provide information about nucleosynthesis processes in AGB stars and how galaxies evolve chemically \citep{2003ARA-vanWincke, 2020JApA-kmath-agb-postagb}.\\
\indent
In recent years, the growth of astronomical datasets, including photometric and spectroscopic catalogs, has increased thanks to ground- and space-based telescopes and surveys. In this way, machine learning methods \citep{2010IJMPD-Data-Mining, 2019arXiv190407248B} have emerged as powerful tools for classifying astronomical objects due to their significant time-saving capabilities. Therefore, the stellar classification using artificial intelligence techniques, including machine learning and deep learning, is becoming increasingly important, and there are many studies regarding this trend \citep{2016AA-singleband-setllar-clssification,2018MNRAS-ml-variablestar,2018Ap&SS-Miettinen-ml-yso,2019AJ-morgan-stellar-classification,2021A&A-Cornu-nuralnetwork,2021ApJ-phtometric-classification, 2021MNRAS-Jacco2021-6822,2022arXiv-ghaziasgar,2022AA-ml-massive-classification,2022MNRAS-jacco2022-m33,2023arXiv-mahdipaper, 2023RSOS-nural_network,2023MNRAS-yso-ml-naive-bayse,ian-10.12688/openreseurope.17023.1,2024MNRAS-ml-classification-china, BAO-2024,Pennock-2025, 2025-ml-stellar-astronomy}.\\
\indent
Machine learning algorithms are commonly categorized into supervised and unsupervised methods, each offering advantages and limitations. In supervised learning, models are trained using labeled datasets to learn patterns from known classifications. This approach can provide accurate results, mainly when reliable labeled data is available \citep{2019arXiv190407248B, 2025-ml-stellar-astronomy}. Supervised learning is limited in identifying new objects because its performance depends on the quality and completeness of the training data \citep{Pennock-2025}. Additionally, the classifier's performance may be affected if the training dataset is imbalanced, meaning that some classes contain significantly more samples than others. Furthermore, supervised learning requires a large amount of labeled data, which is difficult since labeling stellar sources is time-consuming \citep{Pennock-2025,2025-ml-stellar-astronomy}. In contrast, unsupervised learning is employed when no predefined labels exist, making it a powerful tool for discovering hidden structures within the data \citep{2010IJMPD-Data-Mining, 2014sdmm.book.....I, 2019arXiv190407248B, 2022ExA-bigdata, AI-Astronomy-2022}. However, a significant limitation of unsupervised learning is its lack of astrophysical interpretability \citep{2024-unsupervised}, making it less suitable for our dataset. Given the labeled nature of the dataset, supervised learning is the most appropriate choice. Furthermore, since our objective is to refine existing classifications rather than identify new subpopulations, a supervised approach aligns better with the goals of this study.\\
\indent
The dataset used in this study consists of spectroscopically confirmed classifications of LMC and SMC point sources, initially identified photometrically using {\it Spitzer} imaging (see Section~\ref{sec:Data}) \citep{2006AJ-Meixner}. Using the multiwavelength infrared survey, these sources were first identified through the Surveying the Agents of Galaxy Evolution (SAGE) program  \citep{godron-smc-2011AJ....142..102G}. Spectroscopic follow-ups were performed through the targeted SAGE-Spec {\it Spitzer } Legacy program and additional archival IRS observations compiled into the broader SAGE-Spec \citep{2010PASP..122..683K} database. While not all photometrically identified sources in SAGE have spectroscopic follow-up, this study exclusively used sources with confirmed spectroscopic classifications. Ongoing and future spectroscopic observations are expected to further improve the classification and understanding of stellar objects and dusty stars by increasing the volume of labeled data. This will enable more accurate machine learning analyses on dusty stellar populations, allowing models to identify such sources better and enhance classification performance \citep{jones-jwst-nature}. Mid-infrared stellar population studies are typically limited to nearby galaxies; however, the James Webb Space Telescope (JWST) will extend these capabilities to galaxies across the Local Volume \citep{2017ApJ...841...15J}.\\
\indent
In this study, we implemented supervised learning algorithms on the IRS spectral dataset to distinguish young stellar objects (YSOs) and evolved stars. The models were trained on spectroscopically confirmed data and evaluated for classification performance.\\
\indent
The organization of the study is as follows: we delineate the archival dataset for training and testing in Section \ref{sec:Data}. In Section \ref{sec:data preprocessing}, the method designed for preprocessing is proposed. Section \ref{sec:Models} details some models for multiclassification described. We present classification results in Section \ref{sec:Results}. The metallicity effect on classifying dusty stellar sources in the MCs is given in Section \ref{sec:Metallicity Impact}. In Section \ref{sec:photometric data}, we compare the classification results obtained from the newly collected dataset with its corresponding photometric classifications. Finally, in Section \ref{sec:Conclusion}, our findings are summarized.
\section{Data}\label{sec:Data}
The Surveying the Agents of Galaxy Evolution (SAGE) collaboration \citep{2006AJ-Meixner} used the {\it Spitzer} Space Telescope \citep{2004-Werner-Spitzer} to obtain infrared imaging and spectroscopic data for the Magellanic Clouds, using the Infrared Array Camera (IRAC; \citet{2004-Fazio-IRAC} 3.6, 4.5, 5.8, and 8.0 $\mu m$ bands) and the Multiband Imaging Photometer (MIPS; \citet{2004-Rieke-MIPS} 24, 70, and 160 $\mu m$ bands). Observations were conducted in two epochs, three months apart.\\
\indent The spectroscopic data in this study come from the SAGE-Spec \citep{2010PASP..122..683K} database, which includes all {\it Spitzer}/IRS staring-mode observations within the regions covered by SAGE-Spec (for the LMC) and the SAGE-SMC survey. This database combines data from the SAGE-Spec Legacy program \citep{2010PASP..122..683K}, which targeted 197 point sources \citep{2011MNRAS.411.1597W} in the LMC, and additional archival IRS observations from various {\it Spitzer} programs. Nearly 800 point sources were observed in the LMC, resulting in over 1000 IRS spectra \citep{2017MNRAS.470.3250J}. For the SMC, 209 point sources within the SAGE-SMC footprint were observed using IRS staring-mode \citep{2015MNRAS.451.3504R}.\\ 
\indent We used the spectroscopically labeled sources published in the classification catalogs of the LMC and SMC \citep{2011MNRAS.411.1597W, 2015MNRAS.451.3504R, 2017MNRAS.470.3250J}, whose spatial distribution is shown in Fig.~\ref{fig:region}.\\
\indent The spectral classes (SpClass) of dusty stellar objects in both the LMC and SMC catalogs were derived using a structured binary decision tree (yes/no) algorithm, as illustrated in the classification flowcharts by \citep{2011MNRAS.411.1597W, 2015MNRAS.451.3504R, 2017MNRAS.470.3250J}. In multiclassification studies, decision trees expand binary decisions (yes/no branches) into multi-way decisions, and objects are classified starting at the root node and sorted according to their feature values \citep{Murthy1998AutomaticCO, Rokach2005TopdownIO}. This algorithm consists of many steps that bring us closer to the final label, such as redshift, Spitzer infrared spectral features, continuum, spectral energy distribution shape, and bolometric luminosity. For instance, in the top part of the SAGE-Spec classification flowchart, the galaxy and star branches are distinguished according to redshift features. To continue, the stars are subdivided into multiple branches by considering different features and thresholds. The SAGE spectral catalog contains 1071 objects, including stars, galaxies, and planetary nebulae \citep{2011MNRAS.411.1597W, 2017MNRAS.470.3250J}. Among these, 618 objects are classified as dusty stellar sources (see Table~\ref{table:1}), with 486 belonging to the LMC and 132 to the SMC. The remaining objects, such as galaxies and planetary nebulae, are present in the catalog but excluded from our analysis, since this study focuses exclusively on dusty stellar sources.
\begin{figure}[t]
    \centering
    \includegraphics[width=1\linewidth, clip]{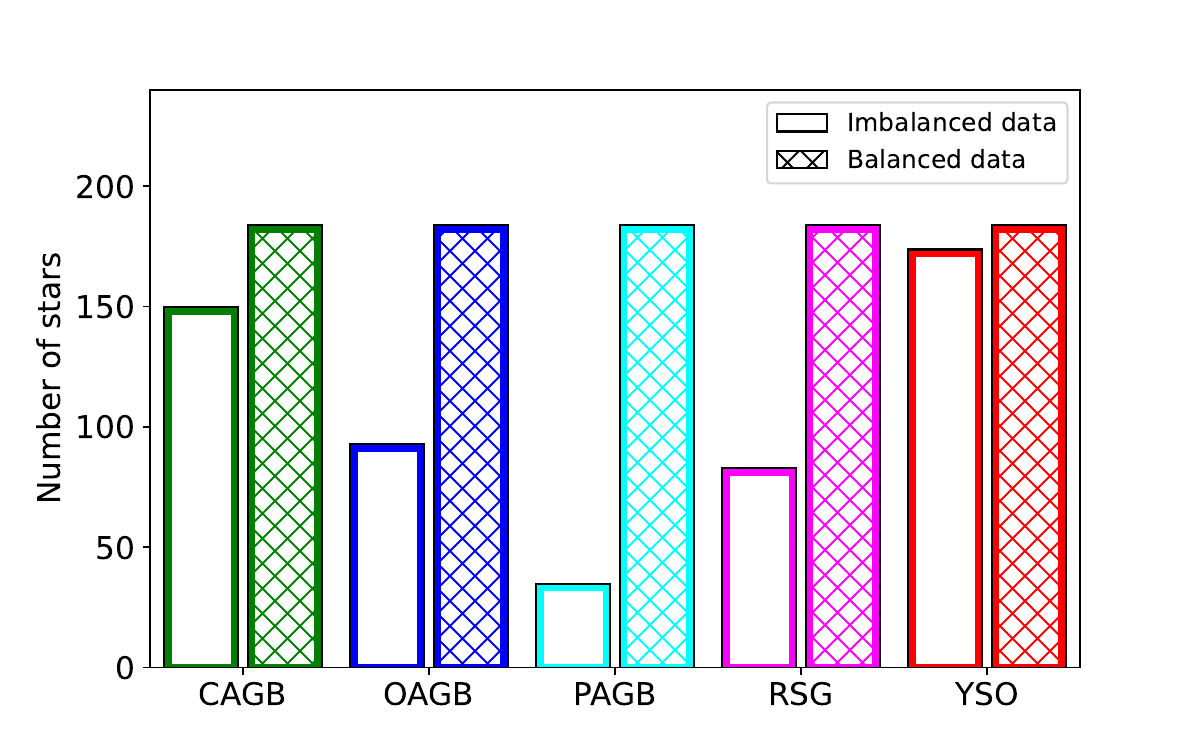}
    \caption {Dusty stellar sources distribution for the training dataset before and after data augmentation within the Synthetic Minority Oversampling Technique (SMOTE).}
    
    \label{fig:SC_hist}
\end{figure}
\begin{table}
\centering
     \caption{This is a list of dusty stars within the Magellanic Clouds, categorized by their spectral classes. *The ``Augmented Data'' column represents the population after data augmentation using the SMOTE method. Each number indicates the sum of the train and test datasets in each class, with SMOTE applied only to the train datasets to equalize their population to the most populated class.}
\label{table:1} 
\begin{tabular}{ l c c c r }
    \hline 
    Classes & LMC & SMC & Total & *Augmented\\
        &   &   &   &Data\\
    \hline  
    CAGB & 136 & 38 & 174 & 200\\
    OAGB & 88  & 19 & 107 & 193 \\
    PAGB & 33  & 4  & 37  & 183\\
    RSG  & 72  & 22 & 94  & 190\\
    YSO  & 157 & 49 & 206 & 206\\
    \hline
    Total & 486 & 132 & 618 & 972\\
    \hline
\end{tabular}
\end{table}
\begin{table*}
\centering
\caption{Sample of spectral properties of detected dusty stellar point sources in the Magellanic Clouds, derived from {\it Spitzer}/IRS observations. The data for the LMC include sources from both the SAGE-Spec {\it Spitzer} Legacy program and archival IRS observations within the SAGE-LMC footprint \citep{2017yCat..74703250J}. For the SMC, sources were selected from the complete IRS archive based on 209 IRS staring-mode observations of IRAC point sources located within the SAGE-SMC region \citep{2015yCat..74513504R}. This table presents five representative dusty stellar sources from the Magellanic Clouds across 12 observational filters, with distance modulus values adopted from \citep{2016ApJ...816...49S,2016AJ....151...88B} and extinction corrections from \cite{2011ApJ-extinction-smc-lmc}. The RA and DEC are included for positional reference and are not used as classification features.}
\small 
\setlength{\tabcolsep}{3.3pt} 
    \begin{tabular}{lcccccccccccccr}
        \hline
        RA &DEC &{\it U} &{\it B} &{\it V} &{\it I} &{\it J} &{\it H} &{\it K$_s$} &[3.6]&[4.5]&[5.8]&[8.0]&[24]&SpClass\\
        (deg)&(deg)&(mag)&(mag)&(mag)&(mag)&(mag)&(mag)&(mag)&(mag)&(mag)&(mag)&(mag)&(mag)&\\
        \hline
        71.6131&-68.7963&1.63&2.23&0.53&-2.14&-3.50&-5.24&-6.84&-8.67&-9.39&-9.81&-10.24&-11.19&CAGB\\
        71.8277&-69.7057&-0.73&-2.88&-4.8&-7.22&-8.42&-9.25&-9.51&-9.65&-9.48&-9.65&-9.76&-10.97&RSG\\
        72.1573&-69.3936&NaN&3.45&2.47&1.14&NaN&NaN&-5.60&-8.15&-9.04&-9.84&-10.65&-13.66&YSO\\
        16.0398&-72.8377&-2.2&-3.75&-5.56&NaN&-8.64&-9.32&-9.59&-9.7&-9.71&-9.92&-10.05&-10.8&OAGB\\
        15.9264&-72.2284&-5.30&-5.16&-5.06&-5.39&-5.89&-6.37&-7.20&-8.99&-9.75&-10.36&-11.42&-13.68&PAGB\\
        \hline
    \end{tabular}
\label{tab:Samples}
\end{table*}
\begin{figure*}[ht!]
	{\hbox{
		\epsfig{figure=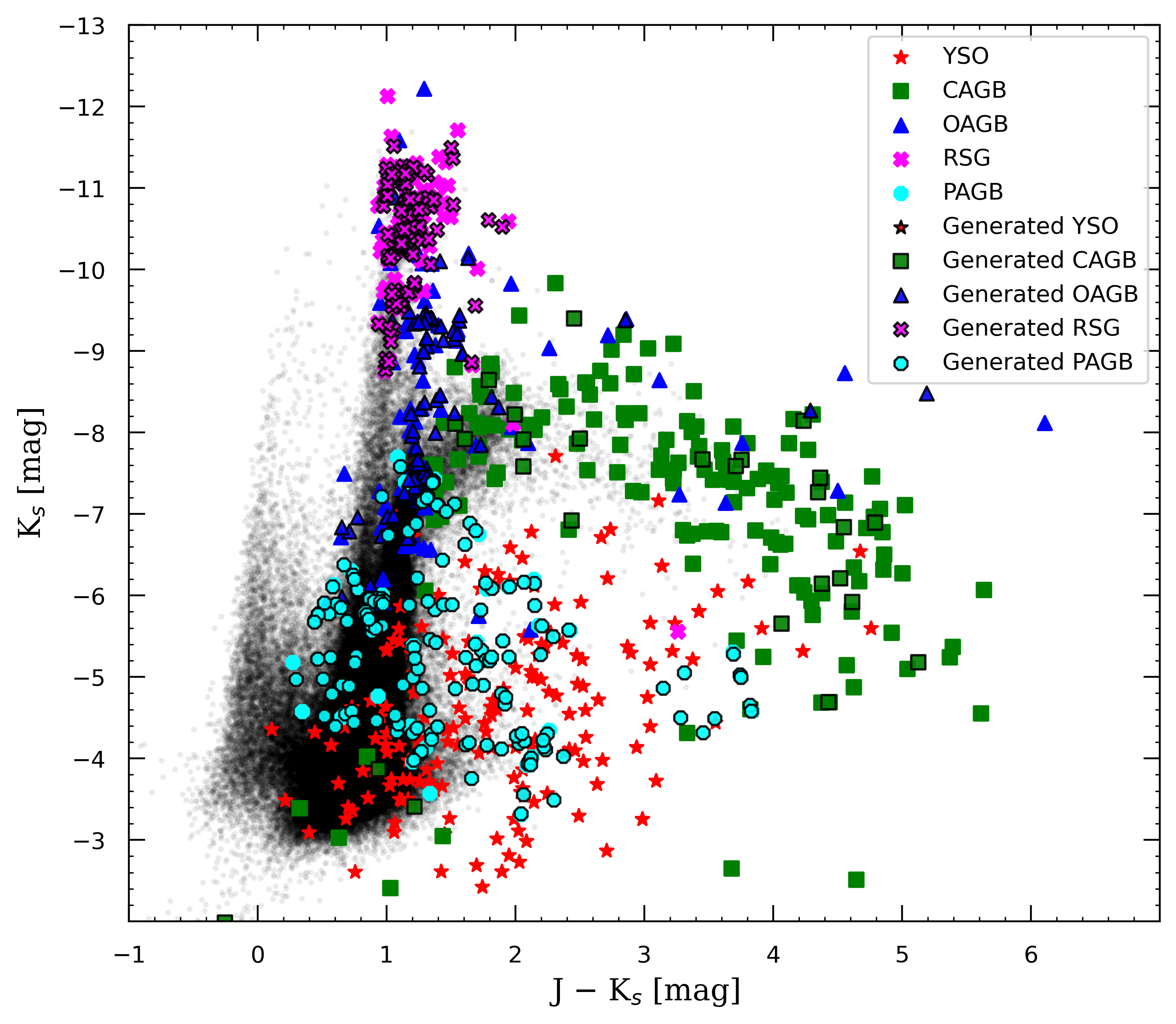,width=90mm,height=90mm}
  
        \epsfig{figure=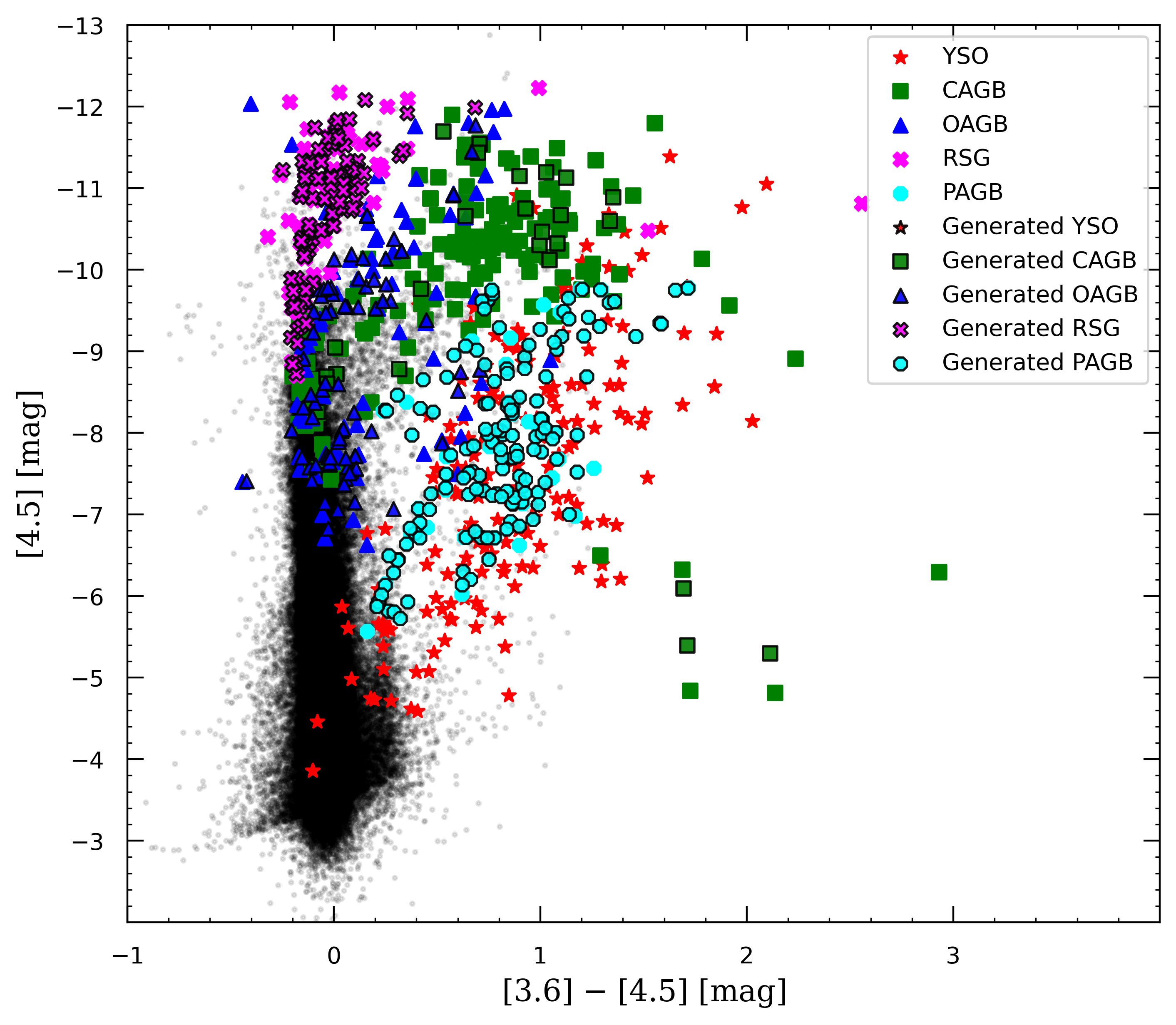,width=90mm,height=90mm}
   \centering
	}}
	\caption{Color-magnitude diagram (CMD) plots the magnitude of stars against their color, representing the locations of each dusty stellar class of object. As shown in this figure, dusty stellar categories are shown along with those generated with the SMOTE method as discussed in Section~\ref{subsec:Imbalanced data}, in the near and mid-infrared bands. The left panel shows near-infrared data from 2MASS ({\it J}, {\it K$_s$)}, and the right panel shows mid-infrared data from {\it Spitzer} ([3.6], [4.5]). The black lines around stellar types indicate data generated by SMOTE, and the black dots in the background represent all the stars that belong to the Magellanic Clouds.}
	\label{fig:CMD_Smote}
\end{figure*}
\section{DATA PREPROCESSING and Exploratory Data Analysis}\label{sec:data preprocessing}
In preprocessing, we perform a series of steps to prepare the data for model training. For classification, we have collected two datasets from LMC and SMC \citep{2017yCat..74703250J,2015yCat..74513504R} that include several columns of data. These columns consist of optical to far-infrared magnitudes gathered from different telescopes, such as 2MASS and Spitzer. Additionally, the dataset consists of other features calculated from various studies. In general, many features provide more information for model training but may increase the complexity of training classifiers. It can also be challenging to handle NaN records due to the many missing values, which require oversampling algorithms to fill the gaps. Hence, we faced these issues and solved them step by step, as presented in the following:
\begin{enumerate}[leftmargin=*]

\item The primary dataset contains 14 features, including UMmag, BMmag, VMmag, IMmag, J2mag, H2mag, Ks2mag, IRAC1, IRAC2, IRAC3, IRAC4, \texttt{[24]}, \texttt{[70]}, \texttt{[160]}.

\item Exploring the data, we found some columns in our dataset containing many NaN records. Consequently, columns with more than 500 NaNs (more than 70\% of all records) have been removed, which includes columns \texttt{[70]} and \texttt{[160]}. As a result, 12 features remain.

\item Drawing the correlation matrix (presented in Appendix~\ref{sec:Correlation Matrix}), we used this matrix to identify correlated features and removed those with a high correlation. Therefore, correlated features do not assist the model in learning and can sometimes complicate the learning process, potentially leading to instability in model parameters \citep{hastie2009elements}. According to Fig.~\ref{fig:Correlation matrix of 12 features}, closely related features (filters) demonstrate high correlation and overlap during preprocessing. This is expected due to the behavior of blackbody radiation of stars, which rises at shorter wavelengths and falls at longer ones \citep{1986rpa..book.....R,Carroll_Ostlie_2017}, resulting in lower correlation where the difference between two features is greater. As a result, all selected features exhibit an acceptable correlation matrix, indicating that no features need to be eliminated.
\item There are still many NaNs in these remaining features, and we need to find a solution. Thus, we neglected stellar sources with more than six NaNs. This threshold is chosen based on the size of the dataset and the improvement in training performance. Finally, after removing some dusty objects, we selected 618 stars with some remaining NaN values. Table~\ref{table:1} presents the final population of each class in the used dataset. In detail, the sample of each class with its features is shown in Table~\ref{tab:Samples}.
\item To fill in the remaining missing values, we applied an IterativeImputer technique \citep{scikit-learn} from the \texttt{scikit-learn} library in Python to handle them. For each missing feature in the dataset, the IterativeImputer constructs a regression model utilizing the remaining features as predictors. This model approximates the missing values for each feature, and the process is repeated until no further approximations can be made. Applying the approximated missing values to the subsequent iteration makes the approximations more precise with each iteration \citep{scikit-learn}.
\item Finally, we applied extinction \citep{2011ApJ-extinction-smc-lmc} and distance modulus (DM) \citep{2016ApJ...816...49S,2016AJ....151...88B} to all 12 filters at the Magellanic Clouds.
\end{enumerate}

\indent After these preprocessing steps to handle and fill in the missing data, the dataset should be split into training and testing sets. Typically, different ratios are used for training and testing to evaluate the model’s performance, and the choice of split ratio depends on the dataset and its distribution. In our case, we partitioned the preprocessed dataset by allocating 85\% of the data to training and 15\% to testing.

\indent
As an optional step, we can use the Exploratory Data Analysis (EDA) approach, which includes statistical and visual tools, to understand the data better. One of these tools is the pairplot (presented in Appendix~\ref{sec:Pairs Plot}), which visualizes features in various colors according to their relationship. In consequence, Fig.~\ref{fig:pair_plot} is drawn based on the multiwavelength features selected for training the model. Since dusty point sources overlap with different classes, they cannot be distinguished by a single or two particular features, making classification challenging.
%
\subsection{Data Augmentation}\label{subsec:Imbalanced data}
The data we used does not have an equal distribution of stars among the five spectral classes. This indicates that we have imbalanced classes of dusty stellar point sources, as illustrated in Fig.~\ref{fig:SC_hist}. We can expect the class distribution to be imbalanced and the skewness to be present when we have actual data \citep{2021arXivLiu-imballanced}.\\
\indent
The imbalanced distribution can pose challenges during model training, as it may bias the models toward more populated classes. This bias occurs because the models learn more from classes with larger populations \citep{Sun2009ClassificationOI}. Thus, classifier models may find it more difficult to effectively learn from the minority class, which contains significantly fewer samples compared to the majority class.\\
\indent
It is possible to solve the problem of imbalanced data distribution in each class and balance samples among classes using a variety of resampling techniques, including undersampling (reducing samples from majority classes), oversampling (adding additional samples to minority classes) \citep{2021arXivLiu-imballanced, 2023PASA-imballanced}, and hierarchical classification \citep{BaderElDen2016HierarchicalCF,2020MNRAS-Imbalancelearning,2023arXiv-mahdipaper}. \\
\indent
Typically, oversampling is performed with the Synthetic Minority Oversampling Technique (SMOTE), which generates random synthetic data from the nearest neighbors of a minority class based on Euclidean distance \citep{2011arXiv1106.1813C}. Through SMOTE, a few samples are linearly interpolated into their neighbors, and a certain number of artificial minority samples are generated to reduce the data imbalance ratio \citep{2021NatSR-SMOTE}.\\ 
\indent
We applied this data augmentation method to the training dataset to address imbalanced distributions within stellar classes, as shown by bars with black diagonal cross lines in Fig.~\ref{fig:SC_hist}. Also, Table~\ref{table:1} shows the final population of dusty stellar objects with imbalanced training and test data in a column named total and balanced data in an augmented column. To understand how SMOTE works and how it increases the number of stars in each stellar class, see the CMDs comparison in near-IR ({\it J}, {\it K$_s$}) and mid-IR [3.6], [4.5] passbands before and after applying SMOTE shown in Fig.~\ref{fig:CMD_Smote}. Therefore, the graph shows that the SMOTE method generates the minority class, namely PAGB stars, based on the population of the YSOs, which is the majority class. As a result, the augmented PAGB stars are generated near the previous PAGB stars, where the YSOs are located. After generating the data, it is clear that there is an overlap between these two classes, YSO (red stars) and PAGB (cyan circles), which may cause issues with the model training.
\section{Models}\label{sec:Models}
In machine learning, the selection of models depends on the nature of the problem, data characteristics, and especially data size \citep{2023RSOS-nural_network}. For instance, neural networks typically require large datasets (with tens of thousands of samples) to learn effectively, due to their high complexity. Since this study deals with a low stellar population, we chose classical algorithms that can provide reliable classification even with sparsely populated data.\\
\indent The models we used as described (see~Appendix \ref{sec:SUPPORTING_INFORMATION}) included Probabilistic Random Forest (PRF) \citep{2019AJ....157...16R,2019arXiv190407248B,2021MNRAS-Jacco2021-6822, 2022MNRAS-jacco2022-m33, Pennock-2025}, Random Forest (RF) \citep{2001MachL..45....5B, 2010ApJ-randomforest-redshift,2017MNRAS-rf-dalya,2019arXiv190407248B}, K-Nearest Neighbor (KNN) \citep{knn1992}, C-Support Vector Classification (SVC) including SVC-poly and SVC-rbf \citep{vapnik95,2019arXiv190407248B}, and Gaussian Naive Bayes (GNB) \citep{2023MNRAS-yso-ml-naive-bayse}.\\
\indent When dealing with training models, each model has hyperparameters, such as the number of estimators in the RF model and neighbors in the KNN model. Therefore, we employed manual and automatic adjustments using the ``Grid Search'' \citep{scikit-learn} from the \texttt{scikit-learn} library to find the optimal hyperparameters of the model. The grid search assigns possible values to each one to determine which hyperparameter performs best.\\
\indent The PRF, which is a developed version of Random Forest (see Appendix~\ref{sec:RF}), generates a forest of decision trees where each tree is trained using a different subset of the data and a random set of features. It assigns probability distributions to each output class, as illustrated in \cite{2019AJ....157...16R}, and can handle noisy datasets. Despite constant hyperparameters, different answers are obtained in the PRF model. By selecting various hyperparameters and finding the best model obtained \cite{Pennock-2025}, we explore several models to address the uncertainty in this problem.\\
\indent In this study, we train and test the models with two kinds of datasets, imbalanced data and balanced data augmented by the SMOTE method, as mentioned in Section~\ref{subsec:Imbalanced data}. In the next section, we will present the results in two approaches, Simple (referred to as imbalanced data) and SMOTE (referred to as balanced data using the augmented technique), named based on the distribution of each of the datasets and shown in Fig.~\ref{fig:SC_hist}.
%
\section{Classification results} \label{sec:Results}
\begin{figure}[t]
    \centering
    \includegraphics[width=0.95\linewidth, clip]{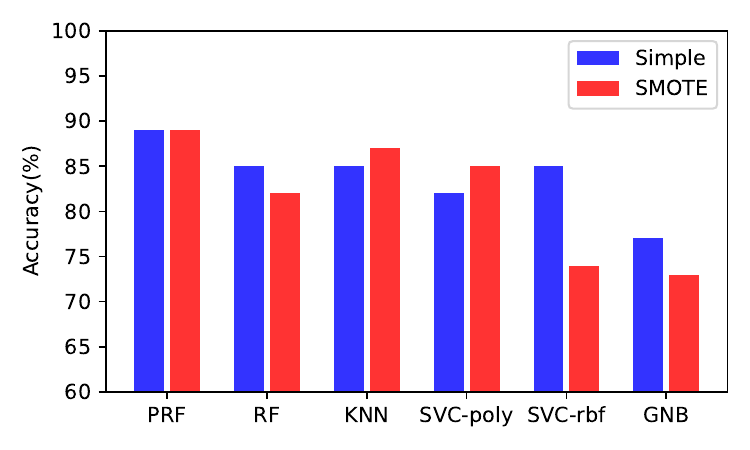}
    \caption {This chart compares the performance of a Simple classifier with a SMOTE classifier based on their respective accuracy scores.}
    \label{fig:hist_model_results}
\end{figure}

This section compares the classification of dusty stellar objects before and after using the SMOTE method, and the classifiers in each approach are evaluated using the performance metrics \citep{2020MetricsFM} and the confusion matrix presented in Appendix \ref{sec:Performance Metric} and \ref{sec:Confusion_Matrix}.\\
\indent As shown in Fig.~\ref{fig:hist_model_results}, the comparison of both Simple and SMOTE approaches in six different classification models indicates that the SMOTE approach outperforms the Simple ones in two of them. In comparison, the situation is reversed for the three other models, including RF, SVC-rbf, and GNB. Using the SMOTE method can sometimes result in misclassification, especially if the original sample point is near the edge of the minority sample distribution, and may not always produce an improved classification \citep{2021NatSR-SMOTE, 2017arXiv-SMOTE2}.

According to Fig.~\ref{fig:CM_Set_1} and Fig.~\ref{fig:CM_Set_2}, we also find that increasing the population of the sparsely populated star class results in misclassification in the class that overlaps with them. In the case of the OAGB star class, which is a third less populated class with 96 stars, applying data augmentation through the SMOTE method results in increased overlaps with the CAGBs (as shown in Fig.~\ref{fig:CMD_Smote}). Based on the comparison of SVC-rbf and GNB models before and after data generation, it can be seen that some CAGBs are classified as OAGBs by models that use augmented data. These misclassifications are reversed among OAGB, PAGB, and RSG stars. As a result, some OAGB stars are classified as PAGBs and RSGs, as shown in the confusion matrix of RF and SVC-rbf models.\\
\begin{figure}[t]
	\centering
     \includegraphics[width=0.95\linewidth, clip]{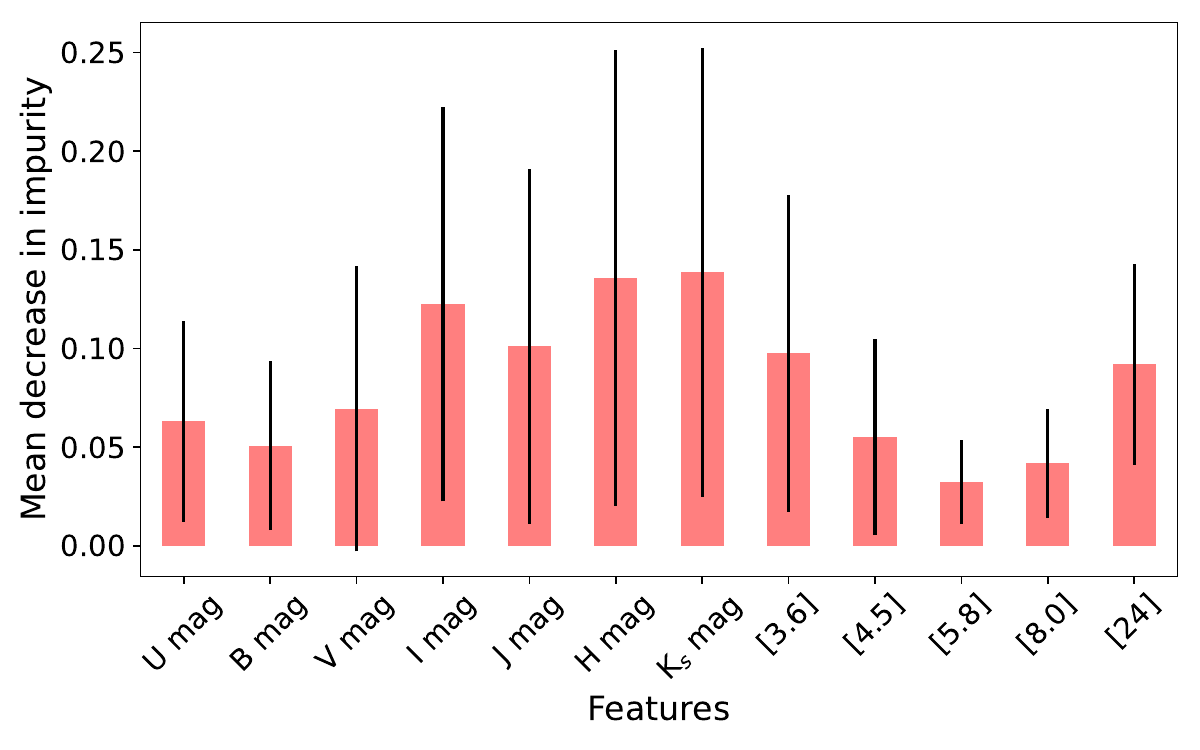}
      \caption {The feature importance diagram ranks the significance of different variables (or features) using the RF model's Mean Decrease in Impurity (MDI) algorithm \ref{sec:RF}. This diagram illustrates how important each of the filters in the SAGE catalog is for making accurate stellar classification predictions. As shown, the importance of the stellar classes and the selected features (12 filters) are all the same, and no feature is preferred.}
      \label{fig:feature_importance_Simple}
\end{figure}
\begin{table}
	\centering
	\caption{PRF models comparison for Simple and SMOTE in terms of accuracy.}
        \centering
	\begin{tabular}{l c c r} 
		\hline
		Number of & Keep- & Accuracy & Accuracy \\
            estimator & probability & (Simple) & (SMOTE) \\
		\hline
            10 & 0.01 & 89 & 87 \\
            10 & 0.1 & 89 & 89 \\  
            10 & 0.3 & 89 & 87 \\
            10 & 0.5 & 87 & 89 \\
            10 & 0.8 & 89 & 85 \\
            10 & 0.9 & 89 & 85 \\
            100 & 0.1 & 85 & 87 \\
		\hline
    \end{tabular}
    \label{tab:PRFs}
\end{table}
Comparing the models in each approach, the results obtained from PRF models were superior to others, with the best record of $89\%$ accuracy. As shown in Table~\ref{tab:PRFs}, the PRF model was trained with different values for the hyperparameters, keep\_probability, and number of estimators due to the uncertainty we found in them. The best PRF model classification reports in each approach are shown in Tables~\ref{tab:PRF_simple} and \ref{tab:PRF_smote}. Considering two models with the highest accuracy in different approaches, we find that the Simple approach shows better results in terms of macro average accuracy at $83\%$. The other comparison method utilizes the confusion matrix, as illustrated in \ref{fig:CM_Set_1}. The accuracy of the three classes, including CAGB, PAGB, and RSG, does not change when switching from the Simple to the SMOTE approach. However, the accuracy of OAGB decreases from $73\%$ to $64\%$ due to classifying some stars as RSGs, which occurs more frequently than in the Simple approach. This could result from data augmentation of other classes, especially the RSG class, which is close to OAGB, as shown in the CMDs in Fig. \ref{fig:CMD_Smote}. In contrast, the accuracy of YSOs increases from $88\%$ to $92\%$, indicating that data augmentation could work better for certain classes. It could be the result that data augmentation does not work well for all classes, and its effectiveness varies depending on the class and the specific problem.\\
As shown in Table~\ref{tab:PRFs}, several PRF models have the same results. Therefore, we selected four of them from both approaches for classification and comparison in Section~\ref{sec:photometric data}.\\
\indent Regarding accuracy, the RF model is one of the accurate models. This model performed well in the Simple approach but not in the SMOTE one. In addition, this model provides a report showing the importance of each feature.
Indeed, identifying the most important features in machine learning helps streamline the problem by eliminating useless ones. The RF classifier can employ the Mean Decrease in Impurity (MDI) algorithm \citep{2001MachL..45....5B, MDIFeatureImportance-2019} to assess the importance of features. The MDI for each feature is calculated by averaging the decrease in impurity over all trees in the ensemble. Features with high MDI values are more important for prediction. As Fig.~\ref{fig:feature_importance_Simple} demonstrates, based on the MDI, on the Y axis, there is no significant difference in the importance of used features; we did not remove any feature based on this calculation. However, it does reveal that the significance of some infrared passbands is greater than that of others, as expected, given that dusty stellar objects exhibit excess in this wavelength range.
\begin{table}
	\centering
	\caption{Classification report, considering the number of estimators: 10, Keep probability: 0.8, Simple PRF.}
\label{tab:PRF_simple}
	\begin{tabular}{l c c r} 
		\hline
		Class & Precision & Recall & F1-score \\
		\hline
            CAGB & 0.95 & 1.00 & 0.97\\
OAGB & 0.80 & 0.73 & 0.76\\  
PAGB & 0.50 & 1.00 & 0.67\\
RSG & 0.78 & 0.88 & 0.82\\
YSO & 0.95 & 0.88 & 0.91\\
     \hline
accuracy & & & 0.89 \\
macro avg & 0.80 & 0.90 & 0.83 \\
weighted avg & 0.89 & 0.89 & 0.89 \\
		\hline
    \end{tabular}
\end{table}
\begin{table}
	\centering
	\caption{Classification report, considering the number of estimators: 10, Keep probability: 0.5, SMOTE PRF.}
	 \label{tab:PRF_smote}
	\begin{tabular}{l c c r} 
		\hline
		Class & Precision & Recall & F1-score \\
		\hline
        CAGB & 0.86 & 1.00 & 0.92\\
        OAGB & 1.00 & 0.64 & 0.78\\  
        PAGB & 0.50 & 1.00 & 0.67\\
        RSG & 0.70 & 0.88 & 0.78\\
        YSO & 1.00 & 0.92 & 0.96\\
        \hline
accuracy & & & 0.89 \\
macro avg & 0.81 & 0.89 & 0.82 \\
weighted avg & 0.91 & 0.89 & 0.89 \\
		\hline
    \end{tabular}
\end{table}
\section{Metallicity Impact on the Classification of Dusty Stellar Sources in the Magellanic Clouds} \label{sec:Metallicity Impact}
\begin{figure}
  \centering
  \includegraphics[width=1\linewidth]{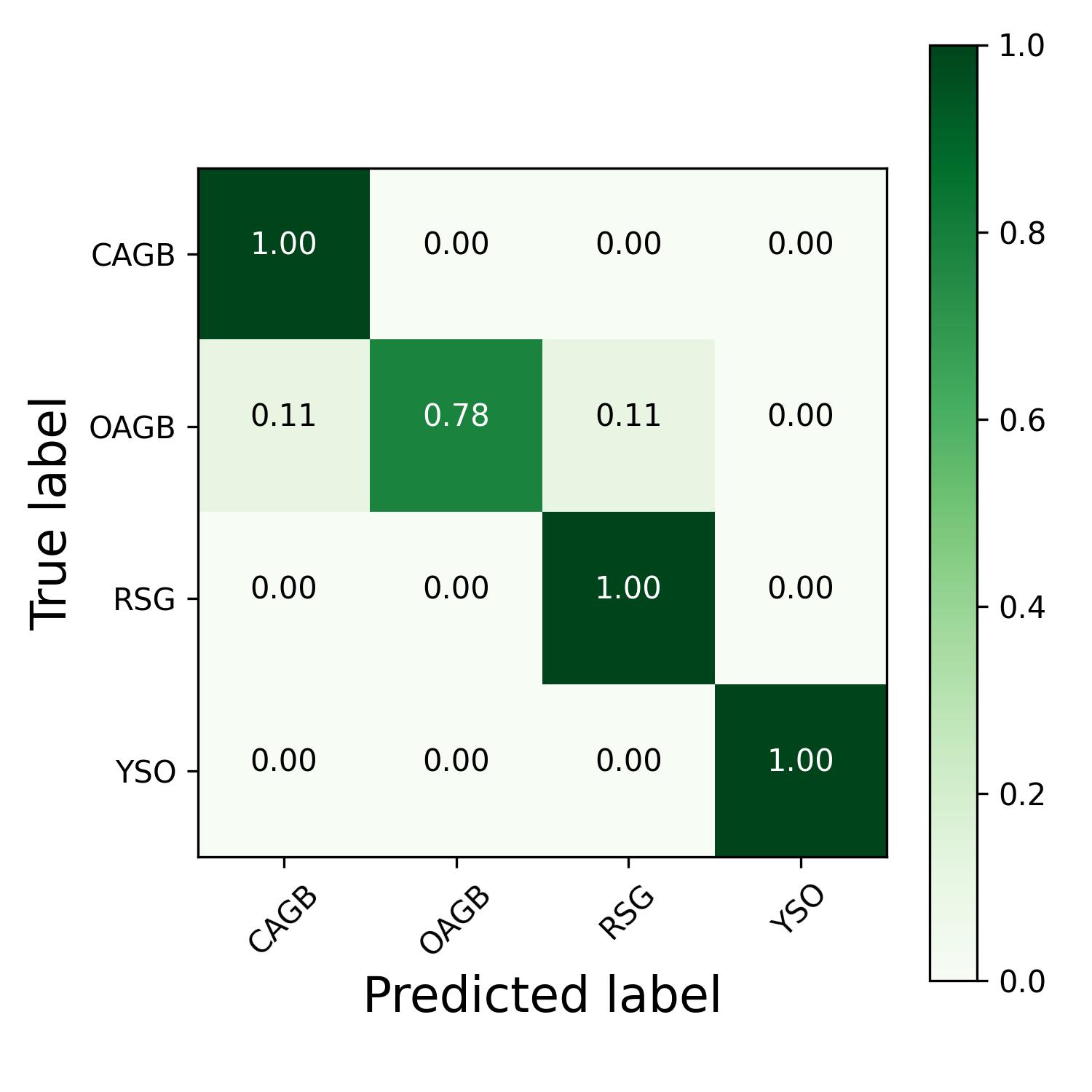}  
  \caption{The confusion matrix obtained from the PRF model trained on the LMC dataset and tested on the SMC dataset, including four populated stellar classes.}
  \label{fig:CM_LMC_train_SMC_test}
\end{figure}

\indent The Large and Small Magellanic Clouds have different metallicity \citep{1992ApJ-metalicity-russel, godron-smc-2011AJ....142..102G}, as mentioned in Section~\ref{introduction}. To assess the impact of these metallicity variations on machine learning-based classification models, we conducted separate training and testing experiments for each galaxy. The PRF model, identified earlier as the best-performing classifier, was used to evaluate classification performance under these conditions. The dataset utilized in this work consists of a stellar sample from the LMC and SMC, as summarized in Table~\ref{table:1}. In general, we consider them as a single dataset. However, in this section, we analyze them separately to assess the impact of metallicity.\\
\indent Beginning with the SMC, the PRF successfully classified four out of five stellar classes. However, PAGB classification failed entirely, with most samples misclassified as YSO. Despite achieving an overall accuracy of 93\%, as shown in Table~\ref{tab:classification_report-SMC} and Appendix~\ref{fig:CM-Metallicity}, the inability to classify PAGB sources correctly limits result interpretability. The primary challenge in SMC classification was the limited number of PAGB samples. With only four PAGB sources available, manual test allocation was necessary, using three for training and one for testing. Additionally, the total sample size in SMC was limited to 132. The SMOTE algorithm, commonly used for data augmentation, was not applicable due to its requirement of at least six samples per class.\\
\indent In contrast, LMC contained 486 dusty stellar sources, providing a more balanced dataset across classes. This reduced some classification difficulties and enabled more stable training compared to SMC. For the LMC, the PRF achieved an accuracy of 88\%, as presented in Table~\ref{tab:classification_report-LMC} and Appendix~\ref{fig:CM-Metallicity}. This result aligns closely with those obtained from the combined LMC and SMC data, indicating that metallicity does not affect the classification. This outcome was anticipated, as the combined sample is predominantly composed of LMC data. However, challenges remained for the classification of less populated classes. While 67\% of PAGB stars were correctly classified, some were misclassified as YSO. \\
\begin{table}
    \centering
   \caption{Classification report for the SMC catalog. The classification was performed using the following settings: number of estimators = 10, Keep probability = 0.8, and Simple PRF.
       \label{tab:classification_report-SMC}}
    \begin{tabular}{lcccc}
        \hline
       Class & Precision & Recall & F1-Score \\
       \hline
        CAGB & 1.00 & 1.00 & 1.00 &  \\
        OAGB & 1.00 & 1.00 & 1.00 &  \\
        PAGB & 0.00 & 0.00 & 0.00 &  \\
        RSG & 1.00 & 1.00 & 1.00 &  \\
        YSO & 0.80 & 1.00 & 0.89 &  \\
        \hline
        accuracy &  &  & 0.93 &  \\
        macro avg & 0.76 & 0.80 & 0.78 &  \\
        weighted avg & 0.88 & 0.93 & 0.90 &  \\
        \hline
   \end{tabular}
\end{table}
\begin{table}
    \centering
    \caption{Classification report for the LMC catalog. The classification was performed using the following settings: number of estimators = 10, Keep probability = 0.8, and Simple PRF.}
      \label{tab:classification_report-LMC}
   \begin{tabular}{lcccc}
        \hline
        Class & Precision & Recall & F1-Score \\
       \hline
      CAGB & 0.94 & 0.88 & 0.91 & \\
        OAGB & 0.72 & 0.90 & 0.86 & \\
        PAGB & 0.67 & 0.67 & 0.67 &  \\
        RSG & 1.00 & 0.90 & 0.95 &  \\
       YSO & 0.80 & 0.80 & 0.88 & \\
       \hline
      accuracy & &  & 0.88 &  \\
      macro avg & 0.84 & 0.85 & 0.84 & \\
       weighted avg & 0.80 & 0.88 & 0.88 &\\
      \hline
   \end{tabular}
\end{table}
\indent As the results indicated, PAGB stars were misclassified due to their low population. To address this and to ensure a stable and fair comparison, we conducted additional assessments excluding PAGB stars. We trained the best models for these assessments using a dataset containing only the four remaining classes.\\
\indent In the first approach, the master dataset was divided into two datasets for the SMC and LMC, and models were trained separately for each. The results, presented in Table~\ref{tab:classification_report_SMC_4class} and Table~\ref{tab:classification_report_LMC_4class}, showed identical classification accuracies of 92\% for the SMC and LMC. Next, we trained and tested the best model using the master dataset, combining data from four LMC and SMC classes. The results remained consistent at 92\%. The close agreement between these results suggests that stellar objects in the four classes can be accurately classified in the MCs, whether the data are analyzed separately or combined. The corresponding confusion matrices are presented in Figure~\ref{fig:CM-Metallicity-4Class}. Additionally, the findings indicate that separating the LMC and SMC datasets based on metallicity differences does not significantly impact the classification of dusty stellar sources. \\
\indent In the second approach, we again used the four-class dataset. We trained a model using the LMC training dataset and evaluated it with the SMC test data. The results, presented in Table~\ref{tab:classification_report_LMC_Train_SMC_test_4class} and Fig~\ref{fig:CM_LMC_train_SMC_test} as a confusion matrix, were compared with those obtained from the model trained on the SMC dataset in the previous approach (Table~\ref{tab:classification_report_SMC_4class}). In this case, the training datasets came from two locations with different metallicities, which should have led to differences in training and, consequently, in evaluation results. Yet, the results remained unchanged, suggesting that metallicity has no significant impact on model training for the LMC and SMC.\\
\indent Finally, we assessed the impact of metallicity on classification and found it insignificant when using the four-class dataset. However, our master dataset includes five classes, one of which—PAGB—has a low population. To address this, we combined the two datasets, as discussed in Section~\ref{sec:data preprocessing}, to enhance the sample size. As an additional verification step, we examined the absolute CMDs of the SMC and LMC to investigate potential differences in stellar populations. The strong agreement observed between their stellar populations suggests that combining data from these two sources is justified.
%
\section{Comparison with Photometric Classification}
\label{sec:photometric data} 
Dusty stellar sources can be classified using photometric techniques \citep{2021ApJ-phtometric-classification,2022AA-ml-massive-classification} that distinguish stellar properties from circumstellar dust. For the photometric classification of such sources, observations are conducted within IR wavelengths because of the presence of infrared excess, a characteristic of dusty stellar environments due to dust grains absorbing ultraviolet and visible light and re-emitting it in the infrared. In this way, it is possible to detect dust disks and envelopes surrounding stars by comparing the observed IR luminosity to what is expected from the star. Therefore, this technique could be useful for identifying young stellar objects (YSOs) and evolved stars.\\
\indent In addition to employing spectral features and emissions from dusty stellar sources, classifying such objects utilizes techniques such as the CMD, narrow-band filters, and blackbody deviations. These methods detect infrared excess evidence of circumstellar dust and distinguish between intrinsic stellar properties and dust influence. The following is a brief explanation of these techniques.
\indent CMD is one of the most popular methods \citep{2006AJ-cmd-sage-lmc}, which plots stars by brightness and color to display their evolutionary stages, which vary among different types of stars. As shown in Fig.~\ref{fig:SC_hist} and Fig.~\ref{fig:CMD_Smote}, it is clear that each class gathers in a distinct region of the CMD; however, there is some overlap.\\
\indent Narrow-band filters have become useful for classifying dusty stellar objects. By isolating specific wavelengths of light, these filters provide information about the properties of such objects. \cite {2004ApJ-narrowband} explores this method, highlighting its effectiveness in discerning crucial details about the composition, temperature, and distribution of dust surrounding these objects.\\
\indent Dusty stellar sources typically show deviations from blackbodies at IR wavelengths that indicate dust emission; the stellar properties of the source can be determined by fitting observed SEDs to models \citep{McDonald-2012MNRAS.427..343M, Boyer-2015ApJ...810..116B, McDonald-2017MNRAS.471..770M, McDonald-2024RASTI...3...89M}; as a result dusty sources can be classified.\\
\indent In summary, photometric methods provide a more straightforward and cost-effective approach compared to spectroscopic methods for classifying stars and galaxies. However, dusty star catalogs and stellar labels are typically photometric, which may not be as reliable as spectroscopic labels. Nonetheless, photometric labeling can provide valuable information for identifying targets, which can be verified using spectral labeling. Additionally, spectroscopic data offer accurate details about the structure and composition of the stars \citep{Riebel2010,2017ApJ...841...15J}.\\
\indent In this section, we collected the LMC and SMC photometric catalogs discussed in the context of dusty stellar categories from various works of literature, which are listed in Table~\ref{tab:testdata-table1} and Table~\ref{tab:testdata-table2}. The common characteristic among these catalogs is the data structure, similar to what we used for training models, as explained in Section~\ref{sec:Data}. However, while the labels in our work are identified by a spectroscopic method, the labels in these catalogs are obtained from the photometric methods explained earlier.\\
\indent Subsequently, we relabeled these data using models trained with spectroscopic labels and compared the new labels with the original photometric labels based on a redefined confusion matrix.\\
\indent In the following, we briefly review the photometric classifications presented in each catalog.
\begin{itemize}[leftmargin=*]
   \item Whitney et al. (2008)\\
   \cite{Whitney2008AJ} identified $\sim$1,000 YSO candidates using {\it Spitzer} photometry (SAGE Point Source Catalog) in the LMC, focusing on young high- or intermediate-mass objects. The study built a color–magnitude grid based on radiative transfer models to select YSO candidates in the LMC using filters including {\it J}, {\it H}, {\it Ks}, [3.6], [4.5], [5.8], [8.0], and [24]. Therefore, the selection criteria were refined to focus on regions of CMD space that are less confused with other IR-bright populations. In addition, the selected list was cross-correlated with other stellar population catalogs and compared with them, resulting in a final list of 1,197 YSO candidates, among which 207 were identified as non-YSOs. The study also found that these YSOs strongly correlate with 24~$\mu$m emission. Physical parameters were derived for 299 YSOs, and their SEDs fit well with radiative transfer models.
\end{itemize} 
\begin{itemize}[leftmargin=*]
    \item Srinivasan et al. (2009)\\
\cite{Srinivasan2009AJ} used the SAGE survey to classify evolved stars in the LMC, which is combined with the optical point source catalogs from 2MASS and the Magellanic Cloud Photometric Survey (MCPS; {\it U}, {\it V}, {\it B}, {\it I}) to construct of spectral SEDs. In total, 16,000 O-rich, 6300 C-rich, and 1,000 extreme sources with 8 $\mu m$ excesses were identified, whereas with 2MASS and IRAC, 4500 O-rich, 5300 C-rich, and 960 extreme sources with 24 $\mu m$ excesses were identified. Their results indicate a distinct increase in infrared excess with luminosity, more notably at 8 $\mu m$, for both oxygen-rich and carbon-rich AGB populations, a sign of circumstellar dust influence. There was also a correlation between greater optical depth and greater infrared excess in extreme AGB candidates. This study suggested the contribution of AGB stars to the LMC's mass loss and dust production, with extreme AGB stars being significant dust contributors. The study quantifies these observations with empirical relations, highlighting the role of AGB stars in the interstellar dust and gas lifecycle.
\end{itemize}
\begin{table*} 
    \centering
    \caption {The photometric distribution of dusty stars from literature reviews of the Magellanic Clouds is presented herein and in Table~\ref{tab:testdata-table2} to make the comparison with our trained classification model.}
    \begin{tabular}{l c c c c c c c r}
        \hline
Name & Location & CAGB & OAGB & PAGB & RSG &YSO &Total & Label Type \\
        \hline
        \cite{Whitney2008AJ} & LMC & & & & &360 &360 &Photometry\\    
        \cite{Srinivasan2009AJ} & LMC &6609 &34584& & & &41193&Photometry\\
        \cite{Gruendl2009} & LMC & & & & & 1090 & 1090 & Photometry \\
        \cite{Riebel2010} & LMC &116 &909 & & & &1025 &Photometry\\
        \cite{Boyer2011b} & SMC &1718 &2457 & &3271& &1025 & Photometry\\
        \cite{Kamath2014MNRAS} & SMC &43 & &14 & &38 &95 &Photometry\\
        \cite{Kamath2015MNRAS} & LMC &51 & &31 & &154 &236 &Photometry\\
        \cite{yang2018AA-lmc} & LMC & & & & 126 & &126 &Photometry\\
        \cite{Yang2019-smc} & SMC & & & & 88 & &88 &Photometry\\
        \cite{yang2020-smc} & SMC & & & & 180 & &180 &Photometry\\  
        \cite{Ya2021} & LMC & & & & 2467 & &2467 &Photometry\\  
        \hline  
    \end{tabular}
    \label{tab:testdata-table1}
\end{table*}
\begin{itemize}[leftmargin=*]
    \item Gruendl \& Chu. (2009)\\
\cite{Gruendl2009} conducted a study to discover young stellar objects (YSOs) with high and intermediate masses in the LMC using Spitzer Space Telescope data. They employed IRAC and MIPS observations to assemble a photometric catalog. This effort led to the identification of 1172 probable YSOs. Their approach included examining mid-infrared photometric data, source shapes, and their surrounding interstellar environment. By comparing their findings with those of the SAGE survey reported \citep{Whitney2008AJ}, they noted differences, such as SAGE's omission of YSOs in more complex regions. Their results underline that both catalogs have strengths and weaknesses, but together, they offer a thorough listing of YSOs in the LMC, which is crucial for studying star formation.
\end{itemize} 
\begin{itemize}[leftmargin=*]
    \item Riebel et al. (2010)\\
    \cite{Riebel2010} focused on $\sim30,000$ AGB candidates comprising oxygen-rich, carbon-rich, and extreme AGB that were identified using photometry. An infrared photometry archive available through the SAGE project and a survey of LMC variability by Massive Compact Halo Objects (MACHO) were combined to create a dataset of variable red sources. The study found that, whereas oxygen-rich and red giant branch stars display a wavelength-independent PL slope, extreme AGB stars display a wavelength-dependent PL slope by exploring the period-luminosity (PL) relationship across multiple infrared wavelengths ({\it J}, {\it H}, {\it Ks} and mid-IR bands). The \textit{Ks} band emerged as the optimal wavelength for PL relationship characterization due to its consistent results for oxygen-rich and carbon-rich AGB stars.
\end{itemize}
\begin{itemize}[leftmargin=*]
    \item Boyer et al. (2011) \\
    \cite{Boyer2011b} examined the SMC's infrared properties of cool evolved stars, focusing on RGBs, RSGs, and AGBs. Utilizing observations from the SAGE-SMC, the survey provides IR coverage [3.6]-[160] of the SMC regions. By combining near-IR and mid-IR photometry, they identified evolved stars and discovered a feature in the mid-IR CMD likely associated with particularly dusty oxygen-rich AGB stars. RSG and AGB stars contribute about 20\% of the total SMC flux at [3.6], highlighting their importance to the integrated flux of distant metal-poor galaxies. Comparisons with the high-metallicity Large Magellanic Cloud (SAGE-LMC) show that SMC's RSG stars produce less dust, as indicated by their [3.6] - [8] color. A higher fraction of carbon-rich stars in the SMC suggests efficient C-rich dust formation. Initial estimates indicate that extreme C-rich AGB stars dominate dust production in both galaxies, whereas oxygen-rich stars may play a more significant role in the LMC.
\end{itemize}
\begin{itemize}[leftmargin=*]
    \item Kamath et al. (2014) \\
    \cite{Kamath2014MNRAS} focused on optically visible candidate sources selection, including PAGB/RGB stars and YSOs in the SMC, through mid-IR observations from the SAGE survey and then assessed the 801 candidates using the low-resolution optical spectra taken through the AAOmega double-beam multifibre spectrograph mounted on the 3.9 m Anglo Australian Telescope (AAT). The final sample comprised 63 post-AGB/RGB candidates of A and F spectral class, of which 42 of these 63 sources were classified by their luminosity as post-red giant branch (post-RGB) candidates, and the remaining 21 were post-AGB candidates. This study also resulted in a new sample of 40 YSOs of A–F spectral type.
\end{itemize}
\begin{itemize}[leftmargin=*]
 \item Kamath et al. (2015)\\
 \cite{Kamath2015MNRAS} classified post-AGBs, post-RGBs, and YSO source candidates in the LMC using photometric and spectroscopic analyses based on mid-IR excess, using the same filters used in \cite[Kamath2014MNRAS], then obtained their optical spectra to confirm the existence of dusty post-RGB stars. This study identified 162 YSO candidates and 35 post-AGB candidates, including 69 hot objects with UV continuum that might be post-AGB or luminous YSO candidates. The classification contributed to understanding stellar objects in the LMC and their evolution.
\end{itemize}
\begin{itemize}[leftmargin=*]
   \item Yang et al. (2018)\\ 
\cite{yang2018AA-lmc} examined RSG stars' infrared characteristics and variability in the LMC, employing data from the ALLWISE and NEOWISE-R projects and a literature review. Based on multiwavelength data analysis, a sample of 773 RSG candidates was refined to 744, revealing a correlation between mid-infrared variability, mass loss rate, and warm dust. This paper provides insights into mass loss mechanisms, the importance of variability and luminosity at mass loss, and the evolutionary stages of RSGs in the LMC. In this study, variable and extinction factors were considered when comparing the identified RSG sample with the theoretical evolutionary models, and reduced differences were observed between observations and models.
\end{itemize}
\begin{itemize}[leftmargin=*]  
  \item Yang et al. (2019)\\
 \cite{Yang2019-smc} compiled the magnitude-limited (IRAC1 or WISE1 $\leq$ 15.0 mag) multiwavelength catalog for the SMC, counting 45,466 massive stars with low metallicity. The catalog provides broad spectral coverage from ultraviolet to far infrared, which combines data from SEIP, VMC, IRSF, AKARI, HERITAGE, and others. The study identified RSG populations using the evolutionary tracks, synthetic photometry from MESA Isochrones and Stellar Tracks, and theoretical $ J-K_s$ color cuts. As a result, candidates were ranked according to the intersection of five CMDs. Therefore, comparing the models with observations demonstrated that RSGs were separated from AGBs and that their initial mass limit was 6-7 $M_{\odot}$. Out of the total number of stars in this catalog, 1405 were identified as RSGs, and the rest as massive supergiants. 
\end{itemize}
\begin{itemize}[leftmargin=*]   
    \item Yang et al. (2020)\\ 
  \cite{yang2020-smc} centered on the RSGs in the SMC, identifying 1239 potential RSGs through a selection process based on a combination of spectroscopic data from the literature and 2MASS CMDs. The study examined infrared color-magnitude diagrams to identify around 1800 potential RSGs by distinguishing them from AGB stars. RSGs were classified based on their variability and brightness, with higher variability associated with greater mass loss. The study also estimated the total output of gas and dust produced by the RSG population and found that the temperature of the RSG population was directly proportional to the color of the $ J-K_s$, corrected for reddening.
\end{itemize}
\begin{itemize}[leftmargin=*]  
    \item Yang et al. (2021)\\ 
 \cite{Ya2021} created the magnitude--limited (IRAC1 or WISE1 $\leq$ 15.0 mag) multiwavelength source catalog for the LMC derived from the {\it Spitzer} Enhanced Imaging Products (SEIP) and Gaia Data Release 2. This catalog contains $\sim19700$ sources in 52 different bands, including 21 optical and 29 infrared bands. The study used the catalog to identify and classify $\sim$2974 red supergiants in the LMC using modified magnitude and color cuts (which represent equivalent evolutionary phases (EEPs) from core helium burning to carbon burning with 7--40 $M_{\odot}$) \cite{Yang2019-smc} in six CMDs. This work significantly contributes to understanding massive star evolution in nearby galaxies.
\end{itemize}
\begin{figure*}[ht!]
	{\hbox{
		\epsfig{figure=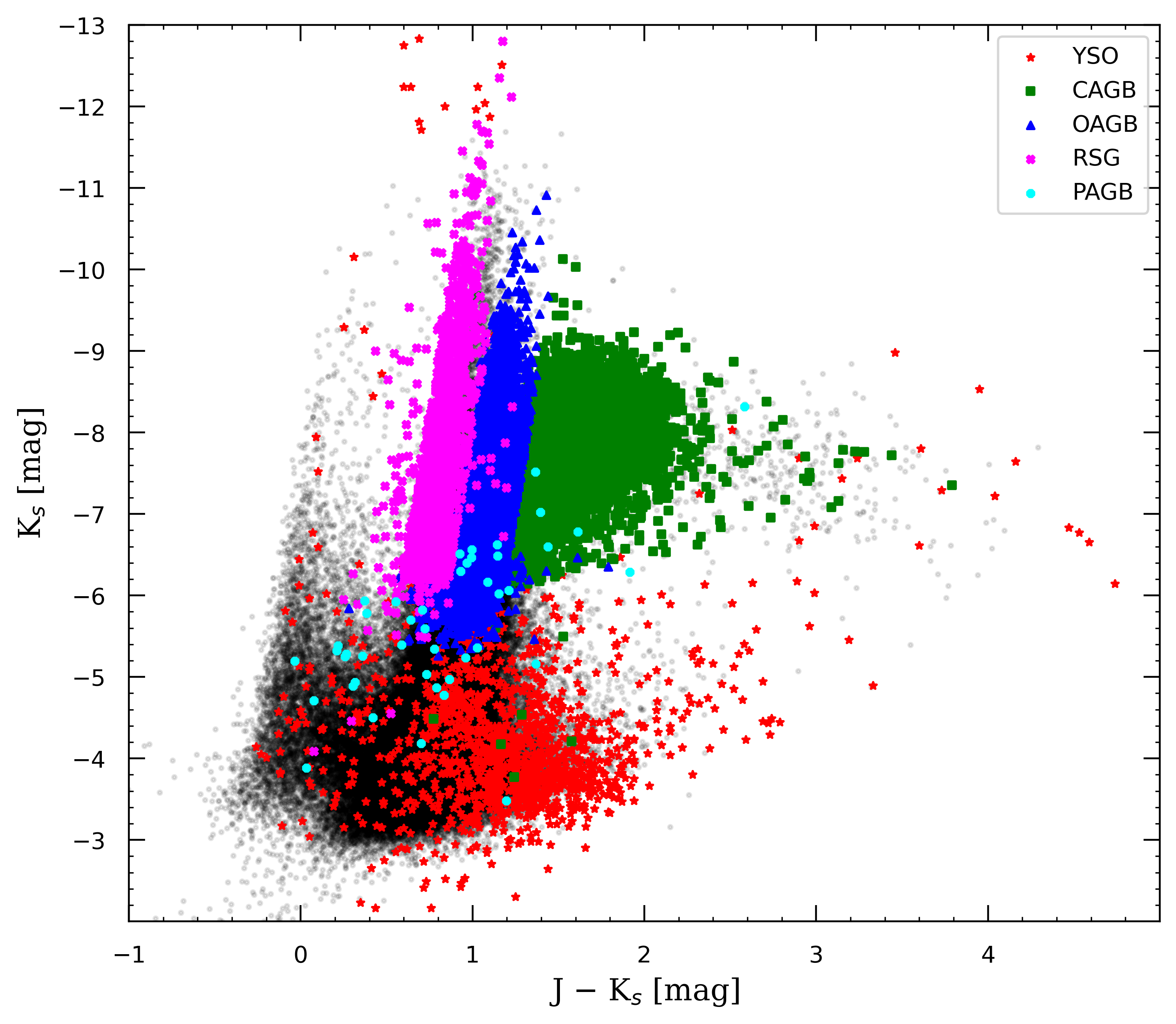,width=90mm,height=90mm}

       \epsfig{figure=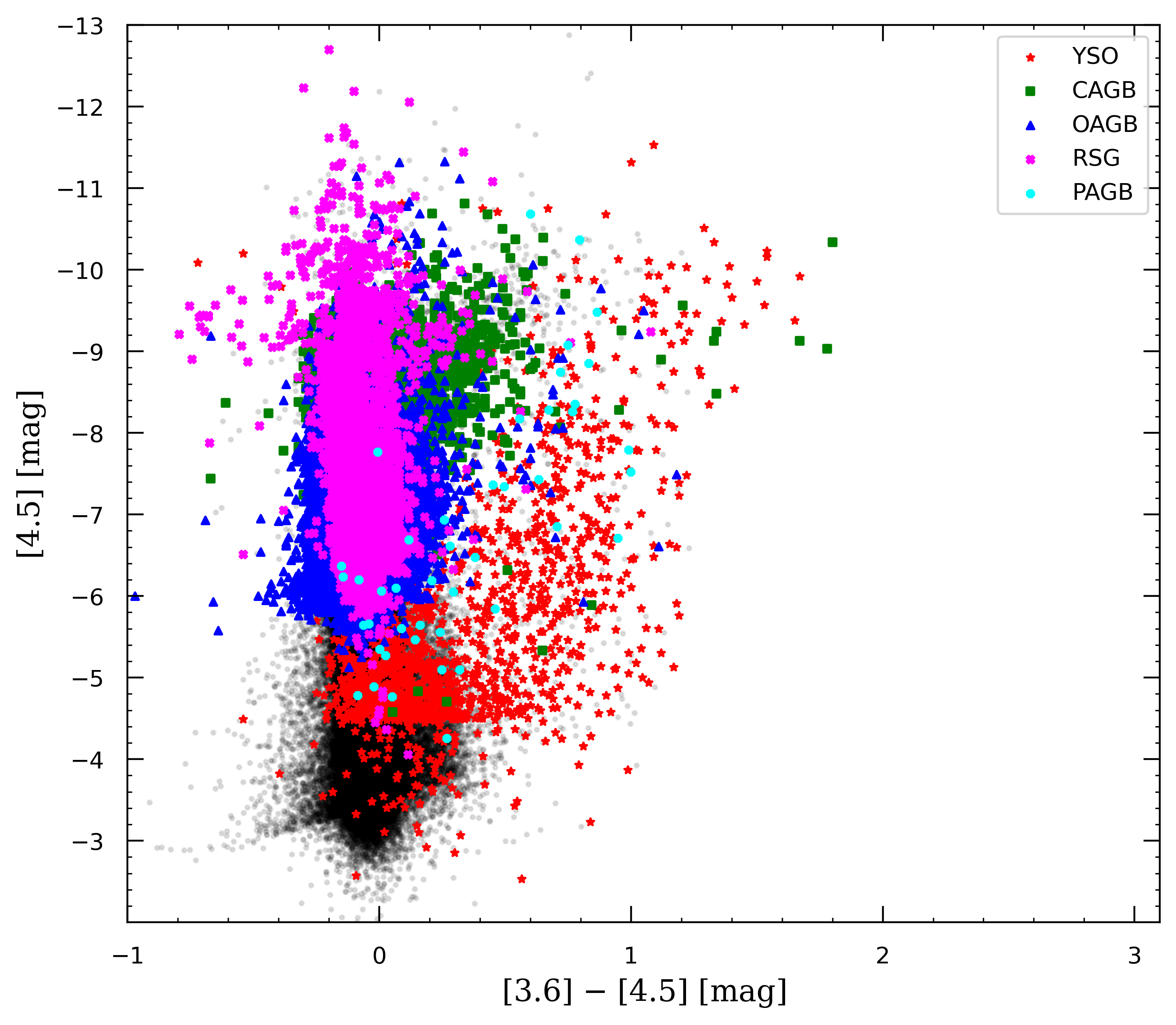,width=90mm,height=90mm}
   \centering
	}
	 \caption{This figure illustrates the distribution of dusty stellar sources throughout the CMDs in the near and mid-infrared bands. As can be seen, there is an overlap between stellar classes if they are located in the same place; therefore, their classification cannot be easily achieved by using just two near-coupled filters, such as near-infrared ({\it J}, {\it Ks}) and mid-infrared ([3.6], [4.5]). In this regard, more filters or different methods may be required for better classification.}
	\label{fig:CMDs_comparison}
    }
\end{figure*}
\begin{table}[t]
\centering
\caption{According to the selected list (see~Table \ref{tab:testdata-table1}), dusty stars within the Magellanic Clouds are categorized by their photometric labeled class after cross-matching all catalogs, thereby excluding those stars having repetition across catalogs.}
\begin{tabular}{lcr}
\hline
Stellar Type & Number \\
\hline
CAGB & 8,537 \\
OAGB & 37,950 \\
PAGB & 45 \\
RSG & 6,132 \\
YSO & 1,642 \\
\hline
Total & 54,306 \\
\hline
\end{tabular}
\label{tab:testdata-table2}
\end{table}
%
\indent

After data collection, we did a cross-match with the spectroscopically labeled data regarding removing repetitive stars. Following that, we applied all steps of the preprocessing method mentioned in Section~\ref{sec:data preprocessing}, providing them as the input of our models.\\
\indent In this section, the best model is applied to the data to determine the labels. According to Section~\ref{sec:Models}, considering the uncertainty and different results obtained in the PRF model with various tuned parameters, we concluded that this model has the best results. By presenting this new label as an output of machine learning models, we can compare it with photometric labels and determine how well they match.\\
\indent To better understand the two kinds of labels, photometric and spectroscopic, we use the confusion matrix to compare the types of dusty stellar objects classified by the photometric method and models trained by spectral types. Therefore, we consider photometric labels as actual labels in the confusion matrix, named in the following comparison matrix. In this comparison, we are not concerned about the correctness or incorrectness of the labels but about whether the labels match the photometric data.\\
\indent
This section's results are provided in the appendix through comparison matrices (Fig.~\ref{fig:Comparison_matrix}). These matrices show that the photometric data do not match some stellar classes across all four models.\\
\indent In machine learning, the best-performing models are introduced as the final model to label new input data based on classification metrics. After tuning the parameters in four modes, we found that PRF was our highest-performing model. There is a slight difference between the four PRF models, and most stars have been assigned the same label. Consequently, we constructed the consensus model, which considers the common results and excludes the rest from the classification, leading to more reliable predictions. For example, when all models consistently classify a star as OAGB, we present OAGB as its label. In contrast, if one predictive label differs from the others, it is excluded.\\
\begin{figure}
  \centering
  \includegraphics[width=1\linewidth]{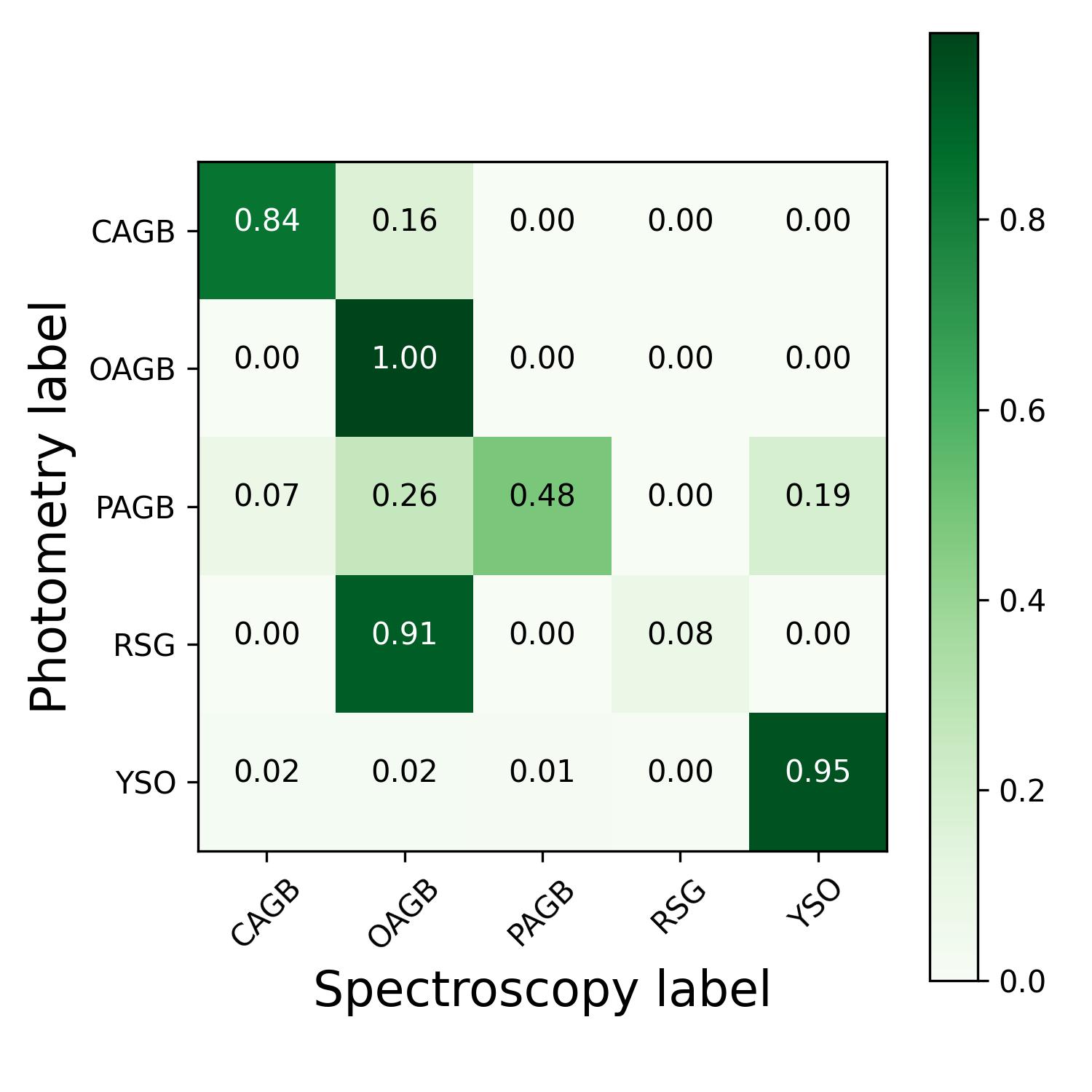}  
  \caption{Comparison matrix of common labels derived from four selected classifiers.} 
  \label{fig:Common_CM_Test}
\end{figure}
After applying this procedure to the stars, the classification resulted in 23,601 objects, categorized as follows: 4,689 CAGBs, 17,124 OAGBs, 38 PAGBs, 409 RSGs, and 1,341 YSOs. We can now redraw and compare the comparison matrix with the photometric data, as illustrated in Fig.~\ref{fig:Common_CM_Test}, which can be interpreted as follows.
\begin{itemize}[leftmargin=*]
\item As much as $16\%$ of CAGBs labeled by models based on the photometric method were oxygen-rich.
\item $100\% $of OAGB stars labeled based on photometric data have been identified with spectroscopic models.
\item $48\%$ of PAGBs are correctly identified, whereas $19\%$ and $26\%$ are classified as YSOs and OAGBs, respectively. These stars cannot be accurately classified because of their small populations and overlap with other classes, as mentioned in Section~\ref{sec:Results}.\
\item Only $8\%$ of the stellar class of RSGs are confirmed by the spectroscopic model, and $91\%$ are predicted as OAGBs.
\item Around $95\%$ of YSOs have been confirmed by models trained using spectroscopic labels, with the rest being subsets of others.\\
\end{itemize}

\indent The results indicate differences between labels predicted by models trained with the spectroscopic label and the photometric method. Thus, having a sufficient population compared to others is crucial for ensuring reliable, trained machine learning models.\\
The comprehensive dataset has been labeled by the best-trained models and released as a catalog in the supplementary material of this paper. The columns and their descriptions are presented in Table~\ref{tab:testdata-table3}. Specifically, this catalog includes 54,306 stars from the works presented in Table~\ref{tab:testdata-table1}, along with their photometric magnitudes used for dusty stellar classification, as described in Section~\ref{sec:data preprocessing} and Table~\ref{tab:Samples}. These stars are classified using the four best models, with their labels provided as columns in the catalog and used in the preceding results, as shown in Fig.~\ref{fig:Comparison_matrix}.

\begin{table}[t]
\centering
\caption{Description of the dataset containing columns from the comprehensive catalogs, including a total of 54,306 stars, as presented in Table~\ref{tab:testdata-table1} and Table~\ref{tab:testdata-table2}. This dataset includes positional information based on Right Ascension (RA) and Declination (Dec), host galaxy classification (LMC and SMC), and multi-band photometric magnitudes. Additionally, it provides photometric predictions derived from other works mentioned in Table~\ref{tab:testdata-table1} and predictions from four of the best-trained models developed using the Simple and SMOTE approaches. The comparison matrices of these models are presented in Fig.~\ref{fig:Comparison_matrix}. The comprehensive catalog of this table is available electronically.}

\begin{tabular}{cl}
\hline
Column No. & Descriptor \\
\hline
1 & Right Ascension (J2000) \\
2 & Declination (J2000) \\
3 & Host\_Galaxy (LMC and SMC) \\
4 & U–band magnitude\\
5 & B–band magnitude \\
6 & V–band magnitude \\
7 & I–band magnitude \\
8 & J–band magnitude \\
9 & H–band magnitude \\
10 & K$_{s}$–band magnitude \\
11 & [3.6] band magnitude \\
12 & [4.5] band magnitude\\
13 & [5.8] band magnitude\\
14 & [8.0] band magnitude\\
15 & [24] band magnitude\\
16 & Photometric prediction (Type)\\
17 & Model \#1 prediction\\
18 & Model \#2 prediction\\
19 & Model \#3 prediction\\
20 & Model \#4 prediction\\
\hline
\end{tabular}
\label{tab:testdata-table3}
\end{table} 

%
\section{Conclusion}
\label{sec:Conclusion}
We trained and tested a model for classifying dusty stellar objects using spectrally labeled data derived from the SAGE survey's multiwavelength filters, including infrared bands.\\
\indent
We identified the most accurate classifier, the PRF model, with accuracies exceeding $89\%$. We found that using the augmentation method to balance the dataset does not improve the results in this study, and it shows different behavior when dealing with various classes.\\
\indent 
In Section~\ref{sec:Metallicity Impact}, we found that metallicity differences did not impact the classification process when datasets from the SMC and LMC were combined. This was evaluated through three approaches, two involving classification using a dataset with four well-populated classes.\\
\indent We collected photometrically labeled data with the same features as spectroscopically labeled data (as explained in Section~\ref{sec:photometric data}). After feeding them into the models, they were classified, and a comprehensive catalog was presented as supplementary material for this paper.\\
\indent In the future, multiwavelength data such as that used in training can be the input of models to determine each object's label. In addition, more multiwavelength and spectroscopic observations, for instance, the JWST, Gaia, WEAVE, 4MOST, MOONS, and ELT MOSAIC catalogs in terms of object classification, are needed to address having more accurate dusty stellar classifiers, significantly in less populated classes, including PAGBs and RSGs.\\
\section*{Acknowledgments}
The authors thank the School of Astronomy at the Institute for Research in Fundamental Sciences (IPM) and the Iranian National Observatory (INO) for supporting this project. This project has received funding from the European Union’s Horizon 2020 research and innovation program under grant agreement No 101004214. We are grateful to the anonymous referee for carefully reading the manuscript and for helpful comments and suggestions, which helped us to improve the quality of the paper.
\newpage
\bibliographystyle{aasjournal}
\bibliography{sample631}

\begin{thebibliography}{}
\expandafter\ifx\csname natexlab\endcsname\relax\def\natexlab#1{#1}\fi
\providecommand{\url}[1]{\href{#1}{#1}}
\providecommand{\dodoi}[1]{doi:~\href{http://doi.org/#1}{\nolinkurl{#1}}}
\providecommand{\doeprint}[1]{\href{http://ascl.net/#1}{\nolinkurl{http://ascl.net/#1}}}
\providecommand{\doarXiv}[1]{\href{https://arxiv.org/abs/#1}{\nolinkurl{https://arxiv.org/abs/#1}}}

\bibitem[{{Abdollahi} {et~al.}(2023){Abdollahi}, {Torabi}, {Raeisi}, \&
  {Rahvar}}]{2023arXiv-mahdipaper}
{Abdollahi}, M., {Torabi}, N., {Raeisi}, S., \& {Rahvar}, S. 2023, arXiv
  e-prints, arXiv:2301.08497, \dodoi{10.48550/arXiv.2301.08497}

\bibitem[{{Adams} {et~al.}(2013){Adams}, {Simon}, {Bolatto}, {Sloan},
  {Sandstrom}, {Schmiedeke}, {van Loon}, {Oliveira}, \&
  {Keller}}]{Adams-2013ApJ...771..112A}
{Adams}, J.~J., {Simon}, J.~D., {Bolatto}, A.~D., {et~al.} 2013, \apj, 771,
  112, \dodoi{10.1088/0004-637X/771/2/112}

\bibitem[{Altman(1992)}]{knn1992}
Altman, N.~S. 1992, The American Statistician, 46, 175,
  \dodoi{10.1080/00031305.1992.10475879}

\bibitem[{Bader-El-Den {et~al.}(2016)Bader-El-Den, Teitei, \&
  Adda}]{BaderElDen2016HierarchicalCF}
Bader-El-Den, M.~B., Teitei, E., \& Adda, M. 2016, 2016 International Joint
  Conference on Neural Networks (IJCNN), 3584.
\newblock \url{https://api.semanticscholar.org/CorpusID:2457143}

\bibitem[{{Ball} \& {Brunner}(2010)}]{2010IJMPD-Data-Mining}
{Ball}, N.~M., \& {Brunner}, R.~J. 2010, International Journal of Modern
  Physics D, 19, 1049, \dodoi{10.1142/S0218271810017160}

\bibitem[{{Baron}(2019)}]{2019arXiv190407248B}
{Baron}, D. 2019, arXiv e-prints, arXiv:1904.07248,
  \dodoi{10.48550/arXiv.1904.07248}

\bibitem[{{Baron} \& {Poznanski}(2017)}]{2017MNRAS-rf-dalya}
{Baron}, D., \& {Poznanski}, D. 2017, \mnras, 465, 4530,
  \dodoi{10.1093/mnras/stw3021}

\bibitem[{{Bhardwaj} {et~al.}(2016){Bhardwaj}, {Kanbur}, {Macri}, {Singh},
  {Ngeow}, {Wagner-Kaiser}, \& {Sarajedini}}]{2016AJ....151...88B}
{Bhardwaj}, A., {Kanbur}, S.~M., {Macri}, L.~M., {et~al.} 2016, \aj, 151, 88,
  \dodoi{10.3847/0004-6256/151/4/88}

\bibitem[{{Blum} {et~al.}(2006){Blum}, {Mould}, {Olsen}, {Frogel}, {Werner},
  {Meixner}, {Markwick-Kemper}, {Indebetouw}, {Whitney}, {Meade}, {Babler},
  {Churchwell}, {Gordon}, {Engelbracht}, {For}, {Misselt}, {Vijh}, {Leitherer},
  {Volk}, {Points}, {Reach}, {Hora}, {Bernard}, {Boulanger}, {Bracker},
  {Cohen}, {Fukui}, {Gallagher}, {Gorjian}, {Harris}, {Kelly}, {Kawamura},
  {Latter}, {Madden}, {Mizuno}, {Mizuno}, {Nota}, {Oey}, {Onishi}, {Paladini},
  {Panagia}, {Perez-Gonzalez}, {Shibai}, {Sato}, {Smith}, {Staveley-Smith},
  {Tielens}, {Ueta}, {Van Dyk}, \& {Zaritsky}}]{2006AJ-cmd-sage-lmc}
{Blum}, R.~D., {Mould}, J.~R., {Olsen}, K.~A., {et~al.} 2006, \aj, 132, 2034,
  \dodoi{10.1086/508227}

\bibitem[{{Boyer} {et~al.}(2015){Boyer}, {McDonald}, {Srinivasan}, {Zijlstra},
  {van Loon}, {Olsen}, \& {Sonneborn}}]{Boyer-2015ApJ...810..116B}
{Boyer}, M.~L., {McDonald}, I., {Srinivasan}, S., {et~al.} 2015, \apj, 810,
  116, \dodoi{10.1088/0004-637X/810/2/116}

\bibitem[{{Boyer} {et~al.}(2011){Boyer}, {Srinivasan}, {van Loon}, {McDonald},
  {Meixner}, {Zaritsky}, {Gordon}, {Kemper}, {Babler}, {Block}, {Bracker},
  {Engelbracht}, {Hora}, {Indebetouw}, {Meade}, {Misselt}, {Robitaille},
  {Sewi{\l}o}, {Shiao}, \& {Whitney}}]{Boyer2011b}
{Boyer}, M.~L., {Srinivasan}, S., {van Loon}, J.~T., {et~al.} 2011, \aj, 142,
  103, \dodoi{10.1088/0004-6256/142/4/103}

\bibitem[{{Breiman}(2001)}]{2001MachL..45....5B}
{Breiman}, L. 2001, Machine Learning, 45, 5, \dodoi{10.1023/A:1010933404324}

\bibitem[{{Brice} \& {Andonie}(2019)}]{2019AJ-morgan-stellar-classification}
{Brice}, M.~J., \& {Andonie}, R. 2019, \aj, 158, 188,
  \dodoi{10.3847/1538-3881/ab40d0}

\bibitem[{{Carliles} {et~al.}(2010){Carliles}, {Budav{\'a}ri}, {Heinis},
  {Priebe}, \& {Szalay}}]{2010ApJ-randomforest-redshift}
{Carliles}, S., {Budav{\'a}ri}, T., {Heinis}, S., {Priebe}, C., \& {Szalay},
  A.~S. 2010, \apj, 712, 511, \dodoi{10.1088/0004-637X/712/1/511}

\bibitem[{Carroll \& Ostlie(2017)}]{Carroll_Ostlie_2017}
Carroll, B.~W., \& Ostlie, D.~A. 2017, An Introduction to Modern Astrophysics,
  2nd edn. (Cambridge University Press)

\bibitem[{{Chawla} {et~al.}(2011){Chawla}, {Bowyer}, {Hall}, \&
  {Kegelmeyer}}]{2011arXiv1106.1813C}
{Chawla}, N.~V., {Bowyer}, K.~W., {Hall}, L.~O., \& {Kegelmeyer}, W.~P. 2011,
  arXiv e-prints, arXiv:1106.1813, \dodoi{10.48550/arXiv.1106.1813}

\bibitem[{Cody {et~al.}(2024)Cody, Scher, McDonald, Zijlstra, Alexander, \&
  Cox}]{ian-10.12688/openreseurope.17023.1}
Cody, S., Scher, S., McDonald, I., {et~al.} 2024, Open Research Europe, 4,
  \dodoi{10.12688/openreseurope.17023.1}

\bibitem[{{Cornu} \& {Montillaud}(2021)}]{2021A&A-Cornu-nuralnetwork}
{Cornu}, D., \& {Montillaud}, J. 2021, \aap, 647, A116,
  \dodoi{10.1051/0004-6361/202038516}

\bibitem[{{Djorgovski} {et~al.}(2022){Djorgovski}, {Mahabal}, {Graham},
  {Polsterer}, \& {Krone-Martins}}]{AI-Astronomy-2022}
{Djorgovski}, S.~G., {Mahabal}, A.~A., {Graham}, M.~J., {Polsterer}, K., \&
  {Krone-Martins}, A. 2022, arXiv e-prints, arXiv:2212.01493,
  \dodoi{10.48550/arXiv.2212.01493}

\bibitem[{{Dorn-Wallenstein} {et~al.}(2021){Dorn-Wallenstein}, {Davenport},
  {Huppenkothen}, \& {Levesque}}]{2021ApJ-phtometric-classification}
{Dorn-Wallenstein}, T.~Z., {Davenport}, J. R.~A., {Huppenkothen}, D., \&
  {Levesque}, E.~M. 2021, \apj, 913, 32, \dodoi{10.3847/1538-4357/abf1f2}

\bibitem[{{Fazio} {et~al.}(2004){Fazio}, {Hora}, {Allen}, {Ashby}, {Barmby},
  {Deutsch}, {Huang}, {Kleiner}, {Marengo}, {Megeath}, {Melnick}, {Pahre},
  {Patten}, {Polizotti}, {Smith}, {Taylor}, {Wang}, {Willner}, {Hoffmann},
  {Pipher}, {Forrest}, {McMurty}, {McCreight}, {McKelvey}, {McMurray}, {Koch},
  {Moseley}, {Arendt}, {Mentzell}, {Marx}, {Losch}, {Mayman}, {Eichhorn},
  {Krebs}, {Jhabvala}, {Gezari}, {Fixsen}, {Flores}, {Shakoorzadeh}, {Jungo},
  {Hakun}, {Workman}, {Karpati}, {Kichak}, {Whitley}, {Mann}, {Tollestrup},
  {Eisenhardt}, {Stern}, {Gorjian}, {Bhattacharya}, {Carey}, {Nelson},
  {Glaccum}, {Lacy}, {Lowrance}, {Laine}, {Reach}, {Stauffer}, {Surace},
  {Wilson}, {Wright}, {Hoffman}, {Domingo}, \& {Cohen}}]{2004-Fazio-IRAC}
{Fazio}, G.~G., {Hora}, J.~L., {Allen}, L.~E., {et~al.} 2004, \apjs, 154, 10,
  \dodoi{10.1086/422843}

\bibitem[{{Fotopoulou}(2024)}]{2024-unsupervised}
{Fotopoulou}, S. 2024, Astronomy and Computing, 48, 100851,
  \dodoi{10.1016/j.ascom.2024.100851}

\bibitem[{{Ghaziasgar} {et~al.}(2024){Ghaziasgar}, {Abdollahi}, {Javadi}, {van
  Loo}, {McDonald}, {Oliveira}, \& {Khosroshahi}}]{BAO-2024}
{Ghaziasgar}, S., {Abdollahi}, M., {Javadi}, A., {et~al.} 2024, Communications
  of the Byurakan Astrophysical Observatory, 71, 377,
  \dodoi{10.52526/25792776-24.71.2-377}

\bibitem[{{Ghaziasgar} {et~al.}(2022){Ghaziasgar}, {Masoudnezhad}, {Javadi},
  {van Loon}, {Khosroshahi}, \& {Khosravaninezhad}}]{2022arXiv-ghaziasgar}
{Ghaziasgar}, S., {Masoudnezhad}, A., {Javadi}, A., {et~al.} 2022, arXiv
  e-prints, arXiv:2211.03403, \dodoi{10.48550/arXiv.2211.03403}

\bibitem[{{Goldman} {et~al.}(2017){Goldman}, {van Loon}, {Zijlstra}, {Green},
  {Wood}, {Nanni}, {Imai}, {Whitelock}, {Matsuura}, {Groenewegen}, \&
  {G{\'o}mez}}]{Goldman-2017MNRAS.465..403G}
{Goldman}, S.~R., {van Loon}, J.~T., {Zijlstra}, A.~A., {et~al.} 2017, \mnras,
  465, 403, \dodoi{10.1093/mnras/stw2708}

\bibitem[{{Gordon} {et~al.}(2011){Gordon}, {Meixner}, {Meade}, {Whitney},
  {Engelbracht}, {Bot}, {Boyer}, {Lawton}, {Sewi{\l}o}, {Babler}, {Bernard},
  {Bracker}, {Block}, {Blum}, {Bolatto}, {Bonanos}, {Harris}, {Hora},
  {Indebetouw}, {Misselt}, {Reach}, {Shiao}, {Tielens}, {Carlson},
  {Churchwell}, {Clayton}, {Chen}, {Cohen}, {Fukui}, {Gorjian}, {Hony},
  {Israel}, {Kawamura}, {Kemper}, {Leroy}, {Li}, {Madden}, {Marble},
  {McDonald}, {Mizuno}, {Mizuno}, {Muller}, {Oliveira}, {Olsen}, {Onishi},
  {Paladini}, {Paradis}, {Points}, {Robitaille}, {Rubin}, {Sandstrom}, {Sato},
  {Shibai}, {Simon}, {Smith}, {Srinivasan}, {Vijh}, {Van Dyk}, {van Loon}, \&
  {Zaritsky}}]{godron-smc-2011AJ....142..102G}
{Gordon}, K.~D., {Meixner}, M., {Meade}, M.~R., {et~al.} 2011, \aj, 142, 102,
  \dodoi{10.1088/0004-6256/142/4/102}

\bibitem[{Grandini {et~al.}(2020)Grandini, Bagli, \& Visani}]{2020MetricsFM}
Grandini, M., Bagli, E., \& Visani, G. 2020, ArXiv, abs/2008.05756.
\newblock \url{https://api.semanticscholar.org/CorpusID:221112671}

\bibitem[{{Gruendl} \& {Chu}(2009)}]{Gruendl2009}
{Gruendl}, R.~A., \& {Chu}, Y.-H. 2009, \apjs, 184, 172,
  \dodoi{10.1088/0067-0049/184/1/172}

\bibitem[{Hastie {et~al.}(2009)Hastie, Tibshirani, \&
  Friedman}]{hastie2009elements}
Hastie, T., Tibshirani, R., \& Friedman, J. 2009, The Elements of Statistical
  Learning: Data Mining, Inference, and Prediction (Springer)

\bibitem[{{Herwig}(2005)}]{2005ARA&A-falk-ahb}
{Herwig}, F. 2005, \araa, 43, 435,
  \dodoi{10.1146/annurev.astro.43.072103.150600}

\bibitem[{{H{\"o}fner} \& {Olofsson}(2018)}]{2018A&AR-massloss-agb-hofner}
{H{\"o}fner}, S., \& {Olofsson}, H. 2018, \aapr, 26, 1,
  \dodoi{10.1007/s00159-017-0106-5}

\bibitem[{{Hony} {et~al.}(2011){Hony}, {Kemper}, {Woods}, {van Loon},
  {Gorjian}, {Madden}, {Zijlstra}, {Gordon}, {Indebetouw}, {Marengo},
  {Meixner}, {Panuzzo}, {Shiao}, {Sloan}, {Roman-Duval}, {Mullaney}, \&
  {Tielens}}]{Hony2011}
{Hony}, S., {Kemper}, F., {Woods}, P.~M., {et~al.} 2011, \aap, 531, A137,
  \dodoi{10.1051/0004-6361/201116845}

\bibitem[{{Hosenie} {et~al.}(2020){Hosenie}, {Lyon}, {Stappers}, {Mootoovaloo},
  \& {McBride}}]{2020MNRAS-Imbalancelearning}
{Hosenie}, Z., {Lyon}, R., {Stappers}, B., {Mootoovaloo}, A., \& {McBride}, V.
  2020, \mnras, 493, 6050, \dodoi{10.1093/mnras/staa642}

\bibitem[{{Houck} {et~al.}(2004){Houck}, {Roellig}, {van Cleve}, {Forrest},
  {Herter}, {Lawrence}, {Matthews}, {Reitsema}, {Soifer}, {Watson}, {Weedman},
  {Huisjen}, {Troeltzsch}, {Barry}, {Bernard-Salas}, {Blacken}, {Brandl},
  {Charmandaris}, {Devost}, {Gull}, {Hall}, {Henderson}, {Higdon}, {Pirger},
  {Schoenwald}, {Sloan}, {Uchida}, {Appleton}, {Armus}, {Burgdorf},
  {Fajardo-Acosta}, {Grillmair}, {Ingalls}, {Morris}, \&
  {Teplitz}}]{IRS2004ApJS}
{Houck}, J.~R., {Roellig}, T.~L., {van Cleve}, J., {et~al.} 2004, \apjs, 154,
  18, \dodoi{10.1086/423134}

\bibitem[{{Ivezi{\'c}} {et~al.}(2014){Ivezi{\'c}}, {Connolly}, {VanderPlas}, \&
  {Gray}}]{2014sdmm.book.....I}
{Ivezi{\'c}}, {\v{Z}}., {Connolly}, A.~J., {VanderPlas}, J.~T., \& {Gray}, A.
  2014, {Statistics, Data Mining, and Machine Learning in Astronomy: A
  Practical Python Guide for the Analysis of Survey Data} (Princeton University
  Press), \dodoi{10.1515/9781400848911}

\bibitem[{{Javadi} \& {van Loon}(2022)}]{2022IAUS..366..210J}
{Javadi}, A., \& {van Loon}, J.~T. 2022, in The Origin of Outflows in Evolved
  Stars, ed. L.~{Decin}, A.~{Zijlstra}, \& C.~{Gielen}, Vol. 366, 210--215,
  \dodoi{10.1017/S1743921322001326}

\bibitem[{{Javadi} {et~al.}(2013){Javadi}, {van Loon}, {Khosroshahi}, \&
  {Mirtorabi}}]{2013MNRAS.432.2824J}
{Javadi}, A., {van Loon}, J.~T., {Khosroshahi}, H., \& {Mirtorabi}, M.~T. 2013,
  \mnras, 432, 2824, \dodoi{10.1093/mnras/stt640}

\bibitem[{{Javadi} {et~al.}(2011{\natexlab{a}}){Javadi}, {van Loon}, \&
  {Mirtorabi}}]{2011MNRAS.411..263J}
{Javadi}, A., {van Loon}, J.~T., \& {Mirtorabi}, M.~T. 2011{\natexlab{a}},
  \mnras, 411, 263, \dodoi{10.1111/j.1365-2966.2010.17678.x}

\bibitem[{{Javadi} {et~al.}(2011{\natexlab{b}}){Javadi}, {van Loon}, \&
  {Mirtorabi}}]{2011MNRAS.414.3394J}
---. 2011{\natexlab{b}}, \mnras, 414, 3394,
  \dodoi{10.1111/j.1365-2966.2011.18638.x}

\bibitem[{Jones {et~al.}(2023)Jones, Nally, Habel, Lenkic, Fahrion, Hirschauer,
  Chu, Meixner, Marchi, Nayak, Robberto, Sabbi, Zeidler, Oliveira, Beck,
  Biazzo, Brandl, Giardino, Jerabkova, \& Soderblom}]{jones-jwst-nature}
Jones, O., Nally, C., Habel, N., {et~al.} 2023, Nature Astronomy, 7, 1,
  \dodoi{10.1038/s41550-023-01945-7}

\bibitem[{{Jones} {et~al.}(2017{\natexlab{a}}){Jones}, {Meixner}, {Justtanont},
  \& {Glasse}}]{2017ApJ...841...15J}
{Jones}, O.~C., {Meixner}, M., {Justtanont}, K., \& {Glasse}, A.
  2017{\natexlab{a}}, \apj, 841, 15, \dodoi{10.3847/1538-4357/aa6bf6}

\bibitem[{{Jones} {et~al.}(2017{\natexlab{b}}){Jones}, {Woods}, {Kemper},
  {Kraemer}, {Sloan}, {Srinivasan}, {Oliveira}, {van Loon}, {Boyer}, {Sargent},
  {McDonald}, {Meixner}, {Zijlstra}, {Ruffle}, {Lagadec}, {Pauly}, {Sewi{\l}o},
  {Clayton}, \& {Volk}}]{2017MNRAS.470.3250J}
{Jones}, O.~C., {Woods}, P.~M., {Kemper}, F., {et~al.} 2017{\natexlab{b}},
  \mnras, 470, 3250, \dodoi{10.1093/mnras/stx1101}

\bibitem[{{Jones} {et~al.}(2017{\natexlab{c}}){Jones}, {Woods}, {Kemper},
  {Kraemer}, {Sloan}, {Srinivasan}, {Oliveira}, {van Loon}, {Boyer}, {Sargent},
  {McDonald}, {Meixner}, {Zijlstra}, {Ruffle}, {Lagadec}, {Pauly}, {Sewilo},
  {Clayton}, \& {Volk}}]{2017yCat..74703250J}
---. 2017{\natexlab{c}}, VizieR Online Data Catalog, J/MNRAS/470/3250

\bibitem[{{Kamath}(2020)}]{2020JApA-kmath-agb-postagb}
{Kamath}, D. 2020, Journal of Astrophysics and Astronomy, 41, 42,
  \dodoi{10.1007/s12036-020-09665-4}

\bibitem[{{Kamath} {et~al.}(2014){Kamath}, {Wood}, \& {Van
  Winckel}}]{Kamath2014MNRAS}
{Kamath}, D., {Wood}, P.~R., \& {Van Winckel}, H. 2014, \mnras, 439, 2211,
  \dodoi{10.1093/mnras/stt2033}

\bibitem[{{Kamath} {et~al.}(2015){Kamath}, {Wood}, \& {Van
  Winckel}}]{Kamath2015MNRAS}
---. 2015, \mnras, 454, 1468, \dodoi{10.1093/mnras/stv1202}

\bibitem[{{Karakas} \& {Lattanzio}(2014)}]{karakas-2014PASA...31...30K}
{Karakas}, A.~I., \& {Lattanzio}, J.~C. 2014, \pasa, 31, e030,
  \dodoi{10.1017/pasa.2014.21}

\bibitem[{{Kemper} {et~al.}(2010){Kemper}, {Woods}, {Antoniou}, {Bernard},
  {Blum}, {Boyer}, {Chan}, {Chen}, {Cohen}, {Dijkstra}, {Engelbracht},
  {Galametz}, {Galliano}, {Gielen}, {Gordon}, {Gorjian}, {Harris}, {Hony},
  {Hora}, {Indebetouw}, {Jones}, {Kawamura}, {Lagadec}, {Lawton}, {Leisenring},
  {Madden}, {Marengo}, {Matsuura}, {McDonald}, {McGuire}, {Meixner}, {Mulia},
  {O'Halloran}, {Oliveira}, {Paladini}, {Paradis}, {Reach}, {Rubin},
  {Sandstrom}, {Sargent}, {Sewilo}, {Shiao}, {Sloan}, {Speck}, {Srinivasan},
  {Szczerba}, {Tielens}, {van Aarle}, {Van Dyk}, {van Loon}, {Van Winckel},
  {Vijh}, {Volk}, {Whitney}, {Wilkins}, \& {Zijlstra}}]{2010PASP..122..683K}
{Kemper}, F., {Woods}, P.~M., {Antoniou}, V., {et~al.} 2010, \pasp, 122, 683,
  \dodoi{10.1086/653438}

\bibitem[{{Kinson} {et~al.}(2021){Kinson}, {Oliveira}, \& {van
  Loon}}]{2021MNRAS-Jacco2021-6822}
{Kinson}, D.~A., {Oliveira}, J.~M., \& {van Loon}, J.~T. 2021, \mnras, 507,
  5106, \dodoi{10.1093/mnras/stab2386}

\bibitem[{{Kinson} {et~al.}(2022){Kinson}, {Oliveira}, \& {van
  Loon}}]{2022MNRAS-jacco2022-m33}
---. 2022, \mnras, 517, 140, \dodoi{10.1093/mnras/stac2692}

\bibitem[{{Kokusho} {et~al.}(2023){Kokusho}, {Torii}, {Kaneda}, {Fukui}, \&
  {Tachihara}}]{2023ApJ-yso-starformation}
{Kokusho}, T., {Torii}, H., {Kaneda}, H., {Fukui}, Y., \& {Tachihara}, K. 2023,
  \apj, 953, 104, \dodoi{10.3847/1538-4357/ace10e}

\bibitem[{{Kuntzer} {et~al.}(2016){Kuntzer}, {Tewes}, \&
  {Courbin}}]{2016AA-singleband-setllar-clssification}
{Kuntzer}, T., {Tewes}, M., \& {Courbin}, F. 2016, \aap, 591, A54,
  \dodoi{10.1051/0004-6361/201628660}

\bibitem[{{Last} {et~al.}(2017){Last}, {Douzas}, \& {Bacao}}]{2017arXiv-SMOTE2}
{Last}, F., {Douzas}, G., \& {Bacao}, F. 2017, arXiv e-prints,
  arXiv:1711.00837, \dodoi{10.48550/arXiv.1711.00837}

\bibitem[{{Levesque}(2010)}]{2010ASPC-Levesque-redsupergiant}
{Levesque}, E.~M. 2010, in Astronomical Society of the Pacific Conference
  Series, Vol. 425, Hot and Cool: Bridging Gaps in Massive Star Evolution, ed.
  C.~{Leitherer}, P.~D. {Bennett}, P.~W. {Morris}, \& J.~T. {Van Loon}, 103,
  \dodoi{10.48550/arXiv.0911.4720}

\bibitem[{{Li} {et~al.}(2025){Li}, {Lu}, {Wang}, \&
  {Wang}}]{2025-ml-stellar-astronomy}
{Li}, G., {Lu}, Z., {Wang}, J., \& {Wang}, Z. 2025, arXiv e-prints,
  arXiv:2502.15300, \dodoi{10.48550/arXiv.2502.15300}

\bibitem[{{Li} {et~al.}(2019){Li}, {Wang}, {Basu}, {Kumbier}, \&
  {Yu}}]{MDIFeatureImportance-2019}
{Li}, X., {Wang}, Y., {Basu}, S., {Kumbier}, K., \& {Yu}, B. 2019, arXiv
  e-prints, arXiv:1906.10845, \dodoi{10.48550/arXiv.1906.10845}

\bibitem[{{Liu} {et~al.}(2021){Liu}, {Wei}, {Wei}, {Yu}, {Jiang}, {Cao},
  {Bian}, \& {Chang}}]{2021arXivLiu-imballanced}
{Liu}, Z., {Wei}, P., {Wei}, Z., {et~al.} 2021, arXiv e-prints,
  arXiv:2111.12791, \dodoi{10.48550/arXiv.2111.12791}

\bibitem[{{Mainzer} {et~al.}(2004){Mainzer}, {McLean}, {Sievers}, \&
  {Young}}]{2004ApJ-narrowband}
{Mainzer}, A.~K., {McLean}, I.~S., {Sievers}, J.~L., \& {Young}, E.~T. 2004,
  \apj, 604, 832, \dodoi{10.1086/382020}

\bibitem[{{Maravelias} {et~al.}(2022){Maravelias}, {Bonanos}, {Tramper}, {de
  Wit}, {Yang}, \& {Bonfini}}]{2022AA-ml-massive-classification}
{Maravelias}, G., {Bonanos}, A.~Z., {Tramper}, F., {et~al.} 2022, \aap, 666,
  A122, \dodoi{10.1051/0004-6361/202141397}

\bibitem[{{Massey} \& {Olsen}(2003)}]{2003AJ-massey-redsuperginat}
{Massey}, P., \& {Olsen}, K.~A.~G. 2003, \aj, 126, 2867, \dodoi{10.1086/379558}

\bibitem[{{McDonald} \& {Zijlstra}(2016)}]{2016ApJ-McDonald}
{McDonald}, I., \& {Zijlstra}, A.~A. 2016, \apjl, 823, L38,
  \dodoi{10.3847/2041-8205/823/2/L38}

\bibitem[{{McDonald} {et~al.}(2012){McDonald}, {Zijlstra}, \&
  {Boyer}}]{McDonald-2012MNRAS.427..343M}
{McDonald}, I., {Zijlstra}, A.~A., \& {Boyer}, M.~L. 2012, \mnras, 427, 343,
  \dodoi{10.1111/j.1365-2966.2012.21873.x}

\bibitem[{{McDonald} {et~al.}(2024){McDonald}, {Zijlstra}, {Cox}, {Alexander},
  {Csukai}, {Ramkumar}, \& {Hollings}}]{McDonald-2024RASTI...3...89M}
{McDonald}, I., {Zijlstra}, A.~A., {Cox}, N. L.~J., {et~al.} 2024, RAS
  Techniques and Instruments, 3, 89, \dodoi{10.1093/rasti/rzae005}

\bibitem[{{McDonald} {et~al.}(2017){McDonald}, {Zijlstra}, \&
  {Watson}}]{McDonald-2017MNRAS.471..770M}
{McDonald}, I., {Zijlstra}, A.~A., \& {Watson}, R.~A. 2017, \mnras, 471, 770,
  \dodoi{10.1093/mnras/stx1433}

\bibitem[{{Meixner} {et~al.}(2006){Meixner}, {Gordon}, {Indebetouw}, {Hora},
  {Whitney}, {Blum}, {Reach}, {Bernard}, {Meade}, {Babler}, {Engelbracht},
  {For}, {Misselt}, {Vijh}, {Leitherer}, {Cohen}, {Churchwell}, {Boulanger},
  {Frogel}, {Fukui}, {Gallagher}, {Gorjian}, {Harris}, {Kelly}, {Kawamura},
  {Kim}, {Latter}, {Madden}, {Markwick-Kemper}, {Mizuno}, {Mizuno}, {Mould},
  {Nota}, {Oey}, {Olsen}, {Onishi}, {Paladini}, {Panagia}, {Perez-Gonzalez},
  {Shibai}, {Sato}, {Smith}, {Staveley-Smith}, {Tielens}, {Ueta}, {van Dyk},
  {Volk}, {Werner}, \& {Zaritsky}}]{2006AJ-Meixner}
{Meixner}, M., {Gordon}, K.~D., {Indebetouw}, R., {et~al.} 2006, \aj, 132,
  2268, \dodoi{10.1086/508185}

\bibitem[{{Miettinen}(2018)}]{2018Ap&SS-Miettinen-ml-yso}
{Miettinen}, O. 2018, \apss, 363, 197, \dodoi{10.1007/s10509-018-3418-7}

\bibitem[{Murthy(1998)}]{Murthy1998AutomaticCO}
Murthy, S.~K. 1998, Data Mining and Knowledge Discovery, 2, 345.
\newblock \url{https://api.semanticscholar.org/CorpusID:207741345}

\bibitem[{{Oliveira} {et~al.}(2013){Oliveira}, {van Loon}, {Sloan},
  {Sewi{\l}o}, {Kraemer}, {Wood}, {Indebetouw}, {Filipovi{\'c}}, {Crawford},
  {Wong}, {Hora}, {Meixner}, {Robitaille}, {Shiao}, \&
  {Simon}}]{oliveira-2013MNRAS.428.3001O}
{Oliveira}, J.~M., {van Loon}, J.~T., {Sloan}, G.~C., {et~al.} 2013, \mnras,
  428, 3001, \dodoi{10.1093/mnras/sts250}

\bibitem[{{Pashchenko} {et~al.}(2018){Pashchenko}, {Sokolovsky}, \&
  {Gavras}}]{2018MNRAS-ml-variablestar}
{Pashchenko}, I.~N., {Sokolovsky}, K.~V., \& {Gavras}, P. 2018, \mnras, 475,
  2326, \dodoi{10.1093/mnras/stx3222}

\bibitem[{Pedregosa {et~al.}(2011)Pedregosa, Varoquaux, Gramfort, Michel,
  Thirion, Grisel, Blondel, Prettenhofer, Weiss, Dubourg, Vanderplas, Passos,
  Cournapeau, Brucher, Perrot, \& Duchesnay}]{scikit-learn}
Pedregosa, F., Varoquaux, G., Gramfort, A., {et~al.} 2011, Journal of Machine
  Learning Research, 12, 2825

\bibitem[{{Pennock} {et~al.}(2022){Pennock}, {van Loon}, {Anih}, {Maitra},
  {Haberl}, {Sansom}, {Ivanov}, {Cowley}, {Afonso}, {Ant{\'o}n}, {Cioni},
  {Craig}, {Filipovi{\'c}}, {Hopkins}, {Nanni}, {Prandoni}, \&
  {Vardoulaki}}]{Pennock-2022MNRAS}
{Pennock}, C.~M., {van Loon}, J.~T., {Anih}, J.~O., {et~al.} 2022, \mnras, 515,
  6046, \dodoi{10.1093/mnras/stac2096}

\bibitem[{{Pennock} {et~al.}(2025){Pennock}, {van Loon}, {Cioni}, {Maitra},
  {Oliveira}, {Craig}, {Ivanov}, {Aird}, {Anih}, {Cross}, {Dresbach}, {de
  Grijs}, \& {Groenewegen}}]{Pennock-2025}
{Pennock}, C.~M., {van Loon}, J.~T., {Cioni}, M.-R.~L., {et~al.} 2025, \mnras,
  537, 1028, \dodoi{10.1093/mnras/staf080}

\bibitem[{{Pietrzy{\'n}ski} {et~al.}(2013){Pietrzy{\'n}ski}, {Graczyk},
  {Gieren}, {Thompson}, {Pilecki}, {Udalski}, {Soszy{\'n}ski}, {Koz{\l}owski},
  {Konorski}, {Suchomska}, {Bono}, {Moroni}, {Villanova}, {Nardetto},
  {Bresolin}, {Kudritzki}, {Storm}, {Gallenne}, {Smolec}, {Minniti}, {Kubiak},
  {Szyma{\'n}ski}, {Poleski}, {Wyrzykowski}, {Ulaczyk}, {Pietrukowicz},
  {G{\'o}rski}, \& {Karczmarek}}]{Pietrzy-lmc-2013Natur.495...76P}
{Pietrzy{\'n}ski}, G., {Graczyk}, D., {Gieren}, W., {et~al.} 2013, \nat, 495,
  76, \dodoi{10.1038/nature11878}

\bibitem[{{Pilbratt} {et~al.}(2010){Pilbratt}, {Riedinger}, {Passvogel},
  {Crone}, {Doyle}, {Gageur}, {Heras}, {Jewell}, {Metcalfe}, {Ott}, \&
  {Schmidt}}]{2010-Herschel}
{Pilbratt}, G.~L., {Riedinger}, J.~R., {Passvogel}, T., {et~al.} 2010, \aap,
  518, L1, \dodoi{10.1051/0004-6361/201014759}

\bibitem[{{Reis} {et~al.}(2019){Reis}, {Baron}, \&
  {Shahaf}}]{2019AJ....157...16R}
{Reis}, I., {Baron}, D., \& {Shahaf}, S. 2019, \aj, 157, 16,
  \dodoi{10.3847/1538-3881/aaf101}

\bibitem[{{Riebel} {et~al.}(2010){Riebel}, {Meixner}, {Fraser}, {Srinivasan},
  {Cook}, \& {Vijh}}]{Riebel2010}
{Riebel}, D., {Meixner}, M., {Fraser}, O., {et~al.} 2010, \apj, 723, 1195,
  \dodoi{10.1088/0004-637X/723/2/1195}

\bibitem[{{Riebel} {et~al.}(2012){Riebel}, {Srinivasan}, {Sargent}, \&
  {Meixner}}]{2012ApJ-lmc-mass-loss}
{Riebel}, D., {Srinivasan}, S., {Sargent}, B., \& {Meixner}, M. 2012, \apj,
  753, 71, \dodoi{10.1088/0004-637X/753/1/71}

\bibitem[{{Rieke} {et~al.}(2004){Rieke}, {Young}, {Engelbracht}, {Kelly},
  {Low}, {Haller}, {Beeman}, {Gordon}, {Stansberry}, {Misselt}, {Cadien},
  {Morrison}, {Rivlis}, {Latter}, {Noriega-Crespo}, {Padgett}, {Stapelfeldt},
  {Hines}, {Egami}, {Muzerolle}, {Alonso-Herrero}, {Blaylock}, {Dole}, {Hinz},
  {Le Floc'h}, {Papovich}, {P{\'e}rez-Gonz{\'a}lez}, {Smith}, {Su}, {Bennett},
  {Frayer}, {Henderson}, {Lu}, {Masci}, {Pesenson}, {Rebull}, {Rho}, {Keene},
  {Stolovy}, {Wachter}, {Wheaton}, {Werner}, \& {Richards}}]{2004-Rieke-MIPS}
{Rieke}, G.~H., {Young}, E.~T., {Engelbracht}, C.~W., {et~al.} 2004, \apjs,
  154, 25, \dodoi{10.1086/422717}

\bibitem[{Rokach \& Maimon(2005)}]{Rokach2005TopdownIO}
Rokach, L., \& Maimon, O. 2005, IEEE Transactions on Systems, Man, and
  Cybernetics, Part C (Applications and Reviews), 35, 476.
\newblock \url{https://api.semanticscholar.org/CorpusID:14808716}

\bibitem[{{Ruffle} {et~al.}(2015{\natexlab{a}}){Ruffle}, {Kemper}, {Jones},
  {Sloan}, {Kraemer}, {Woods}, {Boyer}, {Srinivasan}, {Antoniou}, {Lagadec},
  {Matsuura}, {McDonald}, {Oliveira}, {Sargent}, {Sewi{\l}o}, {Szczerba}, {van
  Loon}, {Volk}, \& {Zijlstra}}]{2015MNRAS.451.3504R}
{Ruffle}, P. M.~E., {Kemper}, F., {Jones}, O.~C., {et~al.} 2015{\natexlab{a}},
  \mnras, 451, 3504, \dodoi{10.1093/mnras/stv1106}

\bibitem[{{Ruffle} {et~al.}(2015{\natexlab{b}}){Ruffle}, {Kemper}, {Jones},
  {Sloan}, {Kraemer}, {Woods}, {Boyer}, {Srinivasan}, {Antoniou}, {Lagadec},
  {Matsuura}, {McDonald}, {Oliveira}, {Sargent}, {Sewilo}, {Szczerba}, {van
  Loon}, {Volk}, \& {Zijlstra}}]{2015yCat..74513504R}
{Ruffle}, P.~M.~E., {Kemper}, F., {Jones}, O.~C., {et~al.} 2015{\natexlab{b}},
  VizieR Online Data Catalog, J/MNRAS/451/3504

\bibitem[{{Russell} \& {Dopita}(1992)}]{1992ApJ-metalicity-russel}
{Russell}, S.~C., \& {Dopita}, M.~A. 1992, \apj, 384, 508,
  \dodoi{10.1086/170893}

\bibitem[{{Rybicki} \& {Lightman}(1986)}]{1986rpa..book.....R}
{Rybicki}, G.~B., \& {Lightman}, A.~P. 1986, {Radiative Processes in
  Astrophysics} (Wiley-VCH)

\bibitem[{{Schlafly} \& {Finkbeiner}(2011)}]{2011ApJ-extinction-smc-lmc}
{Schlafly}, E.~F., \& {Finkbeiner}, D.~P. 2011, \apj, 737, 103,
  \dodoi{10.1088/0004-637X/737/2/103}

\bibitem[{{Scowcroft} {et~al.}(2016){Scowcroft}, {Freedman}, {Madore},
  {Monson}, {Persson}, {Rich}, {Seibert}, \& {Rigby}}]{2016ApJ...816...49S}
{Scowcroft}, V., {Freedman}, W.~L., {Madore}, B.~F., {et~al.} 2016, 816, 49,
  \dodoi{10.3847/0004-637X/816/2/49}

\bibitem[{{Seale} {et~al.}(2009){Seale}, {Looney}, {Chu}, {Gruendl}, {Brandl},
  {Chen}, {Brandner}, \& {Blake}}]{2009ApJ-seale-ys0-class}
{Seale}, J.~P., {Looney}, L.~W., {Chu}, Y.-H., {et~al.} 2009, \apj, 699, 150,
  \dodoi{10.1088/0004-637X/699/1/150}

\bibitem[{{Sen} {et~al.}(2022){Sen}, {Agarwal}, {Chakraborty}, \&
  {Singh}}]{2022ExA-bigdata}
{Sen}, S., {Agarwal}, S., {Chakraborty}, P., \& {Singh}, K.~P. 2022,
  Experimental Astronomy, 53, 1, \dodoi{10.1007/s10686-021-09827-4}

\bibitem[{{Sewilo} {et~al.}(2013){Sewilo}, {Carlson}, {Seale},
  {Indebetouw}, {Meixner}, {Whitney}, {Robitaille}, {Oliveira}, {Gordon},
  {Meade}, {Babler}, {Hora}, {Block}, {Misselt}, {van Loon}, {Chen},
  {Churchwell}, \& {Shiao}}]{Sewiło-2013ApJ...778...15S}
{Sewilo}, M., {Carlson}, L.~R., {Seale}, J.~P., {et~al.} 2013, \apj, 778,
  15, \dodoi{10.1088/0004-637X/778/1/15}

\bibitem[{{Sheets} {et~al.}(2013){Sheets}, {Bolatto}, {van Loon}, {Sandstrom},
  {Simon}, {Oliveira}, \& {Barb{\'a}}}]{Sheets-2013ApJ...771..111S}
{Sheets}, H.~A., {Bolatto}, A.~D., {van Loon}, J.~T., {et~al.} 2013, \apj, 771,
  111, \dodoi{10.1088/0004-637X/771/2/111}

\bibitem[{{Smith} \& {Geach}(2023)}]{2023RSOS-nural_network}
{Smith}, M.~J., \& {Geach}, J.~E. 2023, Royal Society Open Science, 10, 221454,
  \dodoi{10.1098/rsos.221454}

\bibitem[{{Srinivasan} {et~al.}(2009){Srinivasan}, {Meixner}, {Leitherer},
  {Vijh}, {Volk}, {Blum}, {Babler}, {Block}, {Bracker}, {Cohen}, {Engelbracht},
  {For}, {Gordon}, {Harris}, {Hora}, {Indebetouw}, {Markwick-Kemper}, {Meade},
  {Misselt}, {Sewilo}, \& {Whitney}}]{Srinivasan2009AJ}
{Srinivasan}, S., {Meixner}, M., {Leitherer}, C., {et~al.} 2009, \aj, 137,
  4810, \dodoi{10.1088/0004-6256/137/6/4810}

\bibitem[{{Subramanian} \& {Subramaniam}(2009)}]{LMC-SMC-2009A&A...496..399S}
{Subramanian}, S., \& {Subramaniam}, A. 2009, \aap, 496, 399,
  \dodoi{10.1051/0004-6361/200811029}

\bibitem[{{Subramanian} \&
  {Subramaniam}(2011)}]{Subramanian-2011ASInC...3..144S}
{Subramanian}, S., \& {Subramaniam}, A. 2011, in Astronomical Society of India
  Conference Series, Vol.~3, Astronomical Society of India Conference Series,
  144

\bibitem[{{Suh}(2016)}]{2016JASS...33..119S}
{Suh}, K.-W. 2016, Journal of Astronomy and Space Sciences, 33, 119,
  \dodoi{10.5140/JASS.2016.33.2.119}

\bibitem[{{Suh}(2020)}]{2020ApJ...891...43S}
---. 2020, \apj, 891, 43, \dodoi{10.3847/1538-4357/ab6609}

\bibitem[{{Suh}(2021)}]{2021ApJS..256...43S}
---. 2021, \apjs, 256, 43, \dodoi{10.3847/1538-4365/ac1274}

\bibitem[{Sun {et~al.}(2009)Sun, Wong, \& Kamel}]{Sun2009ClassificationOI}
Sun, Y., Wong, A. K.~C., \& Kamel, M.~S. 2009, Int. J. Pattern Recognit. Artif.
  Intell., 23, 687.
\newblock \url{https://api.semanticscholar.org/CorpusID:27118324}

\bibitem[{{van Winckel}(2003)}]{2003ARA-vanWincke}
{van Winckel}, H. 2003, \araa, 41, 391,
  \dodoi{10.1146/annurev.astro.41.071601.170018}

\bibitem[{Vapnik(1995)}]{vapnik95}
Vapnik, V.~N. 1995, The nature of statistical learning theory (Springer-Verlag
  New York, Inc.)

\bibitem[{{Vijh} {et~al.}(2009){Vijh}, {Meixner}, {Babler}, {Block}, {Bracker},
  {Engelbracht}, {For}, {Gordon}, {Hora}, {Indebetouw}, {Leitherer}, {Meade},
  {Misselt}, {Sewilo}, {Srinivasan}, \& {Whitney}}]{2009AJ-vijh-variable-sage}
{Vijh}, U.~P., {Meixner}, M., {Babler}, B., {et~al.} 2009, \aj, 137, 3139,
  \dodoi{10.1088/0004-6256/137/2/3139}

\bibitem[{{Wang} {et~al.}(2021){Wang}, {Dai}, {Shen}, \&
  {Xuan}}]{2021NatSR-SMOTE}
{Wang}, S., {Dai}, Y., {Shen}, J., \& {Xuan}, J. 2021, Scientific Reports, 11,
  24039, \dodoi{10.1038/s41598-021-03430-5}

\bibitem[{{Werner} {et~al.}(2004){Werner}, {Roellig}, {Low}, {Rieke}, {Rieke},
  {Hoffmann}, {Young}, {Houck}, {Brandl}, {Fazio}, {Hora}, {Gehrz}, {Helou},
  {Soifer}, {Stauffer}, {Keene}, {Eisenhardt}, {Gallagher}, {Gautier}, {Irace},
  {Lawrence}, {Simmons}, {Van Cleve}, {Jura}, {Wright}, \&
  {Cruikshank}}]{2004-Werner-Spitzer}
{Werner}, M.~W., {Roellig}, T.~L., {Low}, F.~J., {et~al.} 2004, \apjs, 154, 1,
  \dodoi{10.1086/422992}

\bibitem[{{Whitelock} {et~al.}(2003){Whitelock}, {Feast}, {van Loon}, \&
  {Zijlstra}}]{Whitelock-2003MNRAS.342...86W}
{Whitelock}, P.~A., {Feast}, M.~W., {van Loon}, J.~T., \& {Zijlstra}, A.~A.
  2003, \mnras, 342, 86, \dodoi{10.1046/j.1365-8711.2003.06514.x}

\bibitem[{{Whitney} {et~al.}(2008){Whitney}, {Sewilo}, {Indebetouw},
  {Robitaille}, {Meixner}, {Gordon}, {Meade}, {Babler}, {Harris}, {Hora},
  {Bracker}, {Povich}, {Churchwell}, {Engelbracht}, {For}, {Block}, {Misselt},
  {Vijh}, {Leitherer}, {Kawamura}, {Blum}, {Cohen}, {Fukui}, {Mizuno},
  {Mizuno}, {Srinivasan}, {Tielens}, {Volk}, {Bernard}, {Boulanger}, {Frogel},
  {Gallagher}, {Gorjian}, {Kelly}, {Latter}, {Madden}, {Kemper}, {Mould},
  {Nota}, {Oey}, {Olsen}, {Onishi}, {Paladini}, {Panagia}, {Perez-Gonzalez},
  {Reach}, {Shibai}, {Sato}, {Smith}, {Staveley-Smith}, {Ueta}, {Van Dyk},
  {Werner}, {Wolff}, \& {Zaritsky}}]{Whitney2008AJ}
{Whitney}, B.~A., {Sewilo}, M., {Indebetouw}, R., {et~al.} 2008, \aj, 136, 18,
  \dodoi{10.1088/0004-6256/136/1/18}

\bibitem[{{Wilson} {et~al.}(2023){Wilson}, {Lakeland}, {Wilson}, \&
  {Naylor}}]{2023MNRAS-yso-ml-naive-bayse}
{Wilson}, A.~J., {Lakeland}, B.~S., {Wilson}, T.~J., \& {Naylor}, T. 2023,
  \mnras, 521, 354, \dodoi{10.1093/mnras/stad301}

\bibitem[{{Wood} {et~al.}(1983){Wood}, {Bessell}, \&
  {Fox}}]{1983ApJ-wood-variable}
{Wood}, P.~R., {Bessell}, M.~S., \& {Fox}, M.~W. 1983, \apj, 272, 99,
  \dodoi{10.1086/161265}

\bibitem[{{Wood} {et~al.}(1992){Wood}, {Whiteoak}, {Hughes}, {Bessell},
  {Gardner}, \& {Hyland}}]{1992ApJ-wood-variable-mass-loss}
{Wood}, P.~R., {Whiteoak}, J.~B., {Hughes}, S.~M.~G., {et~al.} 1992, \apj, 397,
  552, \dodoi{10.1086/171812}

\bibitem[{{Woods} {et~al.}(2011){Woods}, {Oliveira}, {Kemper}, {van Loon},
  {Sargent}, {Matsuura}, {Szczerba}, {Volk}, {Zijlstra}, {Sloan}, {Lagadec},
  {McDonald}, {Jones}, {Gorjian}, {Kraemer}, {Gielen}, {Meixner}, {Blum},
  {Sewi{\l}o}, {Riebel}, {Shiao}, {Chen}, {Boyer}, {Indebetouw}, {Antoniou},
  {Bernard}, {Cohen}, {Dijkstra}, {Galametz}, {Galliano}, {Gordon}, {Harris},
  {Hony}, {Hora}, {Kawamura}, {Lawton}, {Leisenring}, {Madden}, {Marengo},
  {McGuire}, {Mulia}, {O'Halloran}, {Olsen}, {Paladini}, {Paradis}, {Reach},
  {Rubin}, {Sandstrom}, {Soszy{\'n}ski}, {Speck}, {Srinivasan}, {Tielens}, {van
  Aarle}, {van Dyk}, {van Winckel}, {Vijh}, {Whitney}, \&
  {Wilkins}}]{2011MNRAS.411.1597W}
{Woods}, P.~M., {Oliveira}, J.~M., {Kemper}, F., {et~al.} 2011, \mnras, 411,
  1597, \dodoi{10.1111/j.1365-2966.2010.17794.x}

\bibitem[{{Yang} {et~al.}(2018){Yang}, {Bonanos}, {Jiang}, {Gao}, {Xue},
  {Wang}, {Lam}, {Spetsieri}, {Ren}, \& {Gavras}}]{yang2018AA-lmc}
{Yang}, M., {Bonanos}, A.~Z., {Jiang}, B.-W., {et~al.} 2018, \aap, 616, A175,
  \dodoi{10.1051/0004-6361/201832833}

\bibitem[{{Yang} {et~al.}(2019){Yang}, {Bonanos}, {Jiang}, {Gao}, {Gavras},
  {Maravelias}, {Ren}, {Wang}, {Xue}, {Tramper}, {Spetsieri}, \&
  {Pouliasis}}]{Yang2019-smc}
---. 2019, \aap, 629, A91, \dodoi{10.1051/0004-6361/201935916}

\bibitem[{{Yang} {et~al.}(2020){Yang}, {Bonanos}, {Jiang}, {Gao}, {Gavras},
  {Maravelias}, {Wang}, {Chen}, {Tramper}, {Ren}, {Spetsieri}, \&
  {Xue}}]{yang2020-smc}
---. 2020, \aap, 639, A116, \dodoi{10.1051/0004-6361/201937168}

\bibitem[{{Yang} {et~al.}(2021){Yang}, {Bonanos}, {Jiang}, {Gao}, {Gavras},
  {Maravelias}, {Wang}, {Chen}, {Lam}, {Ren}, {Tramper}, \&
  {Spetsieri}}]{Ya2021}
{Yang}, M., {Bonanos}, A.~Z., {Jiang}, B., {et~al.} 2021, \aap, 646, A141,
  \dodoi{10.1051/0004-6361/202039475}

\bibitem[{{Zeraatgari} {et~al.}(2024){Zeraatgari}, {Hafezianzadeh}, {Zhang},
  {Mei}, {Ayubinia}, {Mosallanezhad}, \&
  {Zhang}}]{2024MNRAS-ml-classification-china}
{Zeraatgari}, F.~Z., {Hafezianzadeh}, F., {Zhang}, Y., {et~al.} 2024, \mnras,
  527, 4677, \dodoi{10.1093/mnras/stad3436}

\bibitem[{{Zhang} {et~al.}(2023){Zhang}, {Zhang}, {Kang}, {Li}, {Tao}, {Zhao},
  \& {Wu}}]{2023PASA-imballanced}
{Zhang}, J., {Zhang}, Y., {Kang}, Z., {et~al.} 2023, \pasa, 40, e037,
  \dodoi{10.1017/pasa.2023.35}

\end{thebibliography}

\counterwithin{figure}{section}
\counterwithin{table}{section}
\appendix \section{SUPPORTING INFORMATION FOR CLASSIFICATION MODELS}\label{sec:SUPPORTING_INFORMATION}
\subsection{Random Forest (RF)}\label{sec:RF}
The Random Forest (RF) method combines multiple decision trees to classify data. In the algorithm, the prediction made by each tree is averaged or voted on by the majority \citep{2001MachL..45....5B, 2010ApJ-randomforest-redshift,2017MNRAS-rf-dalya,2019AJ....157...16R}.
%
\subsection{K-Nearest Neighbors (KNN)}\label{sec:KNN}
K-Nearest Neighbors (KNN) is a classification algorithm that assigns labels to data based on their nearest neighbors. The KNN algorithm calculates data points by computing their distance from all other data points. The data point can be predicted based on the average value of the K nearest neighbors or the most common classes \citep{knn1992,2001MachL..45....5B}.
\subsection{C-Support Vector Classification (SVC)}\label{sec:SVC}

Support Vector Classification (SVC) is a method that utilizes a hyperplane to separate data into different classes and map data onto a high-dimensional feature space. Support vectors are part of the SVC model and are the closest data points to the hyperplane. Support Vector Machines (SVC) can be adjusted for some hyperparameters to achieve optimal performance. The polynomial kernel SVC (SVC-poly) is one of the most commonly used SVC models. In contrast, an SVC-rbf, according to the Radial Basis Function (RBF) kernel, is the other typically used SVC model (SVC-rbf) \citep{vapnik95,2019arXiv190407248B}.
\subsection{Gaussian Naive Bayes (GNB)}\label{sec:GNB}
Gaussian Naive Bayes executes the Gaussian Naive Bayes algorithm for classification, assuming that the likelihood of the features follows a Gaussian distribution. It includes one key hyperparameter, \texttt{var\_smoothing}, which is optimized using Grid Search.

\begin{equation}
P(x_i \mid y) = \frac{1}{\sqrt{2\pi\sigma^2_y}} \exp\left(-\frac{(x_i - \mu_y)^2}{2\sigma^2_y}\right)
\end{equation}
\
 \section{Performance Metrics} \label{sec:Performance Metric}
The performance metrics of classifiers can be used to predict all spectral classes. At the training stage, especially because the classes have unequal populations, accuracy can be misleading. In addition, the Classification report contains the model’s precision, recall, and F1-score values for each class \citep{2020MetricsFM}. \\
Accuracy is a metric for evaluating the classification models, dividing the number of correct predictions by the total number of predictions.
\begin{equation} \label{eq1}
\textmd{Accuracy} =\frac{\textmd{correct predictions}}{\textmd{all predictions }},
\end{equation}
%
Precision is the fraction of relevant instances among the retrieved ones. 
\begin{equation} \label{eq2}
\textmd{Precision} =\frac{\textmd{True Positive}}{\textmd{False Positive+True Positive}},
\end{equation}
Recall is the fraction of retrieved relevant cases in the following formula.
\begin{equation} \label{eq3}
\textmd{Recall}=\frac{\textmd{True Positive}}{\textmd{False Negative+True Positive}}.
\end{equation} 
The F1-score is a combination of precision and recall. We used performance metrics to evaluate each classifier and defined them as follows (see Fig. \ref{fig:Apendix_Fig_1} for illustration).
\begin{equation}
F_1 = 2 \times \frac{\mathrm{precision} \times \mathrm{recall}}{\mathrm{precision} + \mathrm{recall}}
\end{equation}
The macro average consists of precision, recall, and an F1-score average for all classes (The population of each class in this averaging is ineffective).\\
The weighted average is the average of precision, recall, and F1 scores based on the number of samples for each class.
Precision and recall are weighted based on population weights computed separately for each subclass in the sample and averaged.
%
\section{Confusion Matrix} \label{sec:Confusion_Matrix}
\begin{figure}
    \centering
    \includegraphics[width=0.5\columnwidth]{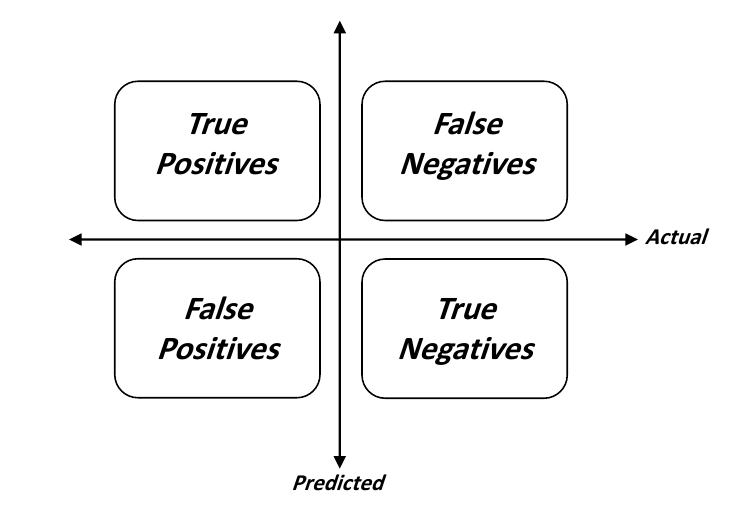}
    \caption{Illustration of the Confusion Matrix for Classification Outcomes. The diagram represents the possible outcomes in the classification task, distinguishing between actual and predicted classifications.}
    \label{fig:Apendix_Fig_1}
\end{figure}

A confusion matrix is a visual representation of the performance of classification algorithms. This matrix displays the number of objects in each class based on the model's predictions. The diagonal elements represent each class's predicted and actual labels. As shown in Fig.~\ref{fig:Apendix_Fig_1}, the confusion matrix consists of true positives, false positives, true negatives, and false negatives. True Positive (TP): The number of samples predicted as positive, which is positive. False Positive (FP): The number of samples predicted as positive but negative. True Negative (TN): The number of samples predicted as negative and actual negative. False Negative (FN): The number of samples predicted to be negative but positive.\\
\section{Correlation Matrix} \label{sec:Correlation Matrix}
The correlation matrix shows the correlation between each feature and the other features. In this matrix, features with high correlation can be removed. A high correlation between two features indicates similar behavior, which can complicate the model. This matrix aims to reduce complexity as much as possible.
\begin{figure*}
\centering
  \includegraphics[width=0.95\linewidth]{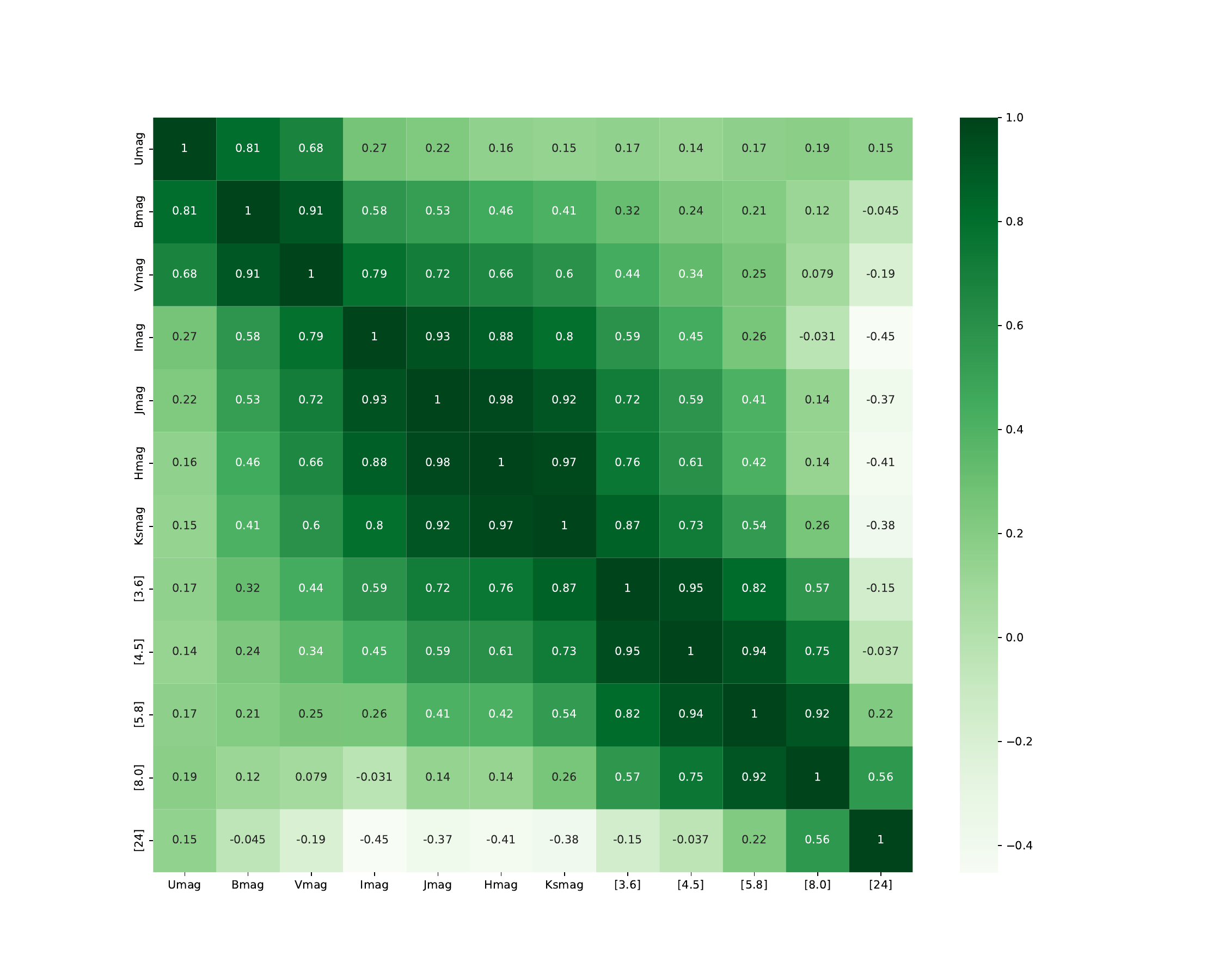} 
  \caption{Feature correlation through different filters from the SAGE spectral catalogs.}
  \label{fig:Correlation matrix of 12 features}
\end{figure*}
%
 \section{Pairplot} \label{sec:Pairs Plot}
The pairplot is a function that creates a grid of scatterplots and histograms for a given dataset derived from a data frame. This visualization allows for the examination of relationships between variables and data pairs. Additionally, it can identify correlations and outliers, which are needed for further analysis.
\begin{figure*}[h]
\centering
	\makebox[1\textwidth][c]
        {\includegraphics[width=1\textwidth]{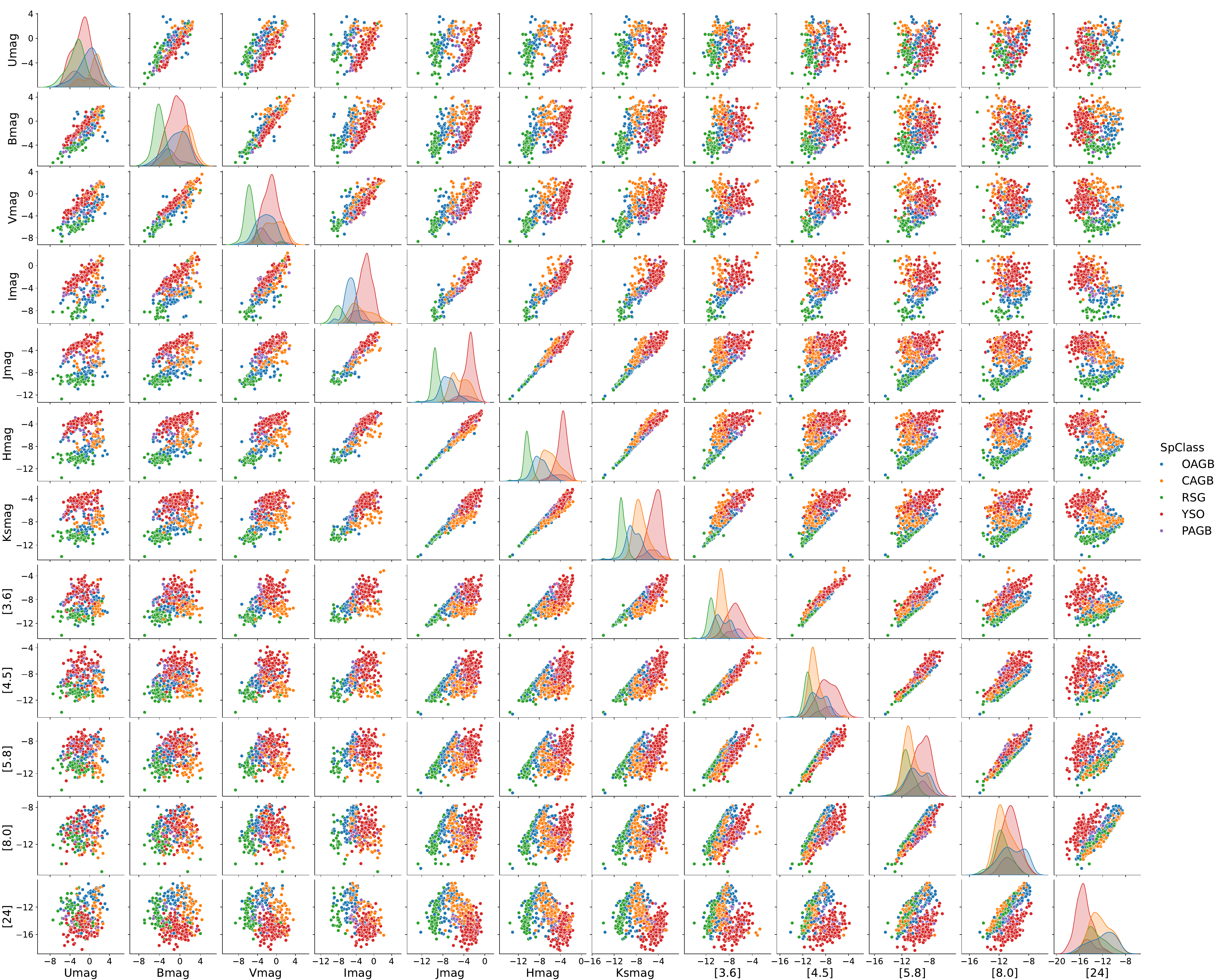}}
	\caption{Feature correlation. This pairplot illustrates the pairwise relationships between each feature. The plot shows the correlation between 12 features for five dusty stellar classes that have been included.}
        \label{fig:pair_plot}
\end{figure*}
 \section{supplemental materials}
\begin{figure*}
    \centering
    \hspace{0.07\linewidth}
    \includegraphics[width=0.35\linewidth,height=0.35\linewidth]{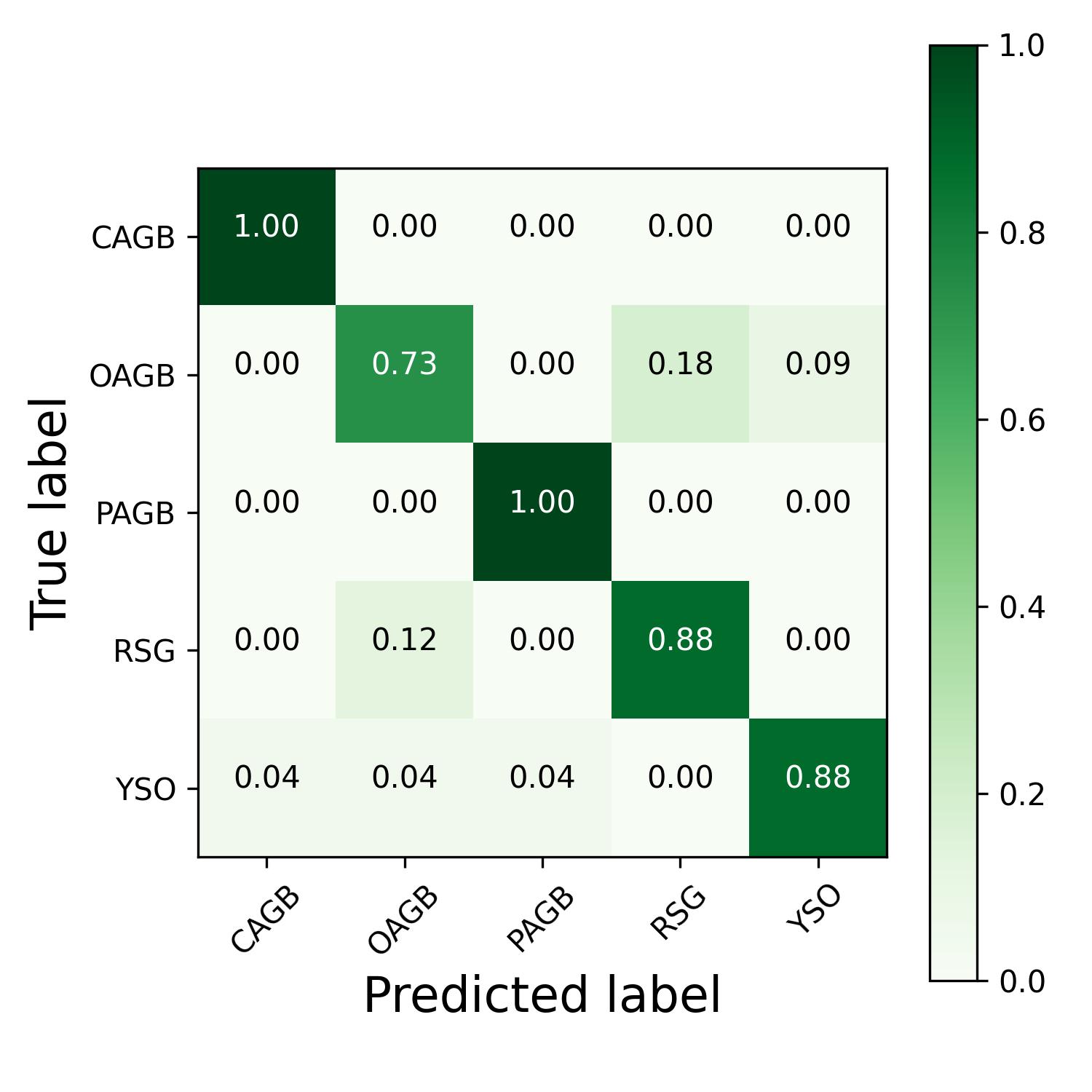}
    \hspace{0.05\linewidth}
    \includegraphics[width=0.35\linewidth,height=0.35\linewidth]{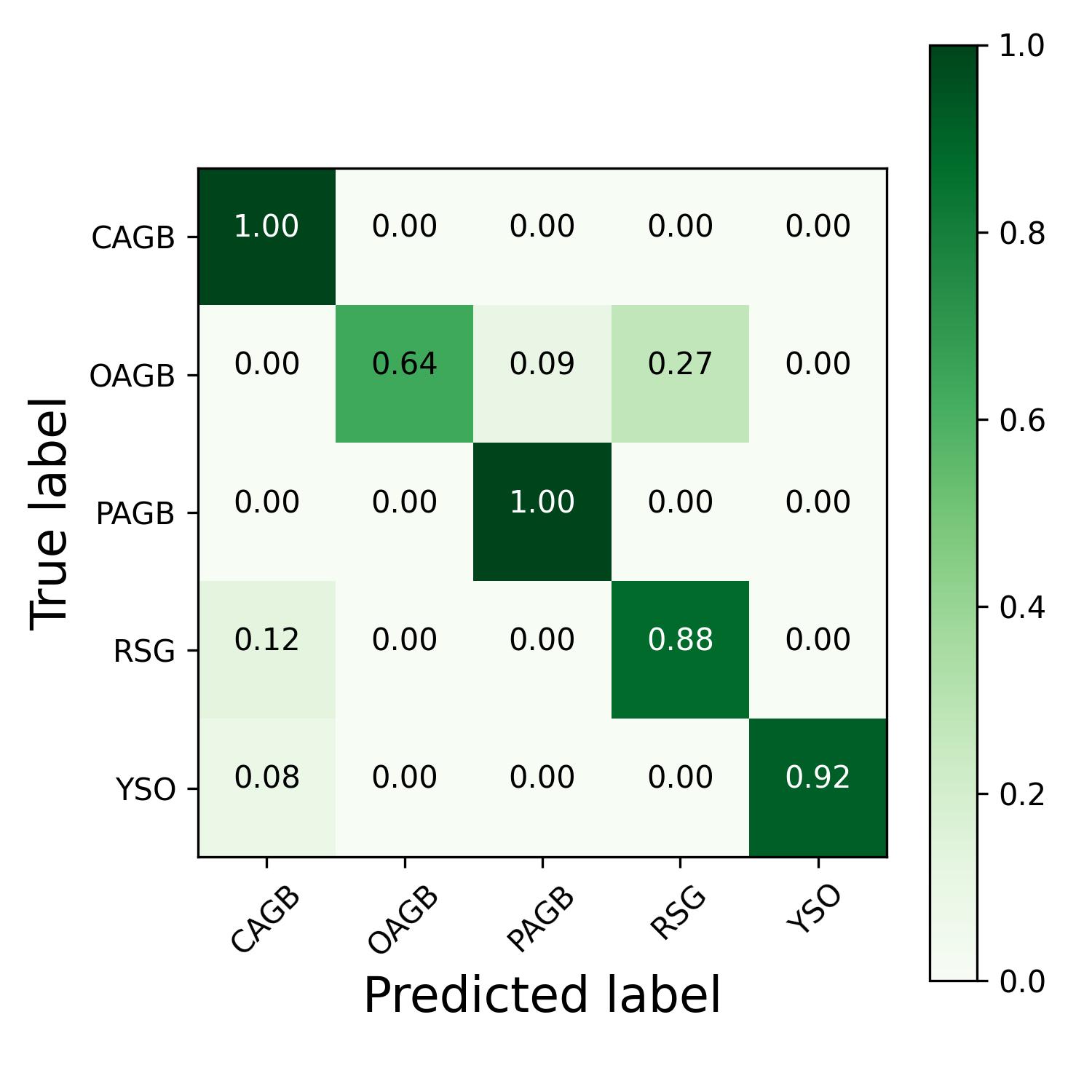}
            \hspace{0.05\linewidth}
    \parbox{0.4\linewidth}{
        \centering
(a) Probabilistic Random Forest (PRF)
    }

    \vspace{1cm} 

    \hspace{0.07\linewidth}
    \includegraphics[width=0.35\linewidth,height=0.35\linewidth]{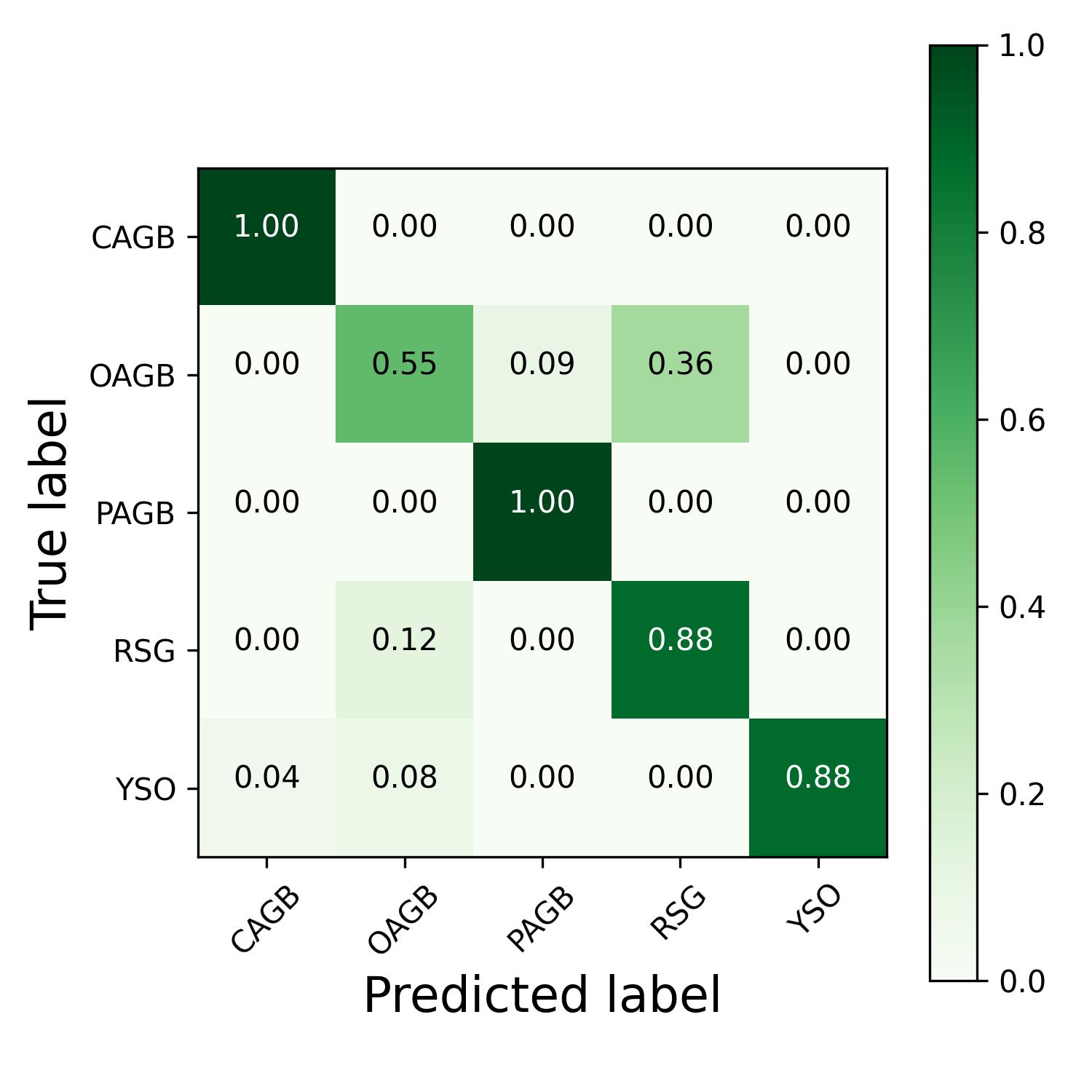}
    \hspace{0.05\linewidth}
    \includegraphics[width=0.35\linewidth,height=0.35\linewidth]{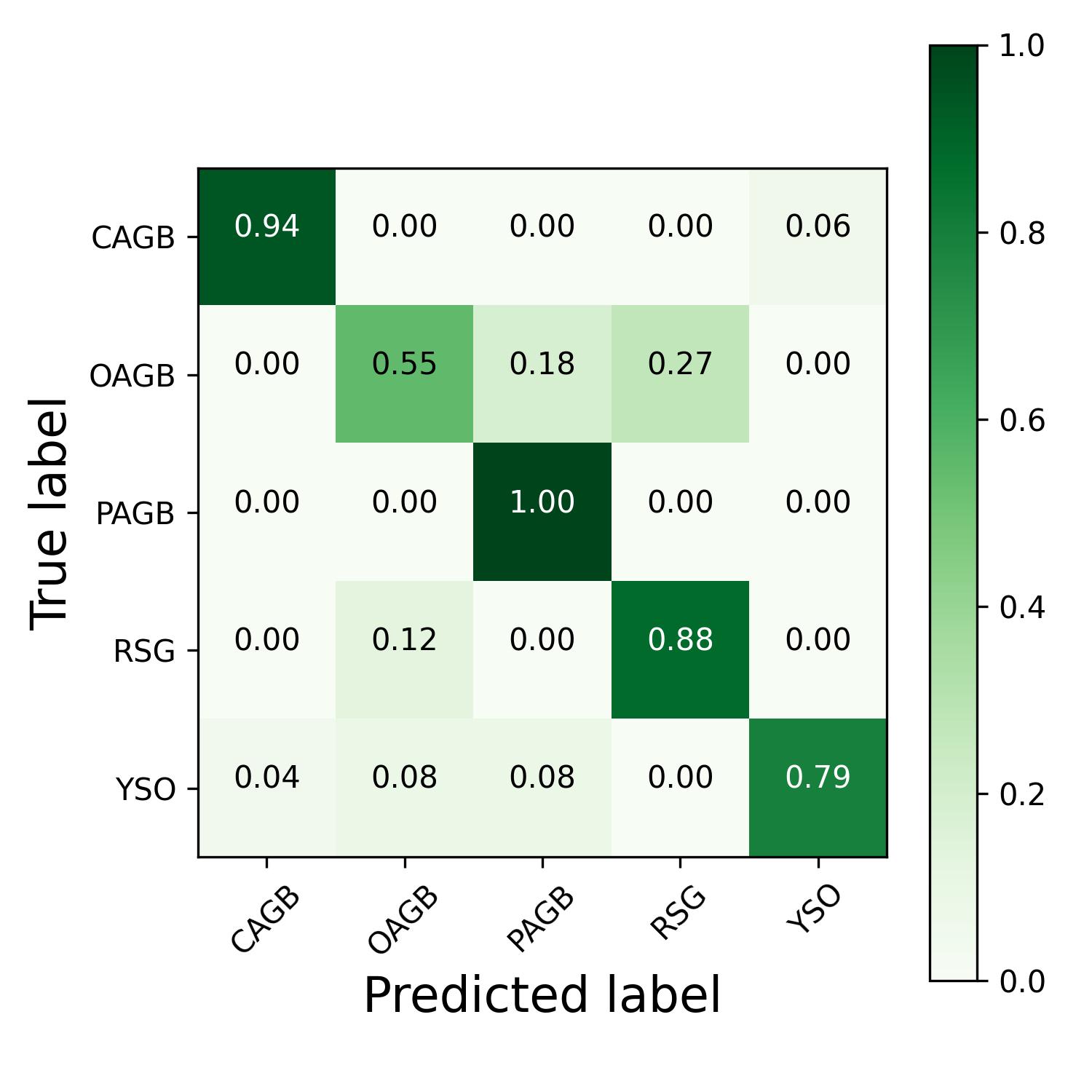}
                \hspace{0.05\linewidth}
    \parbox{0.4\linewidth}{
        \centering
(b) Random Forest (RF)
    }

    \vspace{1cm} 

    \hspace{0.07\linewidth}
    \includegraphics[width=0.35\linewidth,height=0.35\linewidth]{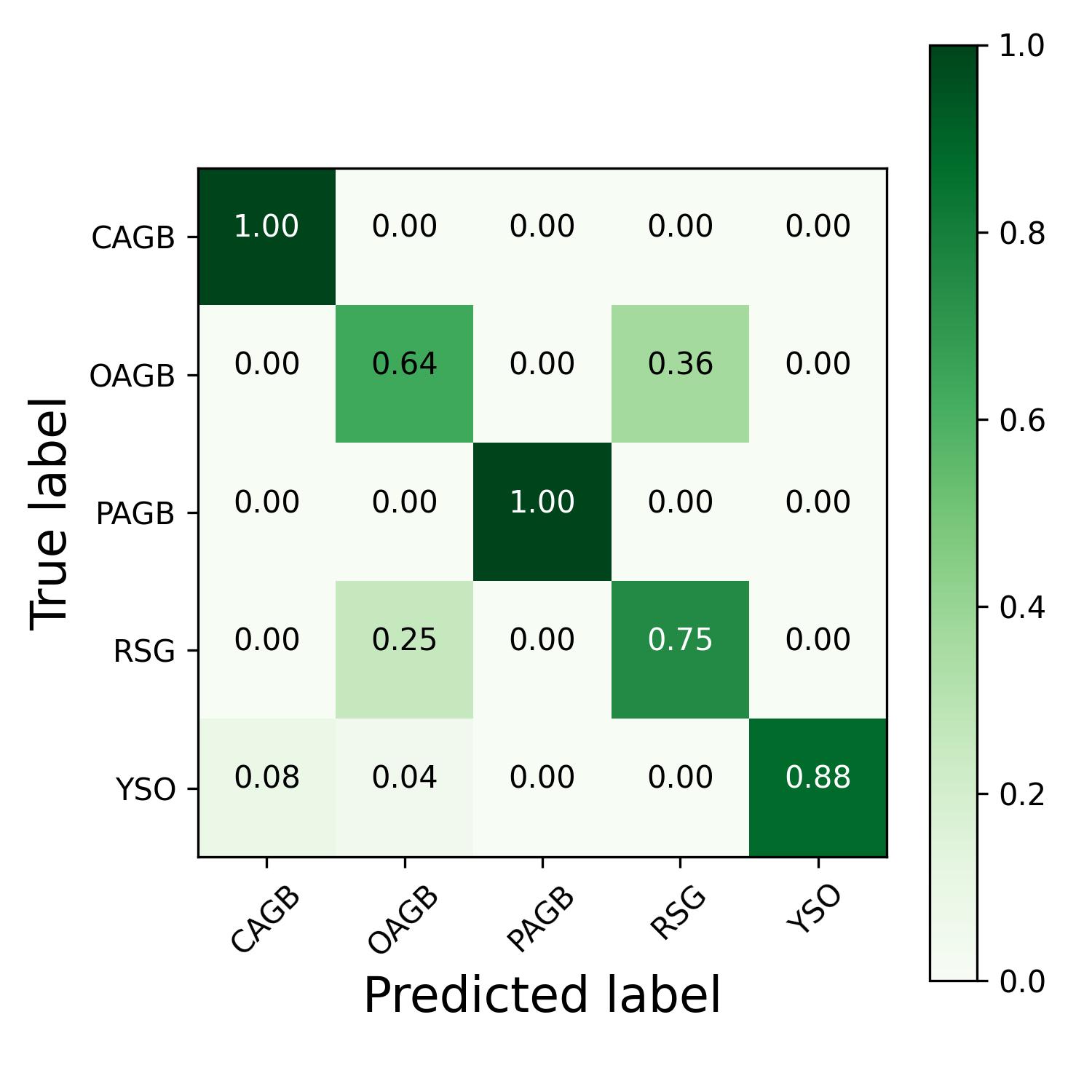}
    \hspace{0.05\linewidth}
    \includegraphics[width=0.35\linewidth,height=0.35\linewidth]{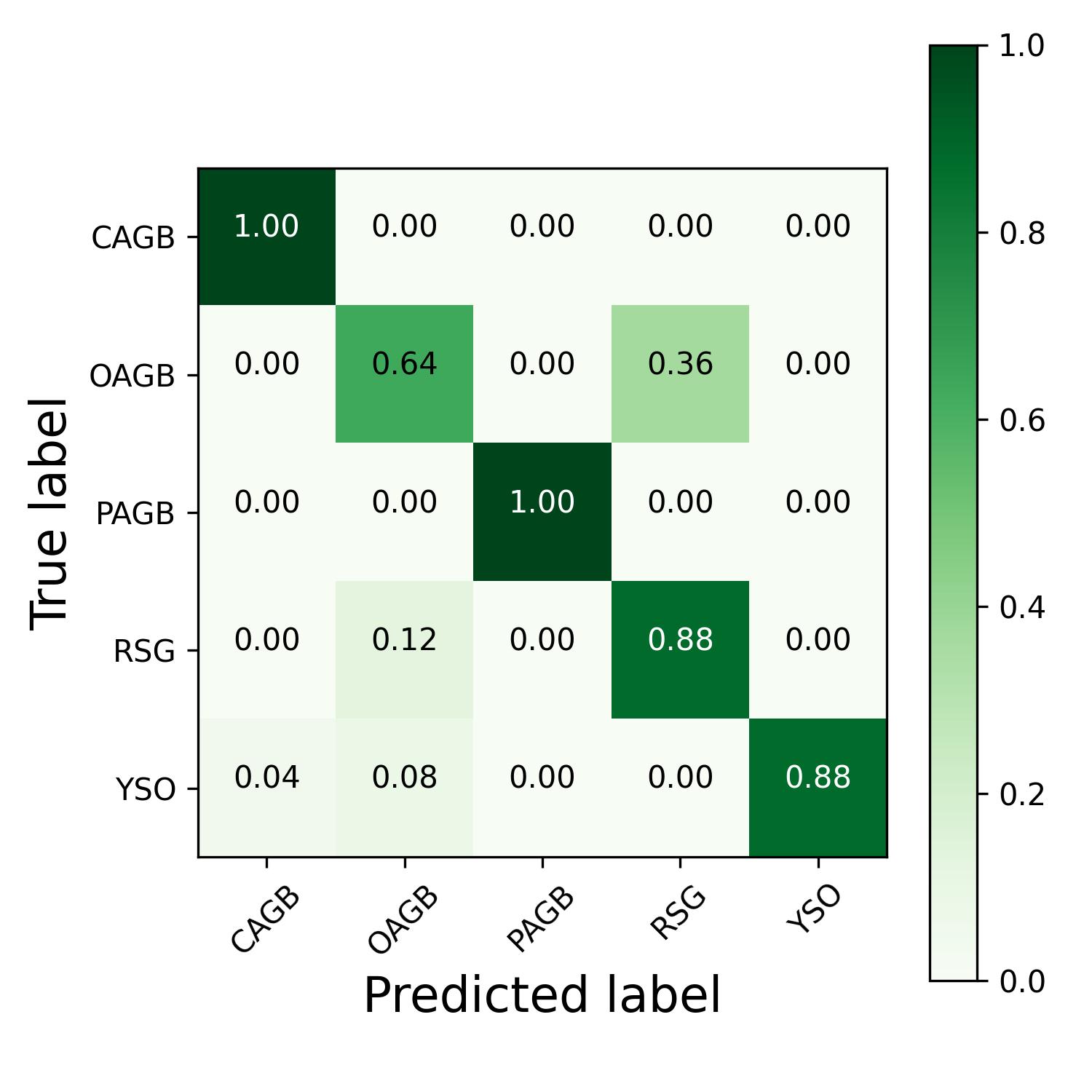}
                \hspace{0.05\linewidth}
    \parbox{0.4\linewidth}{
        \centering
(c) K-Nearest Neighbor (KNN)
    }

    \caption{Confusion matrix for different classifiers before (left panels) and after data augmentation (right panels).}
    \label{fig:CM_Set_1}    
\end{figure*}

\begin{figure*}
    \centering

    \hspace{0.07\linewidth}
    \includegraphics[width=0.35\linewidth,height=0.35\linewidth]{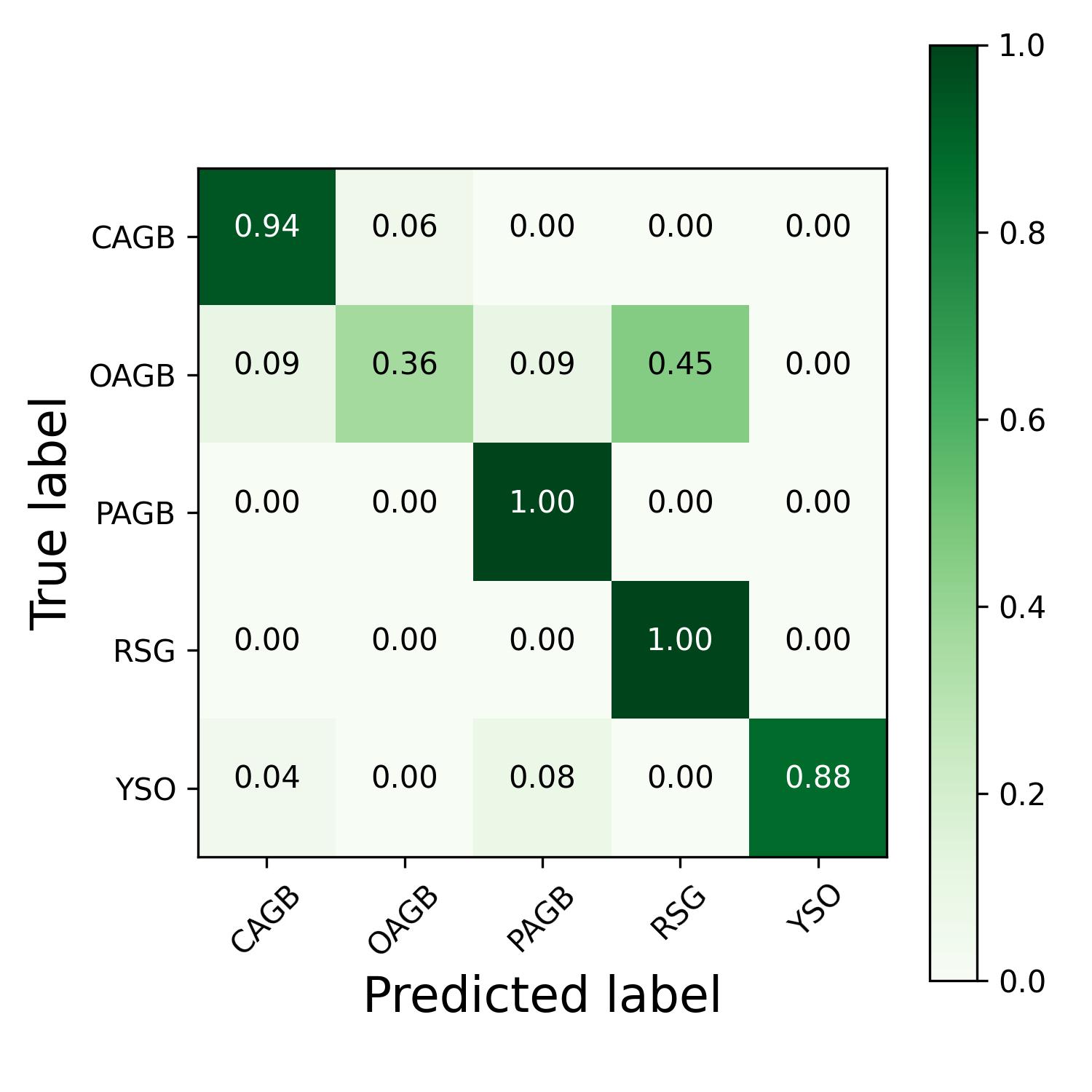}
    \hspace{0.05\linewidth}
    \includegraphics[width=0.35\linewidth,height=0.35\linewidth]{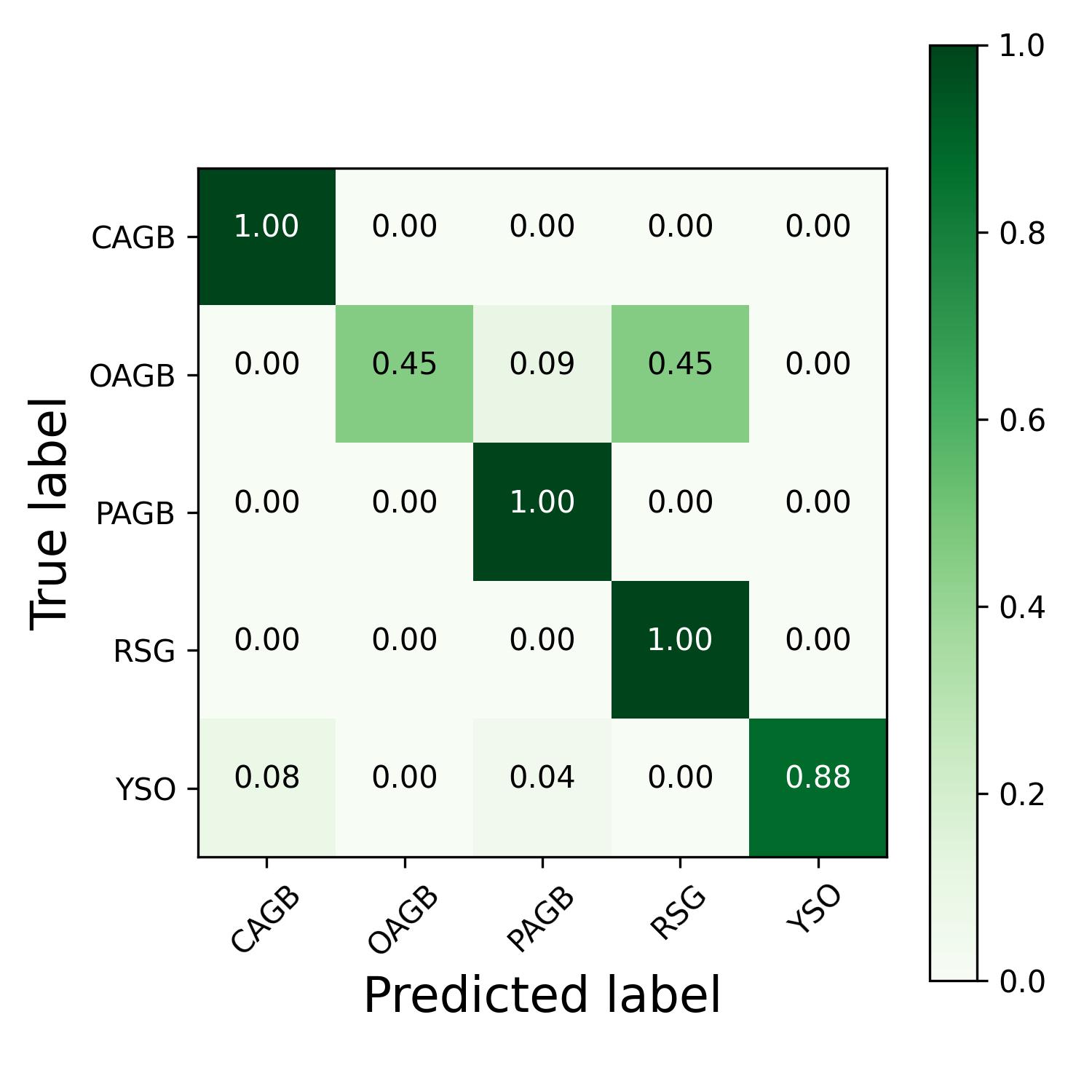}
        \hspace{0.05\linewidth}
    \parbox{0.4\linewidth}{
        \centering
(a) Support Vector Classification (SVC), with ``poly'' kernel.
    }

    \vspace{1cm} 

    \hspace{0.07\linewidth}
    \includegraphics[width=0.35\linewidth,height=0.35\linewidth]{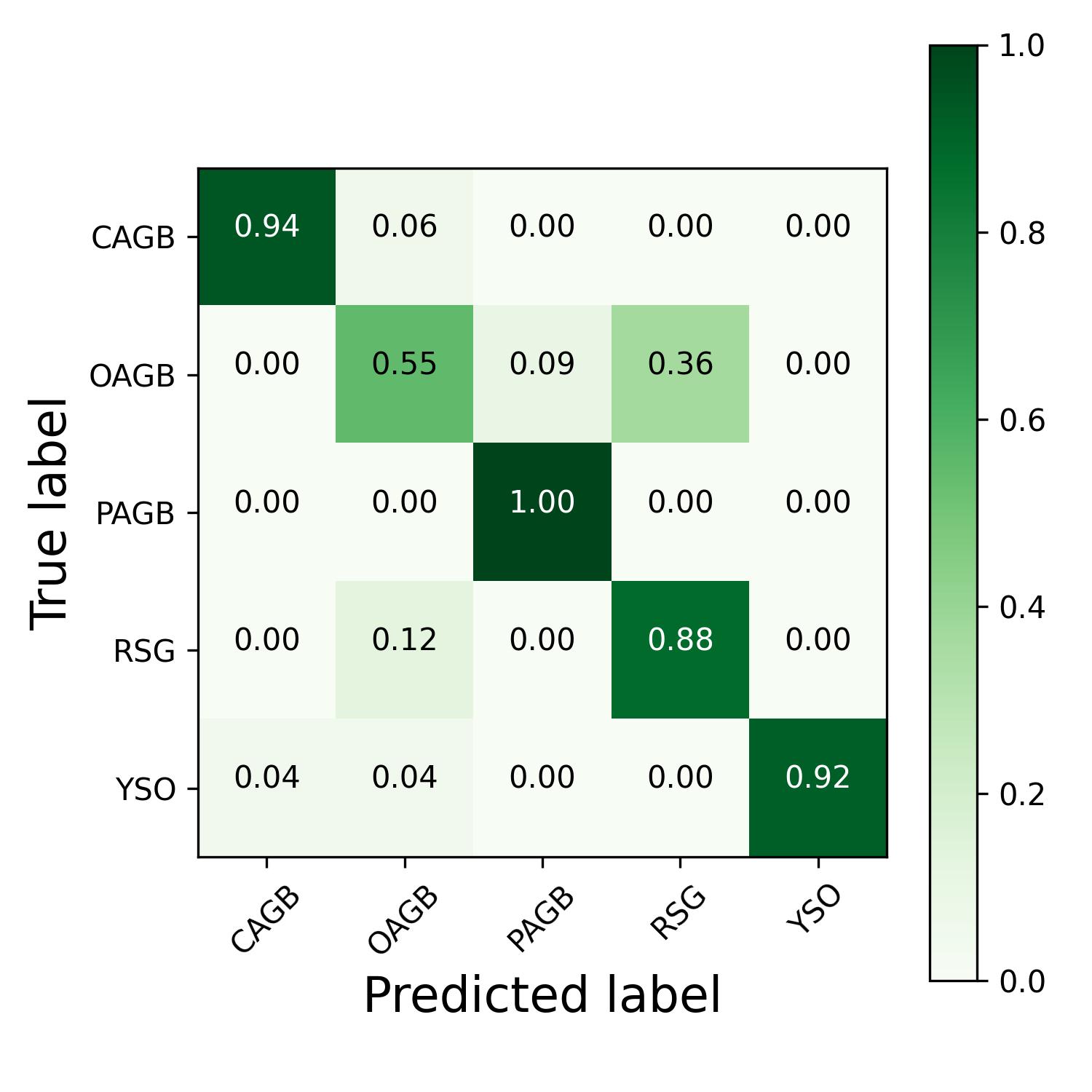}
    \hspace{0.05\linewidth}
    \includegraphics[width=0.35\linewidth,height=0.35\linewidth]{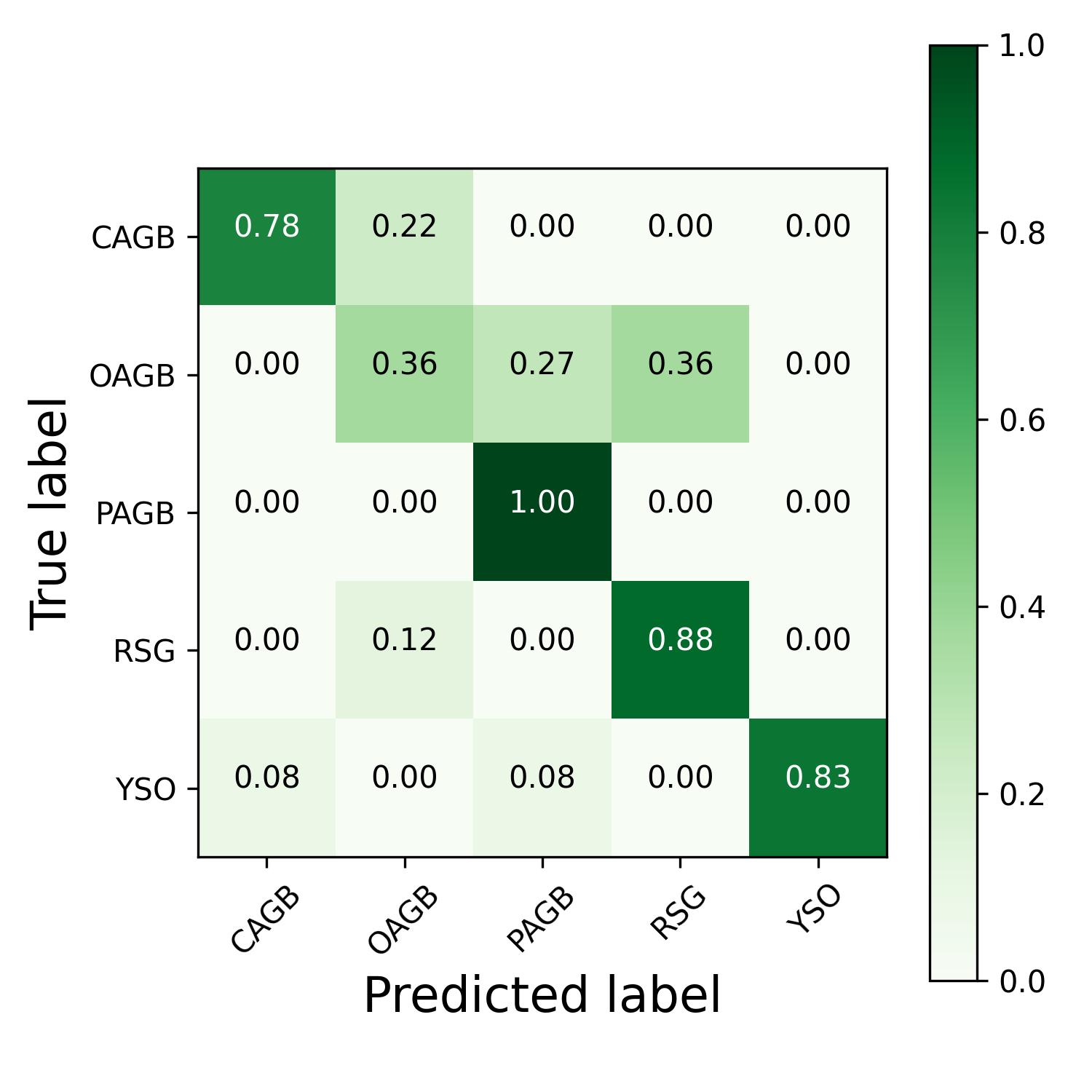}
            \hspace{0.05\linewidth}
    \parbox{0.4\linewidth}{
        \centering
(b) Support Vector Classification (SVC), with ``rbf'' kernel.
    }

    \vspace{1cm} 

    \hspace{0.07\linewidth}
    \includegraphics[width=0.35\linewidth,height=0.35\linewidth]{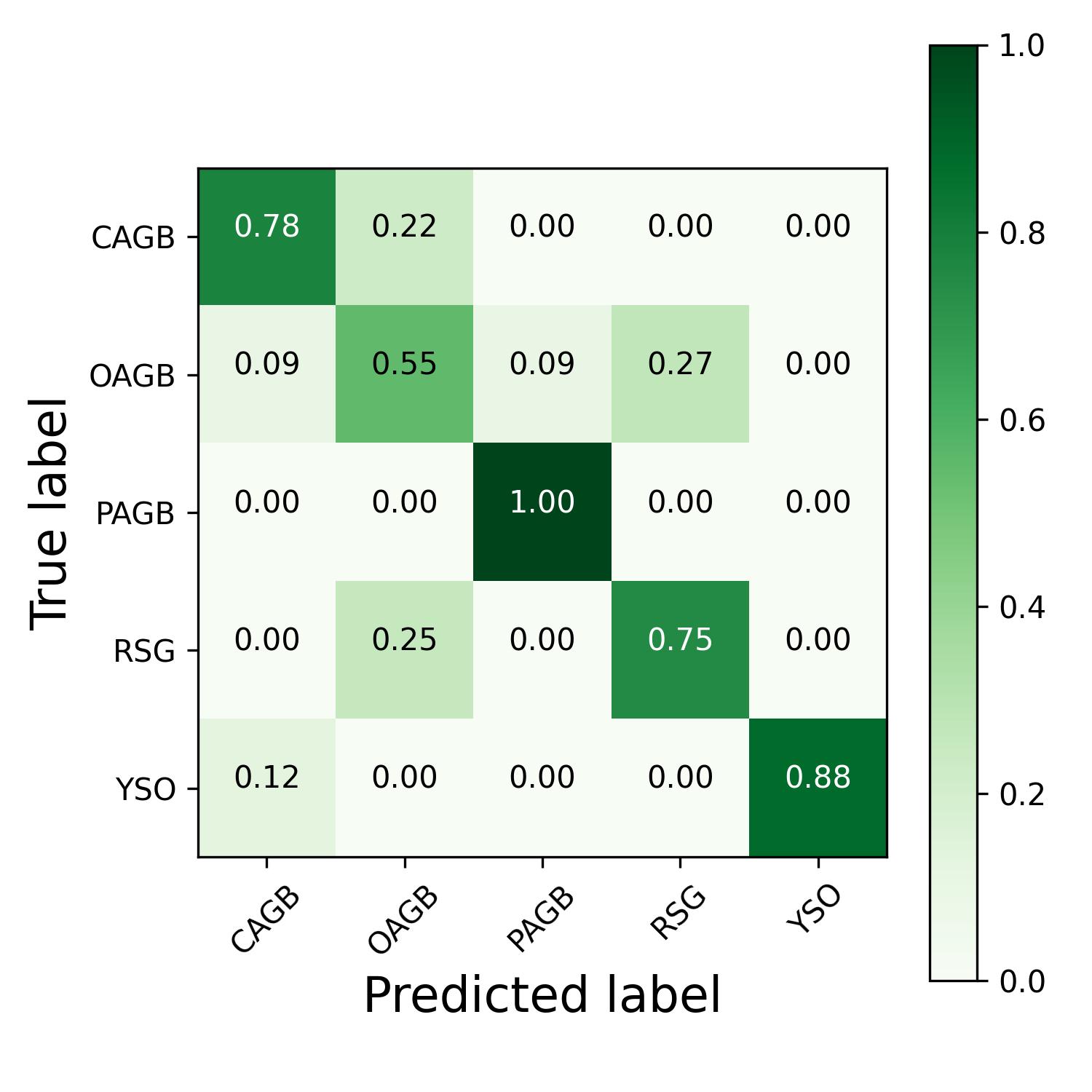}
    \hspace{0.05\linewidth}
    \includegraphics[width=0.35\linewidth,height=0.35\linewidth]{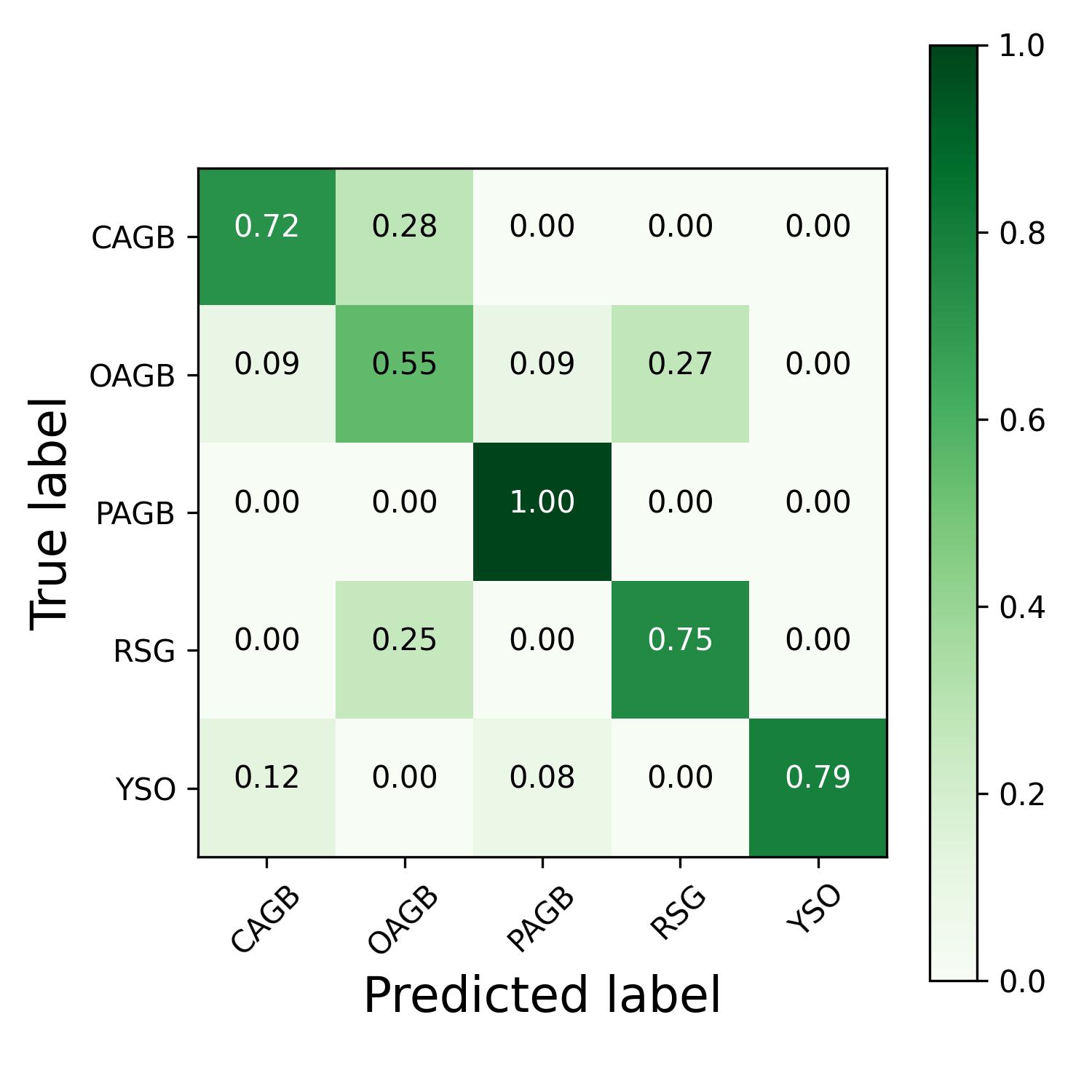}
            \hspace{0.05\linewidth}
    \parbox{0.4\linewidth}{
        \centering
(c) Gaussian Naive Bayes (GNB).
    }

    \caption{Confusion matrix for different classifiers before (left panels) and after data augmentation (right panels).}
    \label{fig:CM_Set_2}
\end{figure*}

\begin{figure*}
    \centering

    \hspace{0.04\linewidth}
    \includegraphics[width=0.4\linewidth]{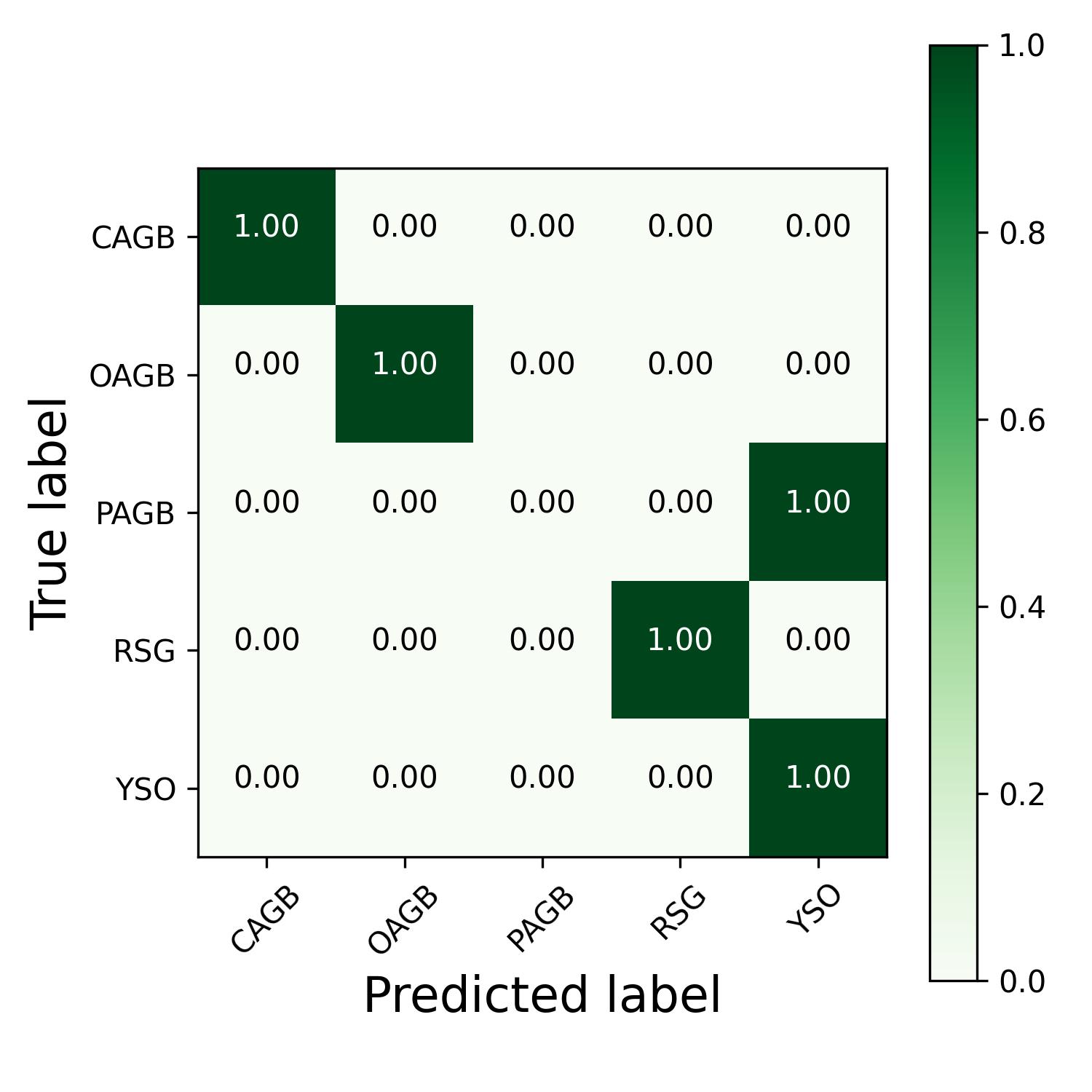}
    \hspace{0.05\linewidth}
    \includegraphics[width=0.4\linewidth]{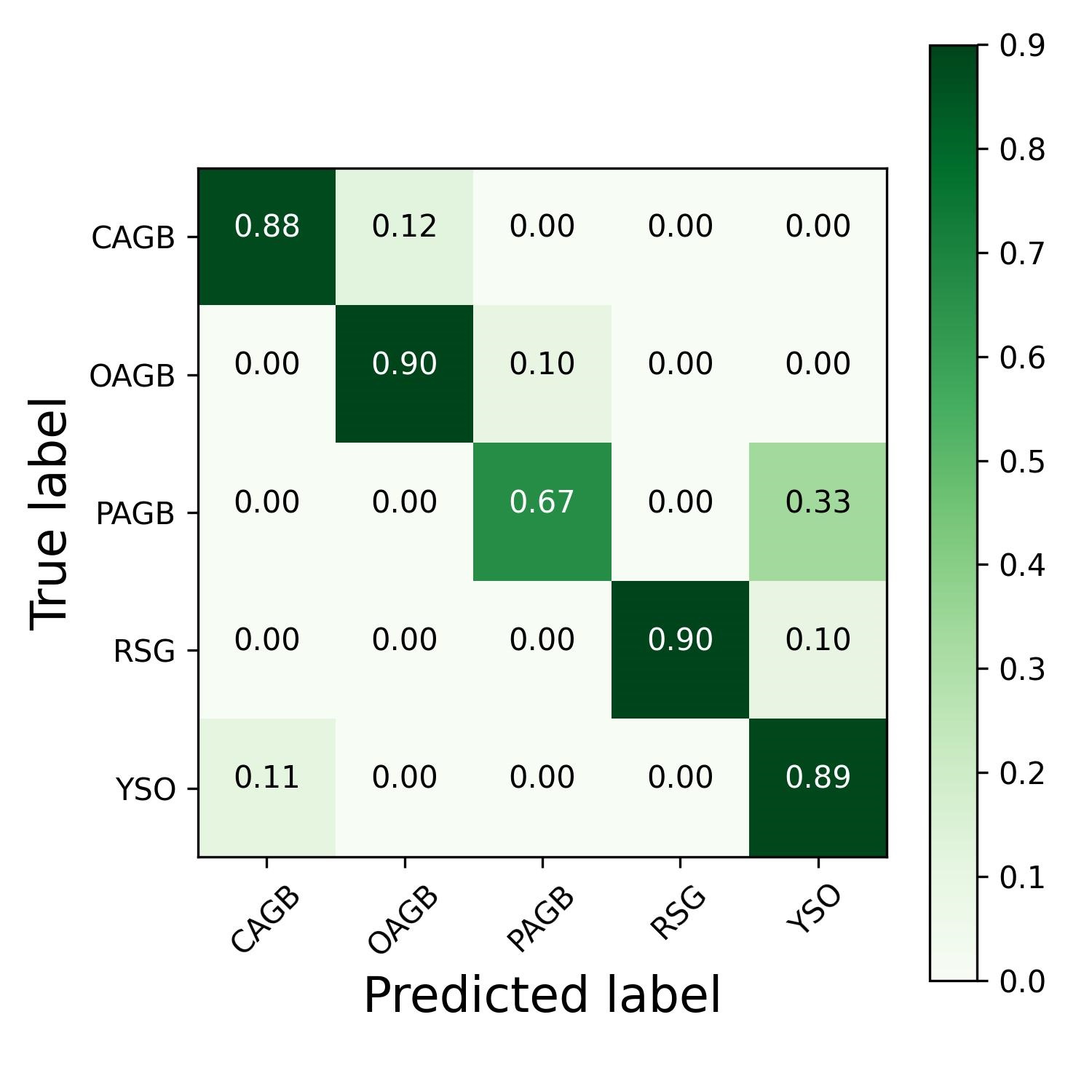}
    \\
    \parbox{0.4\linewidth}{
        \centering
        (a) Location = SMC
    }
    \hspace{0.05\linewidth}
    \parbox{0.4\linewidth}{
        \centering
        (b) Location = LMC
    }
    \caption{Confusion matrices for the best-performing classifier (PRF) applied separately to the SMC (left) and LMC (right) datasets, rather than using a combined dataset, to account for metallicity differences between the two galaxies. This classification was performed considering five stellar classes. The settings for this classification are: Number of Class = 5, Keep\_Probability = 0.8, Number of Estimator = 10, and SMOTE = No.}
    \label{fig:CM-Metallicity}
\end{figure*}

\begin{figure*}
    \centering
    \includegraphics[width=0.33\linewidth]{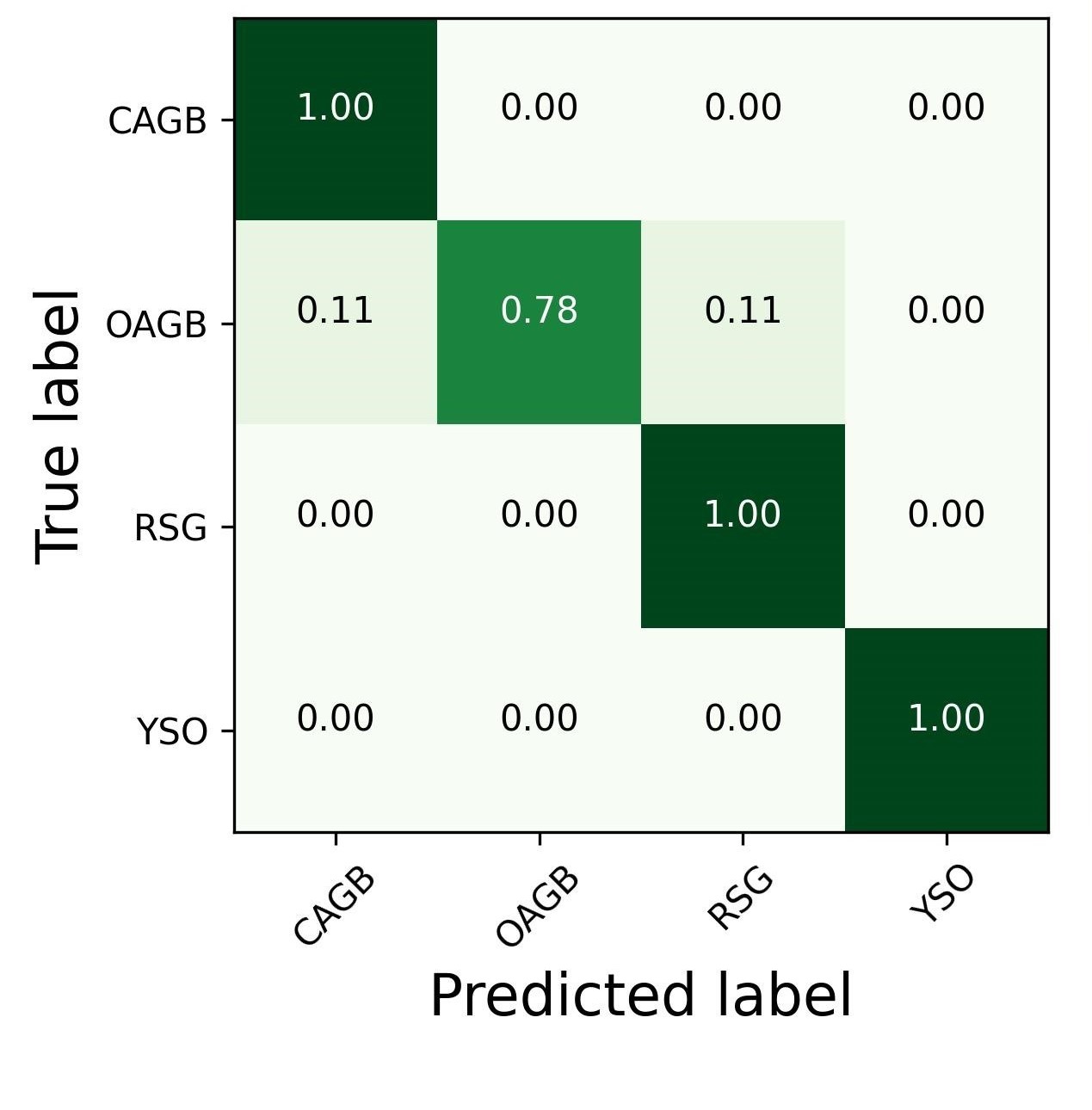}
    \hspace{0.02\linewidth}
    \includegraphics[width=0.3\linewidth]{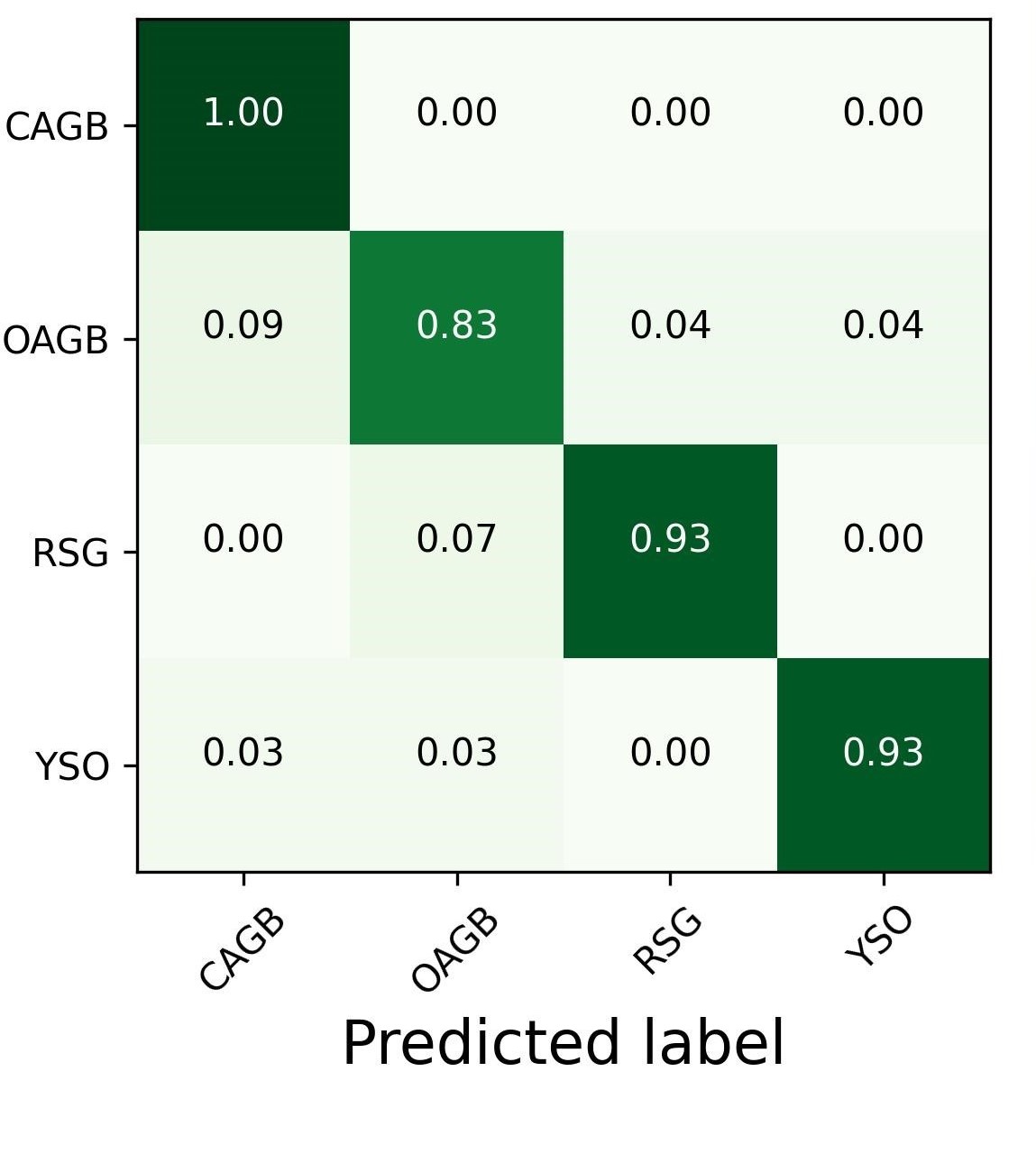}
    \hspace{0.02\linewidth}
    \includegraphics[width=0.3\linewidth]{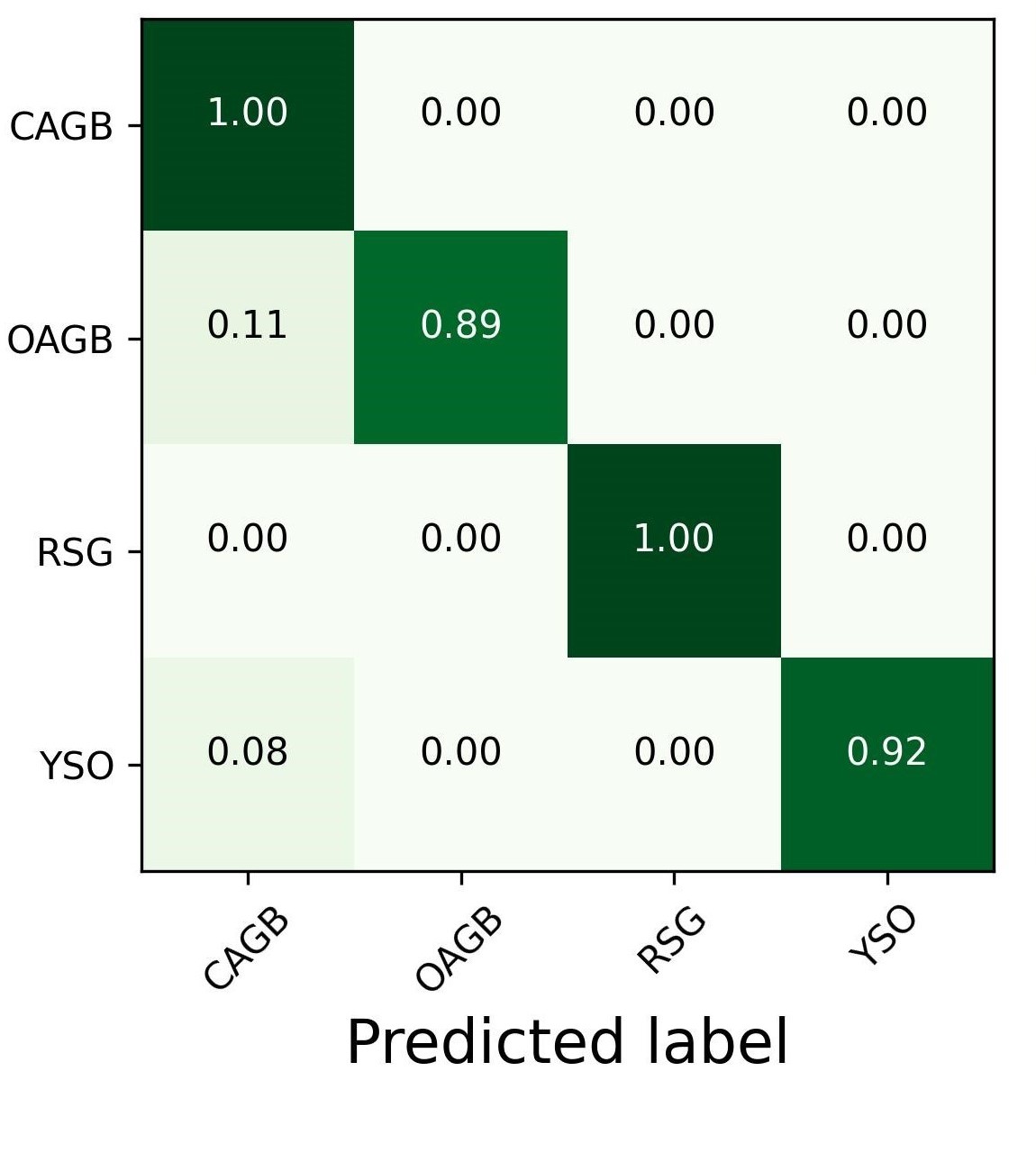}
    \\
    \parbox{0.33\linewidth}{
        \centering
        (a) Location = SMC
    }
    \hspace{0.02\linewidth}
    \parbox{0.3\linewidth}{
        \centering
        (b) Location = LMC
    }
    \hspace{0.02\linewidth}
    \parbox{0.3\linewidth}{
        \centering
        (c) Location = LMC \& SMC
    }
    \caption{Confusion matrices for the best-performing classifier (PRF) applied separately to the SMC (left), LMC (middle), and the combined LMC \& SMC dataset (right) to assess the impact of metallicity differences between the two galaxies. The model was trained with four stellar classes (see Table~\ref{tab:classification_report_SMC_4class}, Table~\ref{tab:classification_report_LMC_4class}, Table~\ref{tab:classification_report_4class} for more details. The settings for this classification are:  Number of Class = 4, Keep\_probability = 0.8, Number of estimator = 10, and SMOTE = No.}
    \label{fig:CM-Metallicity-4Class}
\end{figure*}

\begin{figure*}
    \centering

    \hspace{0.04\linewidth}
    \includegraphics[width=0.4\linewidth]{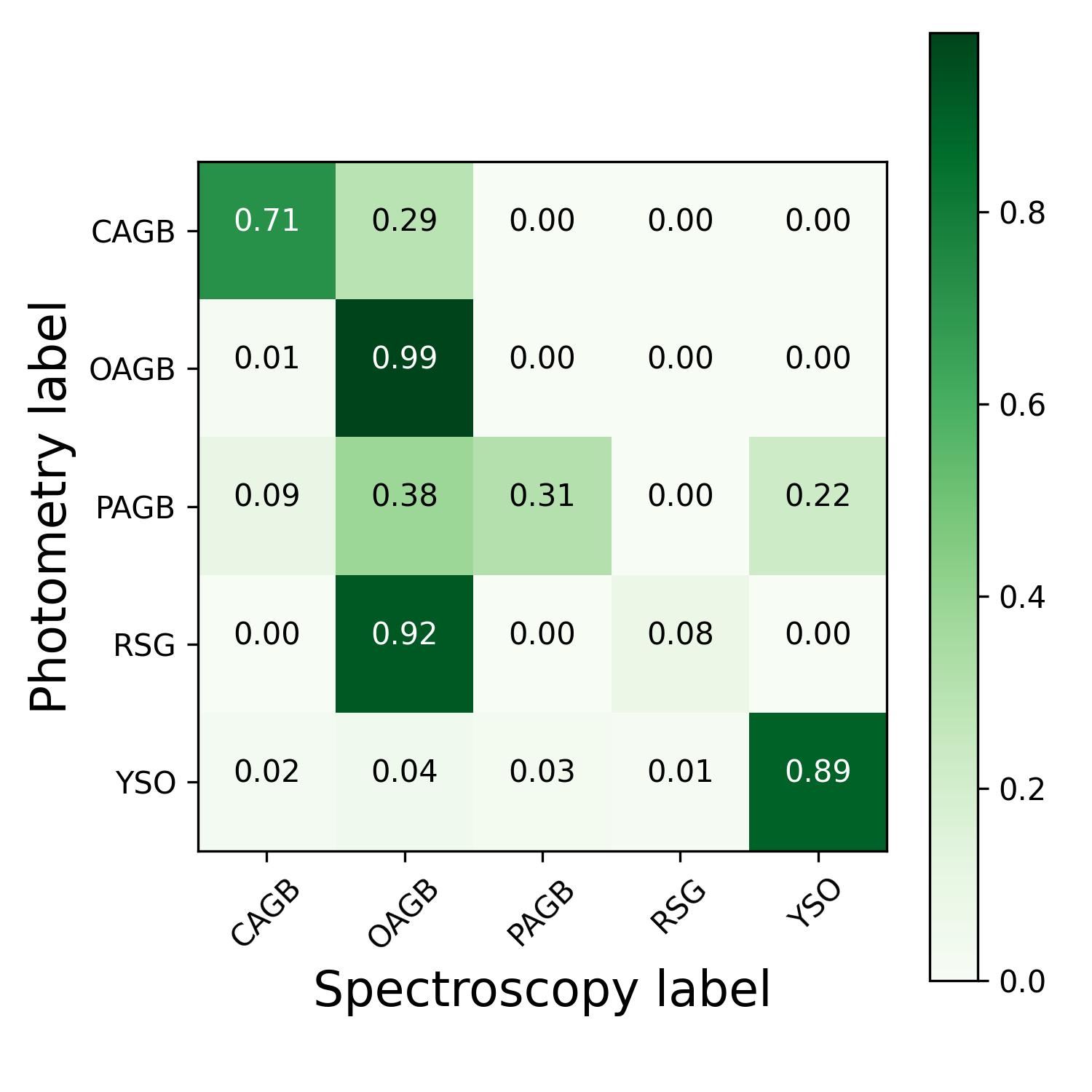}
    \hspace{0.05\linewidth}
    \includegraphics[width=0.4\linewidth]{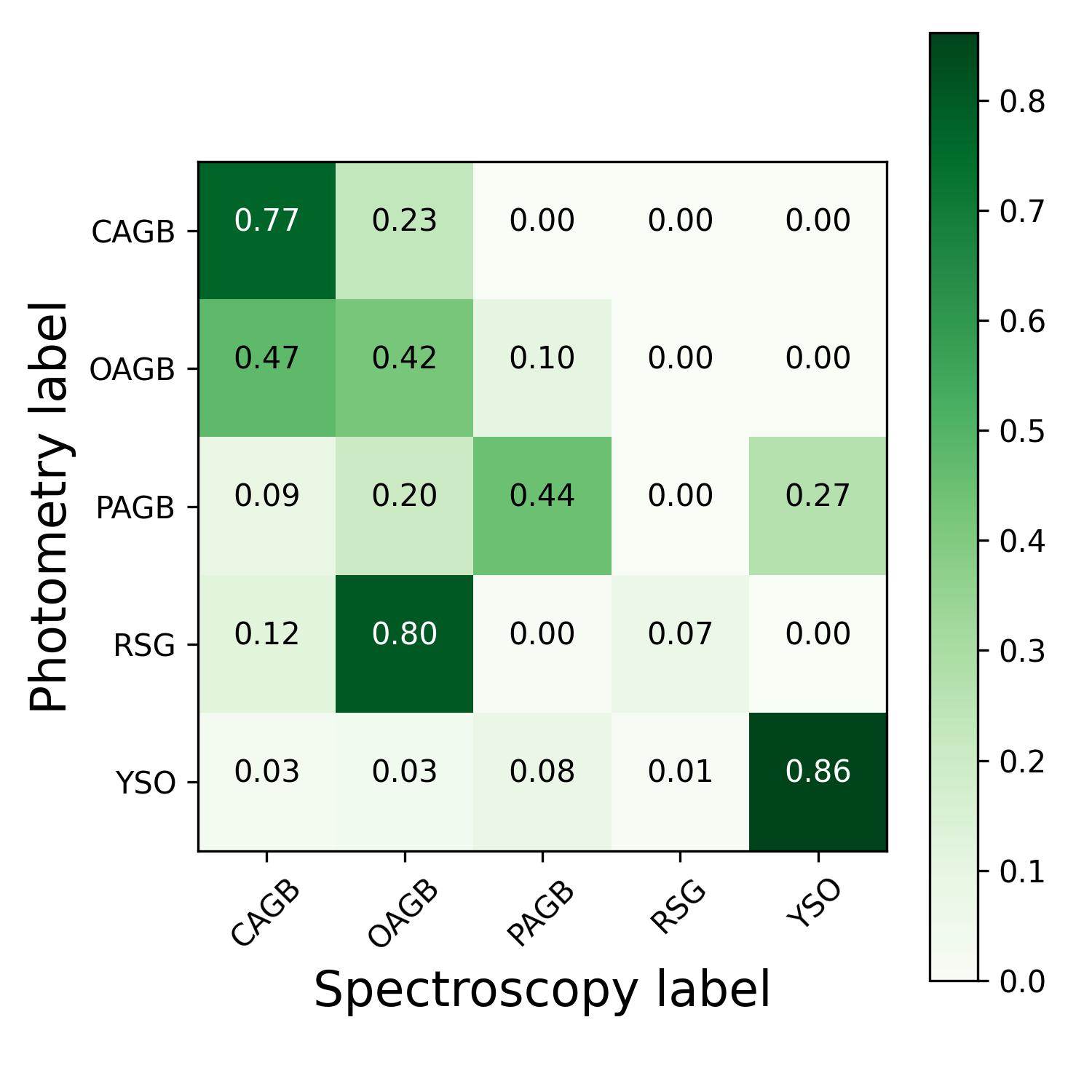}
    \\
    \parbox{0.4\linewidth}{
        \centering
        (a) Keep\_probability = 0.8\\
        Number of estimator = 10\\
        SMOTE = No
    }
    \hspace{0.05\linewidth}
    \parbox{0.4\linewidth}{
        \centering
        (b) Keep\_probability = 0.1\\
        Number of estimator = 10\\
        SMOTE = No
    }

    \vspace{1cm} 

    \hspace{0.04\linewidth}
    \includegraphics[width=0.4\linewidth]{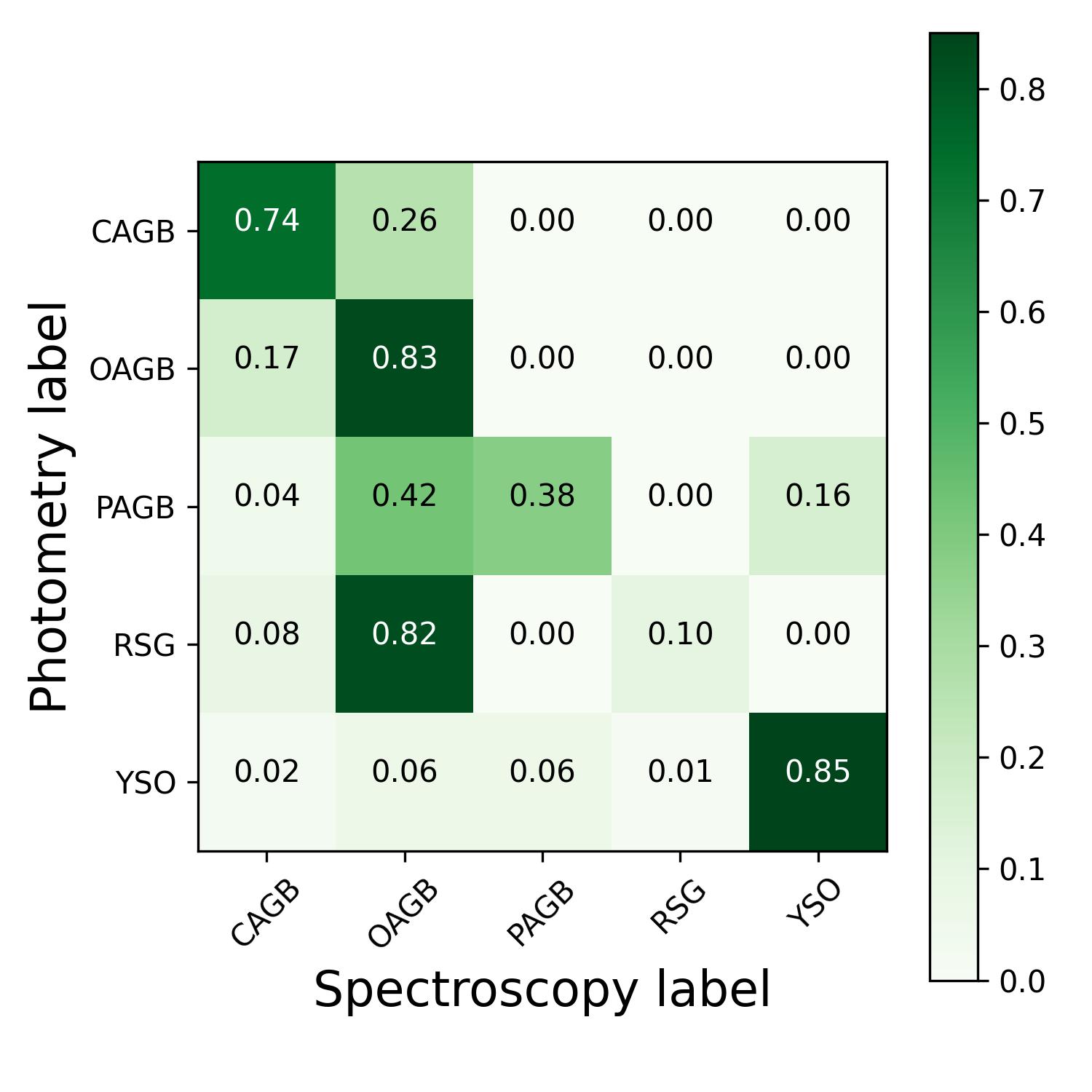}
    \hspace{0.05\linewidth}
    \includegraphics[width=0.4\linewidth]{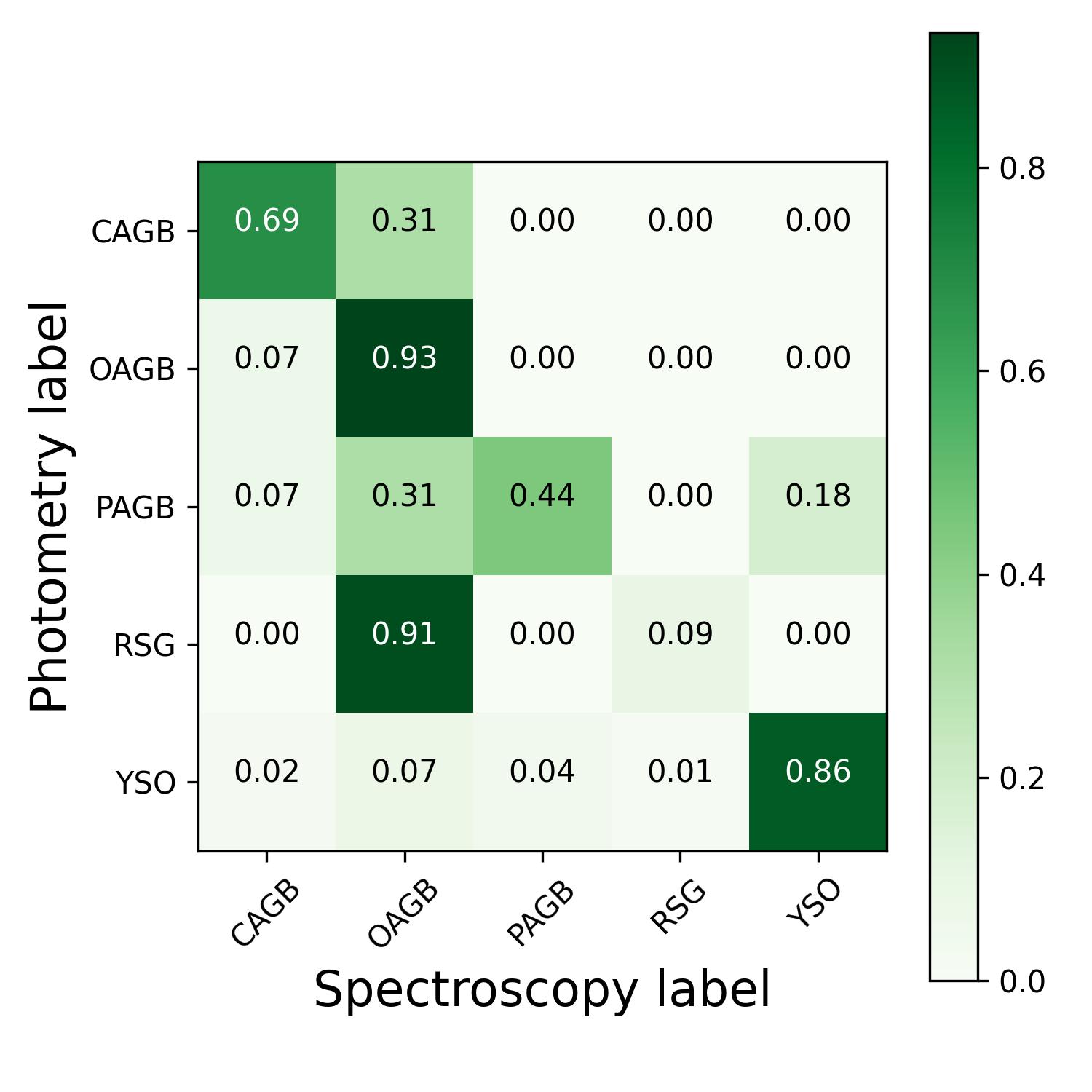}
    \\
    \parbox{0.4\linewidth}{
        \centering
        (c) Keep\_probability = 0.5\\
        Number of estimator = 10\\
        SMOTE = Yes
    }
    \hspace{0.05\linewidth}
    \parbox{0.4\linewidth}{
        \centering
        (d) Keep\_probability = 0.1\\
        Number of estimator = 10\\
        SMOTE = Yes
    }

    \caption{Comparison matrices of selected models for comparison to photometrically labeled data.}
    \label{fig:Comparison_matrix}
\end{figure*}

The results of confusion matrices for training different classifiers before and after augmenting the data using the SMOTE method are presented in Fig.~\ref{fig:CM_Set_1} and Fig.~\ref{fig:CM_Set_2}. The classification reports under different settings, as discussed in Section~\ref{sec:Metallicity Impact}, which examines the effect of metallicity, are presented in Table~\ref{tab:classification_report_SMC_4class}, Table~\ref{tab:classification_report_LMC_4class}, Table~\ref{tab:classification_report_4class}, and Table~\ref{tab:classification_report_LMC_Train_SMC_test_4class}. The corresponding confusion matrices are shown in Fig.\ref{fig:CM-Metallicity} and Fig.\ref{fig:CM-Metallicity-4Class}. In addition, the comparison matrices of selected models for comparison to photometrically labeled data are shown in Fig.~\ref{fig:Comparison_matrix}.
%
%
\begin{table}
    \vspace{2cm}
    \centering
    \label{tab:classification_reports_metallicity}
    \begin{minipage}{0.48\textwidth}
        \centering
        \caption{Classification report for the SMC catalog, including four classes of dusty stellar objects (see the left panel of Fig.~\ref{fig:CM-Metallicity-4Class}). The classification was performed using the following settings: number of estimators = 10, Keep probability = 0.8, and Simple PRF.}
        \label{tab:classification_report_SMC_4class}
        \begin{tabular}{lccc}
            \toprule
            Class & Precision & Recall & F1-Score \\
            \midrule
            CAGB & 0.75 & 1.00 & 0.86 \\
            OAGB & 0.86 & 0.78 & 0.88 \\
            RSG & 0.67 & 1.00 & 0.80 \\
            YSO & 1.00 & 1.00 & 1.00 \\
            \midrule
            accuracy & & & 0.92 \\
            macro avg & 0.85 & 0.94 & 0.88 \\
            weighted avg & 0.95 & 0.92 & 0.92 \\
            \bottomrule 
        \end{tabular}
    \end{minipage}
    \hfill
    \begin{minipage}{0.48\textwidth}
        \centering
        \caption{Classification report for the LMC catalog, including four classes of dusty stellar objects (see the middle panel of Fig.~\ref{fig:CM-Metallicity-4Class}). The classification was performed using the following settings: number of estimators = 10, Keep probability = 0.8, and Simple PRF.}
        \label{tab:classification_report_LMC_4class}
        \begin{tabular}{lccc}
            \toprule
            Class & Precision & Recall & F1-Score \\
            \midrule
            CAGB & 0.88 & 1.00 & 0.94 \\
            OAGB & 0.90 & 0.83 & 0.86 \\
            RSG & 0.93 & 0.93 & 0.93 \\
            YSO & 0.97 & 0.93 & 0.95 \\
            \midrule
            accuracy & & & 0.92 \\
            macro avg & 0.92 & 0.92 & 0.92 \\
            weighted avg & 0.92 & 0.92 & 0.92 \\
            \bottomrule 
        \end{tabular}
    \end{minipage}
    
    \vspace{2cm} 

    \begin{minipage}{0.48\textwidth}
        \centering
        \caption{Classification report for the master dataset (LMC \& SMC), including four classes of dusty stellar objects (see the right panel of Fig.~\ref{fig:CM-Metallicity-4Class}). The classification was performed using the following settings: number of estimators = 10, Keep probability = 0.8, and Simple PRF}
        \label{tab:classification_report_4class}
        \begin{tabular}{lccc}
            \toprule
            Class & Precision & Recall & F1-Score \\
            \midrule
            CAGB & 0.60 & 1.00 & 0.75 \\
            OAGB & 1.00 & 0.89 & 0.94 \\
            RSG & 1.00 & 1.00 & 1.00 \\
            YSO & 1.00 & 0.92 & 0.96 \\
            \midrule
            accuracy & & & 0.92 \\
            macro avg & 0.90 & 0.95 & 0.91 \\
            weighted avg & 0.95 & 0.92 & 0.93 \\
            \bottomrule 
        \end{tabular}
    \end{minipage}
    \hfill
    \begin{minipage}{0.48\textwidth}
        \centering
        \caption{Classification report for the case where the model was trained on LMC data and tested on SMC data, including four classes of dusty stellar objects (see Fig.~\ref{fig:CM_LMC_train_SMC_test}). The classification was performed using the following settings: number of estimators = 10, Keep probability = 0.8, and Simple PRF.}
        \label{tab:classification_report_LMC_Train_SMC_test_4class}
        \begin{tabular}{lccc}
            \toprule
            Class & Precision & Recall & F1-Score \\
            \midrule
            CAGB & 0.75 & 1.00 & 0.86 \\
            OAGB & 0.86 & 0.78 & 0.88 \\
            RSG & 0.67 & 1.00 & 0.80 \\
            YSO & 1.00 & 1.00 & 1.00 \\
            \midrule
            accuracy & & & 0.92 \\
            macro avg & 0.85 & 0.94 & 0.88 \\
            weighted avg & 0.95 & 0.92 & 0.92 \\
            \bottomrule 
        \end{tabular}
    \end{minipage}
\end{table}
\clearpage

\end{document}